\documentclass[11pt]{article}
\usepackage[utf8]{inputenc}
\usepackage{xspace}
\usepackage[english]{babel}
\usepackage{amsmath,amssymb,amsfonts,epic}
\usepackage{textcomp}
\usepackage{graphicx,psfrag,epsf,epstopdf}
\usepackage{caption, subcaption}
\usepackage{fancybox}
\usepackage{amsbsy}
\usepackage{amscd}
\usepackage{amsgen}
\usepackage{amsopn}
\usepackage{amstext}
\usepackage{ifthen}
\usepackage{amsxtra}
\usepackage{color}
\usepackage{fancyhdr}
\usepackage{fullpage}
\usepackage{float}
\usepackage{natbib}
\usepackage{rotating}
\usepackage{hyperref}

\usepackage{authblk}  

\usepackage{setspace}

\newdimen \extrarowheight

\newcommand{\R}{\mathbb{R}}

\newcommand{\I}{\mathcal{I}}

\newtheorem{prop}{Proposition}

\newtheorem{remark}{Remark}

\newenvironment{dem}[1][Proof] {{\textbf{#1. }}}{\hfill $\square$}

\title{Visualizing Outliers in High Dimensional Functional Data for Task fMRI data exploration}

\author[1]{Yasser Alemán-Gómez\thanks{yasseraleman@gmail.com}}
\author[2]{Ana Arribas-Gil\thanks{ana.arribas@uc3m.es}}
\author[3,4]{Manuel Desco\thanks{mdesco@hggm.es}}
\author[5]{Antonio Elías\thanks{antonio.elias@uc3m.es}}
\author[5]{ \hspace*{0.5cm}Juan Romo\thanks{juan.romo@uc3m.es}}
\affil[1]{{\small Medical Image Analysis Laboratory, University of Lausanne, Lausanne, Switzerland}}
\affil[2]{{\small Instituto UC3M-Santander de Big Data, Universidad Carlos III de Madrid, Getafe, Spain}}
\affil[3]{{\small Biomedical Imaging and lnstrumentation Group, Hospital General Universitario Gregorio Marañón, Madrid, Spain}}
\affil[4]{{\small Departamento de Bioingeniería e Ingeniería Aeroespacial, Universidad Carlos III de Madrid, Getafe, Spain}}
\affil[5]{{\small Departmento de Estadística, Universidad Carlos III de Madrid, Getafe, Spain}}

\begin{document}
\maketitle

\begin{abstract}
	Task-based functional magnetic resonance imaging (task fMRI) is a non-invasive technique that allows identifying brain regions whose activity changes when individuals are asked to perform a given task. This contributes to the understanding of how the human brain is organized in functionally distinct subdivisions. Task fMRI experiments from high-resolution scans provide hundred of thousands of longitudinal signals for each individual, corresponding to measurements of brain activity over each voxel of the brain along the duration of the experiment. In this context, we propose some visualization techniques for high dimensional functional data relying on depth-based notions that allow for computationally efficient 2-dim representations of tfMRI data and that shed light on sample composition, outlier presence and individual variability. We believe that this step is crucial previously to any inferential approach willing to identify neuroscientific patterns across individuals, tasks and brain regions. We illustrate the proposed technique through a simulation study and demonstrate its application on a motor and language task fMRI experiment.
\end{abstract}

\section{Introduction}\label{intro}

Functional magnetic resonance imaging (fMRI) is the benchmark neuroimaging technique for measuring brain activity, because of its advantages with respect to other acquisition methods such as PET or EEG among others. Indeed, it is non-invasive and it does not involve radiation, which makes it safe for the subject. It also provides good spatial and temporal resolution (finer for the spatial component). fMRI based on blood-oxygen-level dependent (BOLD) consists in measuring the variation on oxygen consumption and blood flow that occur in brain areas in response to neural activity. That allows to identify those brain regions that take part in specific mental processes. Task fMRI (tfMRI) experiments are conducted by measuring brain activity in this way for a given period of time (several minutes) while the subject is asked to repeatedly perform some task, alternating between task performance and resting periods for the whole length of the experiment. This kind of experiments are often conducted on a relatively small group of patients (around a hundred at most) because of cost and other feasibility reasons. The data set obtained for each one of these subjects is, on the contrary, very large due to the high spatial resolution of the technique. See \cite{tfMRI} for a detailed description of this kind of experiments, including tasks specification.

The analysis of tfMRI data raises different challenges. \cite{tfMRI2} enumerate some of them in the context of an attempt of characterizing task-based and resting state fMRI signals. Among them we focus on three: 1) the high inter-individual variability, in a setting where the number of individuals is relatively small; 2) the high amount of available data for a single experiment, due to the voxel-wise structure of fMRI temporal signals; and 3) the existence of different sources of noise, from individual origin (movement during the experiment, lack of attention, etc) to mechanical nature (scanner instability among others), that induce artifacts and undesirable measurement values in the recorded signals.

In this setting, data visualization and outlier detection tools are of crucial importance to avoid feeding inferential algorithms with low-quality data. Indeed, in any context in which high-dimensionality or complex data structure does not allow for direct visual inspection of the data, the use of dimension reduction visualization tools and robust measures helps shedding light on sample composition. In particular, for functional data, of which task fMRI data can be a particular case, there is a vast literature on robust visualization and outlier detection tools based on depth-measures (see for instance, \cite{SunGenton11, Outliergram}).

In functional data analysis (FDA), individual observations are real functions of time, observed at discrete time points. If several functions of time are observed for each individual, we talk about multivariate FDA. In this setting, the number $p$ of functions observed per individual is small relative to the number $n$ of individuals. Examples of this are the longitudinal patterns of flying, feeding, walking and resting observed over the lifespan of Drosophila flies \citep{flies}, or the $8$-variate signal of electrocardiograph curves \citep{Ieva_Paganoni}.

In the case of task fMRI data, multiple functions of time are observed on each individual, corresponding to the recorded brain activity on each voxel over the duration of the experiment. However, the number of dimensions is given by the number of voxels which, depending on the resolution of the scan, can be of the order of hundred of thousands, whereas the number of individuals is relatively small due to cost and time constraints. We face a new paradigm in multivariate FDA with small $n$ and very large $p$, which we refer to as the high dimensional functional data setting.

The literature on robust multivariate functional data has provided some generalizations of the concept of functional depth to the multivariate functional setting that can be used for visualization and outlier detection purposes. One of the first proposals was done by \cite{Ieva_Paganoni} who defined a multivariate functional depth measure as a weighted sum of the functional band depths \citep{BandDepth} computed over the marginal functional data sets. Later, \cite{Claeskens_JASA14} proposed a different definition consisting in the integration over the time domain of any multivariate depth measure computed on the $p$-dimensional sample of points observed at each time instant. Based on these definitions, several visualization tools have been proposed with the aim of allowing for data inspection and detection of outlying observations. On the one hand, 
\cite{Ieva_Paganoni2} extended the \emph{outliergram} \citep{Outliergram} to the multivariate framework for component-wise outlier detection. On the other hand, given an integrated multivariate functional depth measure or its outlyingness counterpart, several visualization tools have been defined as a two dimensional graphical representation of its average value (over the time domain) versus some measure of its variability (also over time). This allows to distinguish typical observations (low depth/outlyingness variability over time) from magnitude outliers (low average depth, resp. high average outlyingness) and from shape outliers (high depth/outlyingness variability). Examples of this are the \emph{centrality-stability plot} (CS-plot) of \cite{Huberetal2015} and its modification proposed by \cite{NietoCuesta_DiscussionHuber}.
Recently, two new approaches follow this same line based on different \emph{directional outlyingness} notions that are computationally more efficient than the previously existing ones, and propose the corresponding graphical representations, the \emph{functional outlier map} (FOM) \citep{Roussetal18} and the \emph{magnitude-shape plot} (MS-plot) \cite{DaiGenton18}. They both offer interesting results. However, they have been designed for the low dimensional multivariate functional setting.\\

In this article we consider a different approach. We propose a methodology for reducing the high dimensional functional problem to a functional problem by keeping, and not averaging, depth values both over dimensions (voxels) and time. The use of computationally efficient depth-based measures allows us to do this even for very large $p$. Analysis of the resulting functional data sets, namely depths over dimensions and time, provides insight on sample composition and outlier presence. In particular, we focus on the identification of joint outliers across dimensions, since marginal outliers can be detected by the means of standard functional data techniques applied on each component A graphical two dimensional representation of the data, the DepthGram, is also proposed.\\

The rest of the paper is as follows. 
In Section \ref{method} we discuss the taxonomy of atypical observations in multivariate and high-dimensional functional data sets and we introduce the proposed depth-based visualization techniques, providing the properties of the functional depth measures that are the basis for the methodology. In Section \ref{simus} we show the performance of our visualization tools through a simulation study in the high dimensional setting and assess its computationally efficiency in comparison with existing methods for low multivariate functional data. In section \ref{fMRIdata} we demonstrate the application of the proposed approach with a motor and language task fMRI experiment conducted on $100$ individuals. We conclude the article with a discussion in Section \ref{disc}.

%
%

\section{Visualization of high dimensional functional data}\label{method}
A general setting in FDA is to consider that  observations are i.i.d. realizations of some stochastic process $\mathbf{X}$, taking values in the space of continuous functions defined from some compact real interval $\cal{I}$ into $\R^p$, ${\cal C} ({\cal I}, \mathbb{R}^p)$. That is, a sample of size $n$ of functional data is a collection of $n$ i.i.d. continuous functions $X_i:{\cal I}
\longrightarrow \R^p$, $i=1,\ldots,n$, whose realizations $x_i(t)$, $t\in {\cal I}$, are observed on a time grid of $N$ points, $\{t_1,\ldots,t_N\}\subseteq {\cal I}$. We denote $x_i^j:{\cal I}\longrightarrow \R$ the $j$-th component of the $i$-th observation. If $p=1$, we are in the univariate functional data setting where each individual has associated one curve, whereas if $p>1$ we are in the multivariate functional data setting where for each individual we observe several processes over time. In this article, we consider the case $p>>n$ and refer to it as the high dimensional functional data setting. In particular, in the context of tfMRI experiments, $n$ is of the order $10^2$ to $10^3$ whereas $p$ is of the order $10^5$ to $10^6$. 

\subsection{Outliers in multivariate and high dimensional functional data}
In FDA outlying observations are generally classified as being \emph{magnitude outliers}, if they are curves with values lying outside the range of the majority of the data or \emph{shape outliers} if they are curves that exhibit a different shape from the rest of the sample. Magnitude outliers are also referred to as \emph{shift outliers}, and some times the distinction between \emph{isolated} and \emph{persistent outliers}, is done, where the first term refers to curves that have an outlying behavior during a very short time interval and the second one to observations with an outlying pattern on the whole observation domain (or at least on a large part of it). See \cite{Huberetal2015, Ana_DiscussionHuber} for a detailed discussion on a taxonomy for functional outliers.

\begin{figure}
	\begin{center}
		\includegraphics[width=16cm]{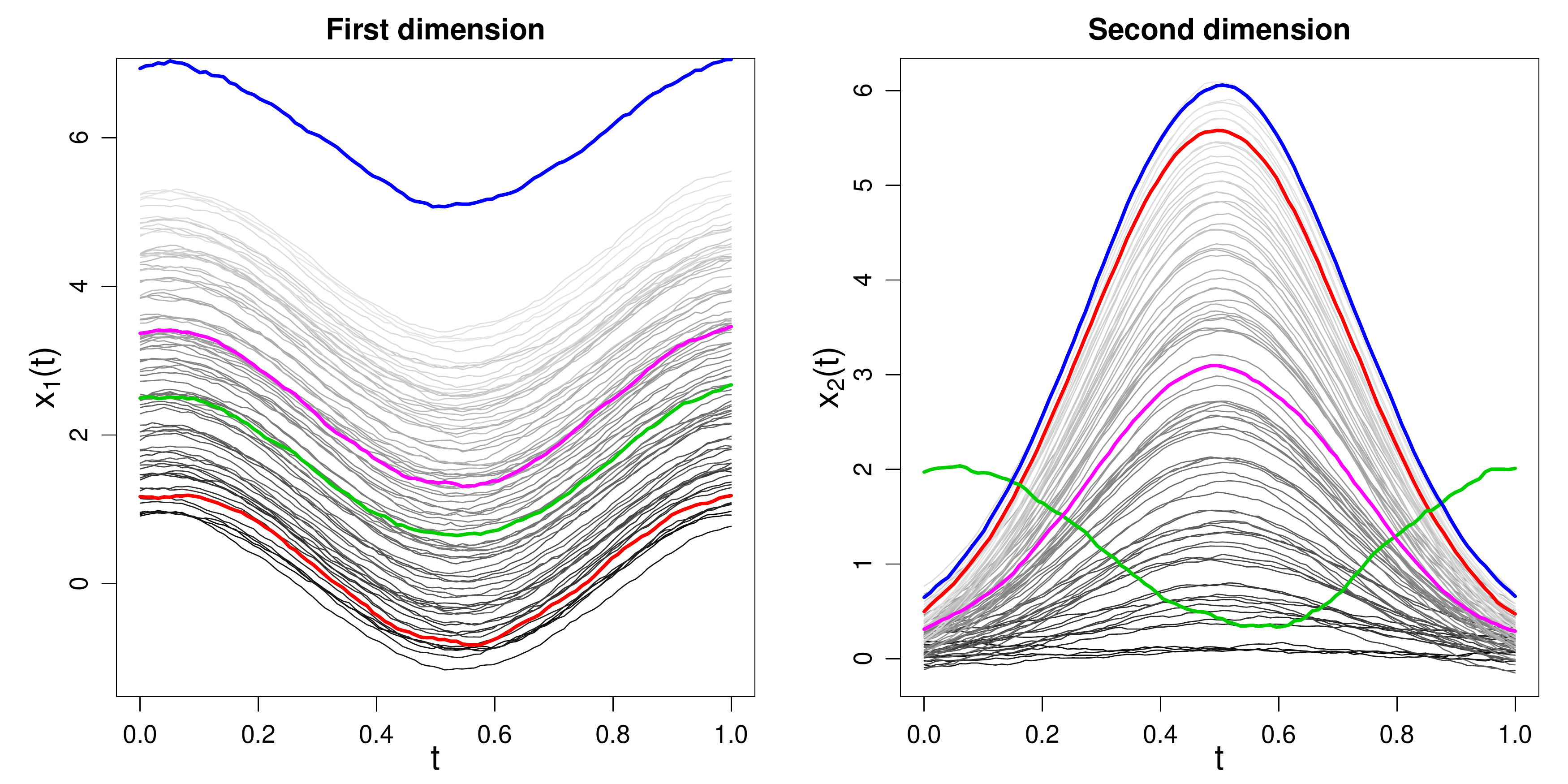}
	\end{center}
	\caption{Bivariate functional sample where the same color is used to draw both components of the same observation. We can appreciate how observations with higher values in the first component tend to also take higher values in the second component. We can distinguish three outlying observations. The observation coded in green is a marginal shape outlier in the second dimension. The observation coded in blue is a marginal magnitude outlier in the first dimension. Finally, the observation coded in red is a joint outlier, since low values in the first dimension are associated to high values in the second dimension. However, none of its marginals is outlying in any of the dimensions. The magenta curve represents a central curve in both dimensions.}\label{marg_joint}
\end{figure}

All these notions apply to a set of observed curves over the same time interval, that is, to univariate functional data. 
In the multivariate FDA framework, we need to consider a higher hierarchy to distinguish between \emph{marginal outliers} and \emph{joint outliers}. Marginal outliers would be observations whose marginal components fall in some of the above mentioned categories in one or several dimensions, whereas joint outliers would be observations with non-outlying marginals but joint outlying behavior. Indeed, when thinking of task fMRI data, we can imagine  individuals for which brain activity patterns are standard in every voxel, but relationships across brain regions are atypical. Figure \ref{marg_joint} illustrates the outlier categorization in a bivariate synthetic data set.

Thus, outlier detection for multivariate functional data needs to rely on procedures that are able to detect outliers jointly over dimensions and not only marginally. The methods proposed in the literature to define population depth/outlyingness measures for multivariate functional data mainly rely on two approaches: 1) Given a functional depth measure $d_{F}:  {\cal C} ({\cal I}, \mathbb{R}) \rightarrow [0,1]$, define a multivariate functional depth measure as
$$d_{F}:  {\cal C} ({\cal I}, \mathbb{R}^p) \rightarrow [0,1] \quad \quad  d_{MF}(\mathbf{X};  F_{\mathbf{{\cal X}}} ) \equiv \sum_{j=1}^{p}\omega_j d_F(X^{j}; F_{{\cal X}^j} ) $$
as proposed in \cite{Ieva_Paganoni}; and 2) Given a multivariate depth measure  
$d_{M}:  \mathbb{R}^p \rightarrow [0,1]$, define a multivariate functional depth measure as
$$d_{MF}:  {\cal C}({\cal I}, {\R}^p) \rightarrow [0,1]\quad \quad  d_{MF}(\mathbf{X}; F_{\mathbf{{\cal X}}} )  \equiv \int_{{\cal I}} \omega(t) d_M(\mathbf{X} (t) ; F_{\mathbf{{\cal X}}(t)}) dt$$
where in both cases $\omega_j's, \omega(\cdot)$ are suitable weights/weight functions that can be chosen to calibrate the relative contribution of dimensions/time points to the global depth value, and $F_{\mathbf{{\cal X}}}, F_{{\cal X}^j}$ and $F_{\mathbf{{\cal X}}(t)}$ denote the probability distributions of the corresponding $p$-variate random process, univariate random process and $p$-variate random vector. Outlier detection methods based on depth measures or their outlyingness counterparts defined under the second approach will be able to identify joint outliers are soon as they rely on depth (outlyingness) measures for multivariate data that are designed to do so in finite dimensional spaces. This is the case for some recent approaches as \cite{Huberetal2015, NietoCuesta_DiscussionHuber, Roussetal18,DaiGenton18}.
However, multivariate depth functions are in general computationally expensive, or even unfeasible for moderate dimension, and even efficient alternative such as the random Tukey depth (as in \cite{NietoCuesta_DiscussionHuber}) or the \emph{directional outlyingness} of \cite{Roussetal18} may fail to provide a computationally efficient method in a high dimensional functional setting. 
To remedy this situation, we propose to work with highly efficient depth notions based on the concept of band depth \citep[see][]{FastBD}, that apply to both functional and multivariate spaces. Moreover, instead of considering any of the two approaches described above, mainly depth average over dimensions or depth integration over the time domain, we embrace the \emph{depth of depths} approach, which allows us to better characterize different types of observations and outlying behaviors.

\subsection{Depth-based tools for high dimensional functional data}

We now present some depth notions for functional data that will be incorporated into our procedure. Let us recall the modified band depth (MBD) and modified epigraph index (MEI), a depth measure and a depth-based index defined in for the analysis of functional data in \cite{BandDepth} and \cite{HalfRegion11}, respectively. Both measures provide an idea of how central or deep a curve is with respect to a sample of curves (indeed, we introduce the sample versions of these measures here). Let $x_1,\dots, x_n$ be $n$ continuous functions defined on a given compact real interval  $\I$. For any $x\in \{x_1,\dots,x_n\}$, its modified band depth is
$$MBD_{\{x_1,\dots,x_n\}} (x) =\binom{n}{2}^{-1} \sum_{i=1}^n \sum_{j=i+1}^n \dfrac{\lambda\left(\left\{t\in {\cal I} \,\middle|\, \min( x_i(t),x_j(t)) \leq x(t) \leq \max( x_i(t),x_j(t)) \right\}\right)}{\lambda(\I)},$$
where $\lambda(\cdot)$ stands for the Lebesgue measure on $\R$. If for each pair of curves $x_i$ and $x_j$ in the sample we consider the band that they define in $\I\times\R$ as $\left\{(t,y)\,\middle|\,t\in \I, \, \, \min( x_i(t),x_j(t)) \leq y \leq \max( x_i(t),x_j(t))\right\}$, then $MBD_{\{x_1,\dots,x_n\}} (x)$ represents the mean over all possible bands of the proportion of time that $x(t)$ spends inside a band. The modified band depth is an extension of the original band depth that accounts for the proportion of bands in which a curve is entirely contained \citep[see][for details]{BandDepth}.

The modified epigraph index of  $x\in \{x_1,\dots,x_n\}$ is defined as
$$MEI_{\{x_1,\dots,x_n\}} (x) =\dfrac{1}{n} \sum_{i=1}^n \dfrac{\lambda\left(\left\{t\in {\cal I} \,\middle|\, x_i(t) \geq x(t) \right\}\right)}{\lambda(\I)}$$
and it stands for the mean proportion of time that $x$ lies below the curves of the sample. As in the case of the MBD, the MEI is a generalization of the epigraph index that accounts for the proportion of curves that lie entirely above $x$ \citep{HalfRegion11}.

In \cite{Outliergram}, a quadratic relationship between these two quantities was established which allowed to define a procedure to detect shape outliers. Indeed, since the MBD of a curve is highly dependent on its location in the sample (in the sense of vertical position), and the MEI provides a measure of this location, the conditional observation of the MBD given the MEI provides an accurate shape descriptor which allows to identify shape outliers, that is, curves with low MBD values for relatively high MEI values. In particular it is shown that
\begin{equation}\label{par1}
MBD_{\{x_1,\dots,x_n\}} (x) \leq f_n\left(MEI_{\{x_1,\dots,x_n\}} (x)\right), \quad  x\in  \{x_1,\dots,x_n\}
\end{equation}
where $f_n:[0,1]\rightarrow \mathbb{R}$ is the parabola defined by $f_n(z)=a_0+a_1z+a_2 n^2 z^2$ and $a_0=a_2=-2/n(n-1)$, $a_1=2(n+1)/(n-1)$ (notice the dependence on the sample size). The equality in (\ref{par1}) holds if and only if none of the curves in the sample cross each other.\\

Let us introduce some notation in the context of multivariate functional data . Given a sample $\mathbf{x}=\{x_i=(x_i^j(t_k))^{j=1,\ldots,p}_{k=1,\ldots,N},\,i=1,\ldots,n\}$ of $p$-variate functions observed at discrete time points we will consider the $n\times p$ matrix
$\mathbf{MBD}_d(\mathbf{x})$ whose $ij$ element corresponds to the modified band depth of curve $x^j_i$ with respect to the $j$-th marginal sample, that is, $\mathbf{MBD}_d(\mathbf{x})_{ij}=MBD_{\{x^j_1,\dots,x^j_n\}} (x^j_i)$. The subindex $d$ of $\mathbf{MBD}_d(\mathbf{x})$ stands for dimensions, meaning that the MBD is computed on each dimension of the data set. Indeed, the columns of $\mathbf{MBD}_d(\mathbf{x})$ are the MBDs on each marginal. Equivalently, we will denote
%
$\mathbf{MEI}_d(\mathbf{x})$ the $n\times p$ matrix whose columns are the MEIs on each marginal of the sample, that is, $\mathbf{MEI}_d(\mathbf{x})_{ij}=MEI_{\{x^j_1,\dots,x^j_n\}} (x^j_i)$, $i=1,\ldots,n$, $j=1,\ldots,p$.

In a similar way we will denote by $\mathbf{MBD}_t(\mathbf{x})$ the  $n\times N$ matrix whose columns are the MBDs on each time point of the sample, that is, $\mathbf{MBD}_t(\mathbf{x})_{ij}=MBD_{\{x_1(t_j),\dots,x_n(t_j)\}} (x_i(t_j))$, $i=1,\ldots,n$, $k=1,\ldots,N$, where $x_i(t_j)=(x^1_i(t_j),\ldots,x^p_i(t_j))$. Here $t$ stands for time, since MBD is computed on time point of the observation domain across dimensions. Finally, let $\mathbf{MEI}_t(\mathbf{x})$ be the  $n\times N$ matrix whose columns are the MEIs on each time point, that is, $\mathbf{MEI}_t(\mathbf{x})_{ij}=MEI_{\{x_1(t_j),\dots,x_n(t_j)\}} (x_i(t_j))$, $i=1,\ldots,n$, $k=1,\ldots,N$.\\
Now, each one of these four matrices can be understood as a set of functional observations, indexed by dimensions or time. That is, MBD and MEI (or other depth functions) can be computed on them as we will see in the next section. 
%

\subsection{Relationship between MEI(\textbf{MBD}) and MBD(\textbf{MEI})}

The modified band depth of a curve with respect to a univariate functional sample of size $n$ is bounded in $(0, 1/2+3/2n]$. Values close to $0$ stand for observations that can either have an atypical shape but are located at the center of the functional sample, or are placed at the ends (upper or lower) of the curves cloud, with both typical or atypical shapes. That is, it provides a center-outwards ordering of the sample of observations.\\
The modified epigraph index is bounded in $(0,1]$, with low values corresponding to curves placed above (in terms of function values) the majority of the curves in the sample, high values corresponding to curves lying at the bottom part of the curves cloud, and values close to $0.5$ corresponding to curves placed in the middle of the sample. That is, it induces a bottom-up ordering of the sample of observations. Of course, this would be most useful when such an order is naturally present in the observed process. If curves in the sample do not exhibit such an order, i.e., there are many crossings between curves due to noise, presence of phase variation or \emph{time warping} or by the nature of the underlying process, the range of $MEI$ would be narrowed to some interval around $0.5$. This is due to the average over time points that leads to the definition of $MEI$. Equivalently, if different shapes co-exist in a sample and no curve is significantly contained in more/less bands than any other curve in the sample, the range of $MBD$ values will also be narrowed.\\
In any case, the \emph{most central individuals} according to the modified band depth correspond to the overall highest values of $MBD$ whereas the \emph{most central individuals} according to the modified epigraph index correspond to the overall most central values (close to $0.5$) of $MEI$, and both quantities are related by equation (\ref{par1}).\\

When switching to the multivariate functional framework, one would expect that those individuals that are persistently \emph{central} over dimensions according to $MBD$, will get an $MBD$ curve (row) in the matrix
$\mathbf{MBD}_d(\mathbf{x})$ with higher values than most of the rest of the individuals. On the contrary, those individuals that are persistently far from the center of the sample across dimensions, in the $MBD$ sense, would get a low $MBD$ curve (row) in the matrix $\mathbf{MBD}_d(\mathbf{x})$. That is, if we apply the modified epigraph index on the \emph{$MBD$ sample}, we should get low values for the central individuals and high values for the non-central individuals.
Now, if we consider the $MEI$ matrix $\mathbf{MEI}_d(\mathbf{x})$, those curves with central values across dimensions will get a central curve in the matrix, whereas individuals whose corresponding curves take low or high values across dimensions, will tend to get high and low $MEI$ curves in $\mathbf{MEI}_d(\mathbf{x})$. Again if we now apply the modified band depth on the \emph{$MEI$ sample}, we should get high $MBD$ values for those individuals with curves that take central values in every dimension. \\
This expected behavior is summarized in the following result, where the relationship between $MBD\left(\mathbf{MEI}_d(\mathbf{x})\right)$ and $MEI\left(\mathbf{MBD}_d(\mathbf{x})\right)$ is established. The proof is included in the Supplementary Materials.

\begin{prop}\label{prop}
Let $\mathbf{x}=\{x_i=(x_i^j(t_k))^{j=1,\ldots,p}_{k=1,\ldots,N},\,i=1,\ldots,n\}$ be a sample of $p$-variate continuous  functions observed at $N$ discrete time points. If \begin{itemize}
\item[a)] $(x^j_i(t_{k_1})-x^j_h(t_{k_1}))(x^j_i(t_{k_2})-x^j_h(t_{k_2}))>0$, $k_1,k_2 \in 1,\ldots,N$, $i\neq h$, for all $j=1,\ldots,p$
\end{itemize} holds then
\begin{equation}\label{par2}
MBD\left(\mathbf{MEI}_d(x)\right) \leq g_n\left(1-MEI\left(\mathbf{MBD}_d(x)\right) \right), \quad x\in  \{x_1,\dots,x_n\},
\end{equation}
where $g_n:[0,1]\rightarrow \mathbb{R}$ is the parabola defined by $g_n(z)=\alpha_0+z+\alpha_2 z^2$ and $\alpha_0=2/n$, $\alpha_2=-n/2(n-1)$.\\ Moreover if
\begin{itemize}
\item[b)] $(x^j_i(t_{k})-x^j_h(t_{k}))(x^\ell_i(t_{k})-x^\ell_h(t_{k}))>0$, $k=1,\ldots,N$, $i\neq h$, $j\neq \ell$
\end{itemize} 
also holds then
\begin{equation}\label{par3}
MBD\left(\mathbf{MEI}_d(x)\right) = g_n\left(1-MEI\left(\mathbf{MBD}_d(x)\right) \right), \quad  x\in  \{x_1,\dots,x_n\}.
\end{equation}

\end{prop}

\begin{remark} Notice that assumptions \emph{a)} and \emph{b)} of the previous result won't hold in practice, since they require a perfectly ordered set of curves, both inside each dimension and across dimensions. However, Proposition \ref{prop} establishes the theoretical relationship between the two measures of interest, justifying their combination. Indeed, although it would have been possible to use MBD and MEI computed over the same depth set, either $\mathbf{MEI}_d(\mathbf{x})$ or $\mathbf{MBD}_d(\mathbf{x})$, on the basis of the results established in \cite{Outliergram}, the cross-use of MBD and MEI will allow for a more nuanced description of the sample, as will be shown.
\end{remark}

\begin{remark} The result of Proposition \ref{prop} also holds for $MBD\left(\mathbf{MEI}_t(x)\right)$ and $MEI\left(\mathbf{MBD}_t(x)\right)$ under the same assumptions (in reverse order), and the proof follows the same lines.
\end{remark}

\begin{figure}
	\begin{center}
		\includegraphics[width=16cm]{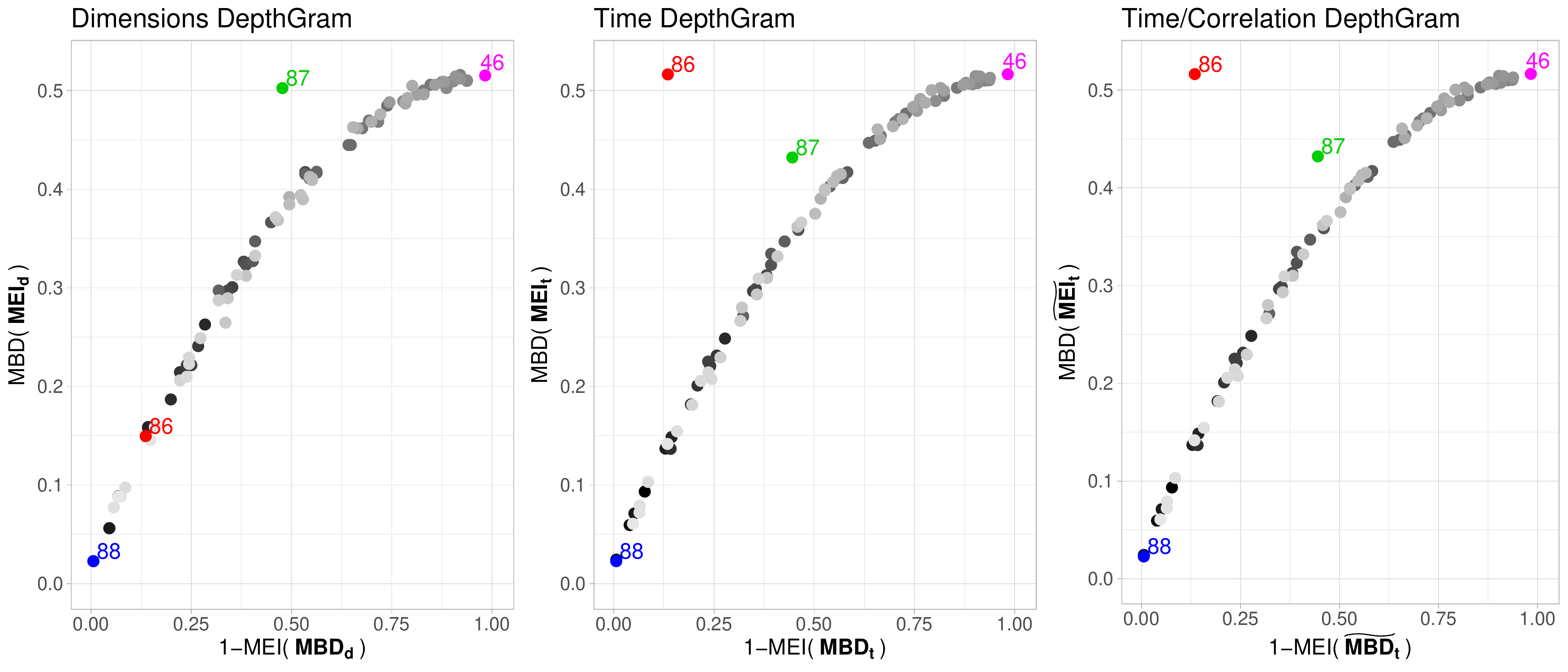}
	\end{center}
	\caption{2-dimensional representations of the bi-variate sample of Figure \ref{marg_joint} through the relationship between $MBD(\mathbf{MEI})$ and $MEI(\mathbf{MBD})$. In the first graphic, the depths over dimensions are used, whereas for the second one, the modified band depth and modified epigraph index are first computed over time points. We can appreciate how the most central observation is found at the right top corner in both graphics. Also, the observation corresponding to a magnitude outlier in the first dimension, and also taking high values in the second dimension is found in the left bottom corner of both plots. The joint outlier is isolated in the time graphic, whereas the shape outlier is isolated in the dimensions graphic.}\label{DG1_example}
\end{figure}

\begin{remark} Unlike in (\ref{par1}), we are unable to establish a bound on the relationship (\ref{par2}) that would hold under general conditions. Indeed, as will be illustrated through several examples, the parabola tends to be a lower bound for $MBD\left(\mathbf{MEI}_d(x)\right)$ (see Figure \ref{DG1_example}, for example). However, this is not always the case.\\
It would be unusual for an individual with a high $1-MEI\left(\mathbf{MBD}_d(x)\right)$ value, thus standing for overall central curves in terms of MBD across all dimensions, to have a low $MBD\left(\mathbf{MEI}_d(x)\right)$ value, that is, having globally non-central curves in terms of MEI across dimensions. However, if there is high variability within each dimension and curves tend to cross, no relationship between $\mathbf{MBD}_d(x)$ and $(\mathbf{MEI}_d(x)$ can be established and then even less of an structured pattern is to be expected for $1-MEI\left(\mathbf{MBD}_d(x)\right)$ and $MBD\left(\mathbf{MEI}_d(x)\right)$. This is what happens when the result is applied to the depths computed on each time point in most of the examples presented in the paper. In that case, the reference data set for the first depth calculation is the one composed by the values of all the curves evaluated on the same time point across dimensions. If no clear association pattern exists between dimensions, each of these \emph{pseudo univariate functional data sets} will be highly unstructured with the resulting \emph{curves} crossing many times, which will yield to a very spread 2-dimensional representation of the corresponding quantities, as shown in Figure \ref{DG1_example}.
\end{remark}

The two dimensional representation of $1-MEI(\mathbf{MBD}_d(\mathbf{x}))$ and $MBD (\mathbf{MEI}_d(\mathbf{x}))$ allows to identify different types of observations. See Figure \ref{DG1_example} for an example. Note that we use $1-MEI$ and not $MEI$ in the $x$-axis so that for both axis high values stand for \emph{central} observations and low values stand for \emph{non-central} observations. In this representation, the most central observations will appear in the right top corner, whereas the least central observations will be shown in the left bottom corner. This includes magnitude outliers which will be found at the left bottom corner of the plot but not necessarily separated from the rest of the data points (as it is also the case in the \emph{outliergram}). Shape outliers will tend to appear in the left upper part of the graphic, above the bulk of the majority of points of the sample. 

Let us now consider the representation of $1-MEI(\mathbf{MBD}_t(\mathbf{x}))$ versus $MBD (\mathbf{MEI}_t(\mathbf{x}))$. The interpretation of extreme values in both axes is similar to the previous one, but here joint outliers play the role of shape outliers. Indeed, joint outliers are observations that have an association pattern across dimensions (that is shape, when the functional data is considered as a function of dimensions for a fixed time point) different from that of the majority of the observations.\\

\begin{figure}
	\begin{center}
\includegraphics[width=16cm]{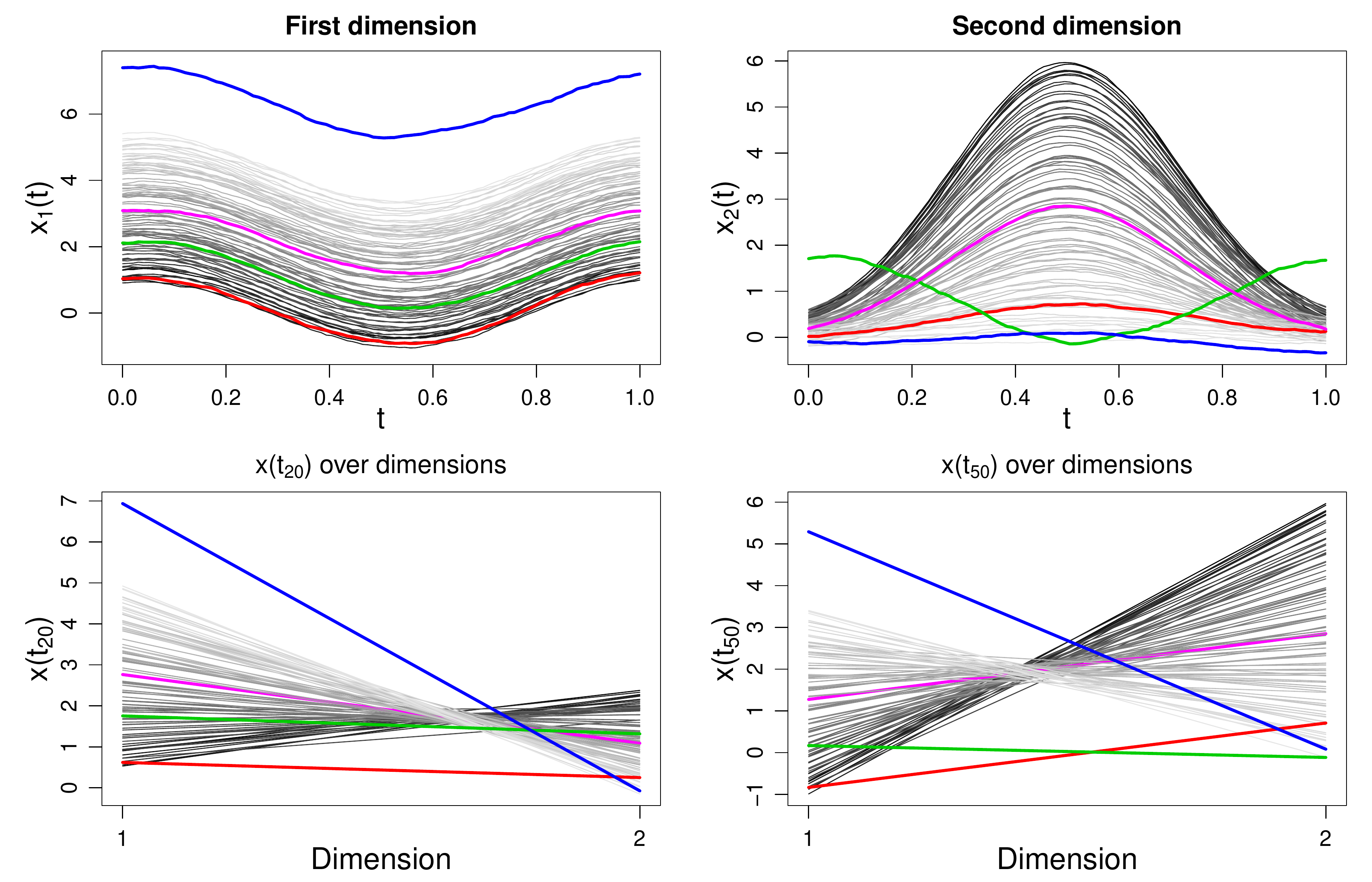}\\
\includegraphics[width=16cm]{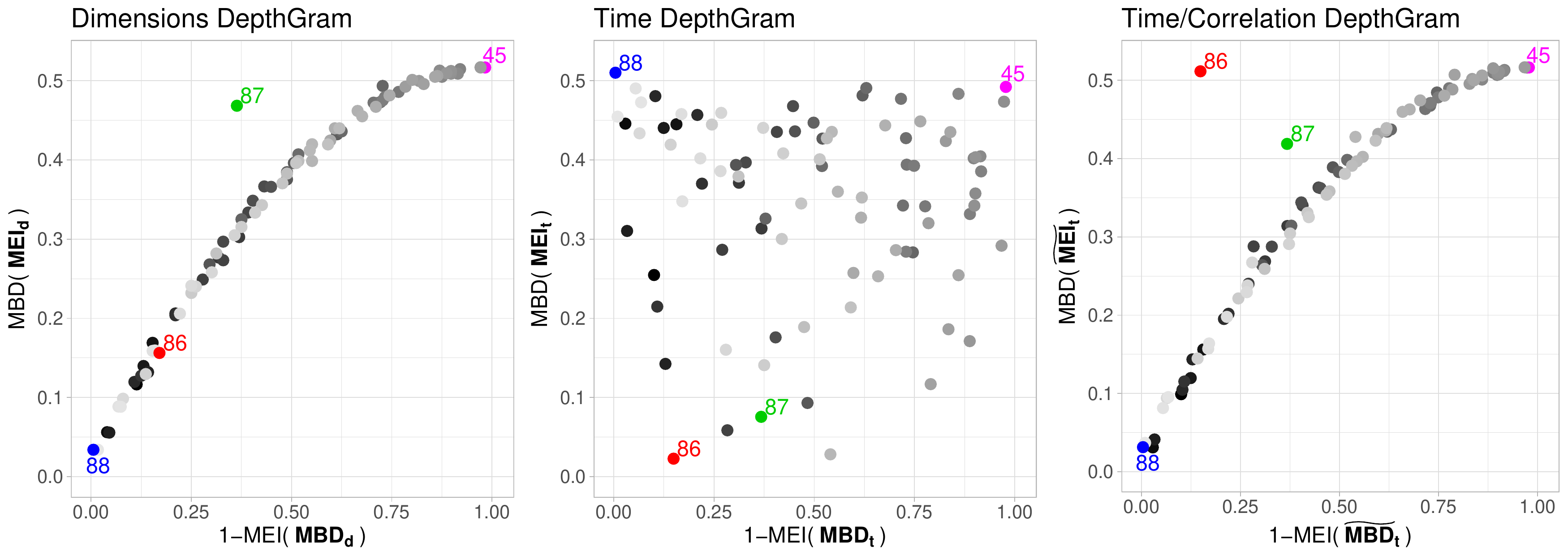}
	\end{center}
	\caption{In the first row, a bi-variate functional data set with reversed curve ordering between the two dimensions is presented. The same color is used to draw both components of the same observation. The observation coded in red is a joint outlier, since low values in the first dimension are associated to low values in the second dimension, where it is the opposite for the rest of the curves in the sample. In the second row, two time points are chosen, $t_{20}=0.2$ and $t_{50}=0.5$ ($N=100$), and the corresponding multivariate observations are represented using a parallel coordinate plot.
In the third row the three DepthGrams, on dimensions, on time and on time/correlation are presented. Since there are changes in the sign of the correlation among dimensions, the third DepthGram version is most suitable to identify joint outliers.}\label{Ex_Neg_Corre}
\end{figure}

In fact, this is only true if the association across dimensions is positive, that is the ordering of the curves is preserved from one dimension to the next one. If this is not the case, that is, if negative association exists between the different components of the process, the identification of an observation whose components exhibit a different association behavior is not as straight forward. 
In that case at each time point $t$, the corresponding multivariate observation $\mathbf{x}_i(t)=(x_i^1(t),\ldots,x_i^p(t))$ will be nearly constant across dimensions whereas the observations of the rest of the sample will oscillate from a medium-high to a medium-low value between dimensions. This will result in a sample of pseudo-functional observations with many different shapes, in which the one corresponding to observation $i$ might not stand out as a shape outlier since there is no a single common pattern from which it differs. This is illustrated in Figure \ref{Ex_Neg_Corre}. Indeed, note that here we treat $\mathbf{x}_i(t)=(x_i^1(t),\ldots,x_i^p(t))$ as if it was a functional observation and we apply functional depth tools to its parallel coordinate representation \citep{ParaCoord}.\\
In such a setting, multivariate depth tools would be more useful at detecting joint outliers at each time point,
but they can be unfeasible for high dimensional settings. Instead, we propose to remedy this situation by considering a transformation of the data that will allow to \emph{correct} the ordering reversion along dimensions. Indeed, the idea is very simple and consists on inverting (by multiplying by -1) the univariate functional data sets corresponding to the dimensions in which the order of curves is reversed with respect to the precedent dimensions. This is detailed in the next section.\\

\begin{figure}
	\begin{center}
\hspace*{-0.4cm}\includegraphics[width=16.75cm]{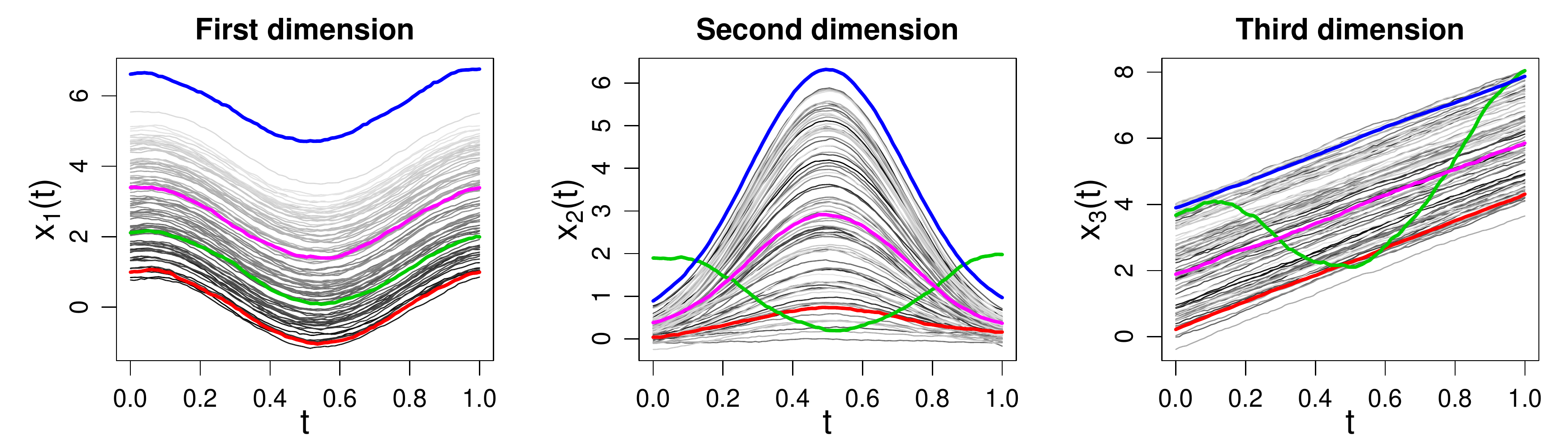}\\
\includegraphics[width=16cm]{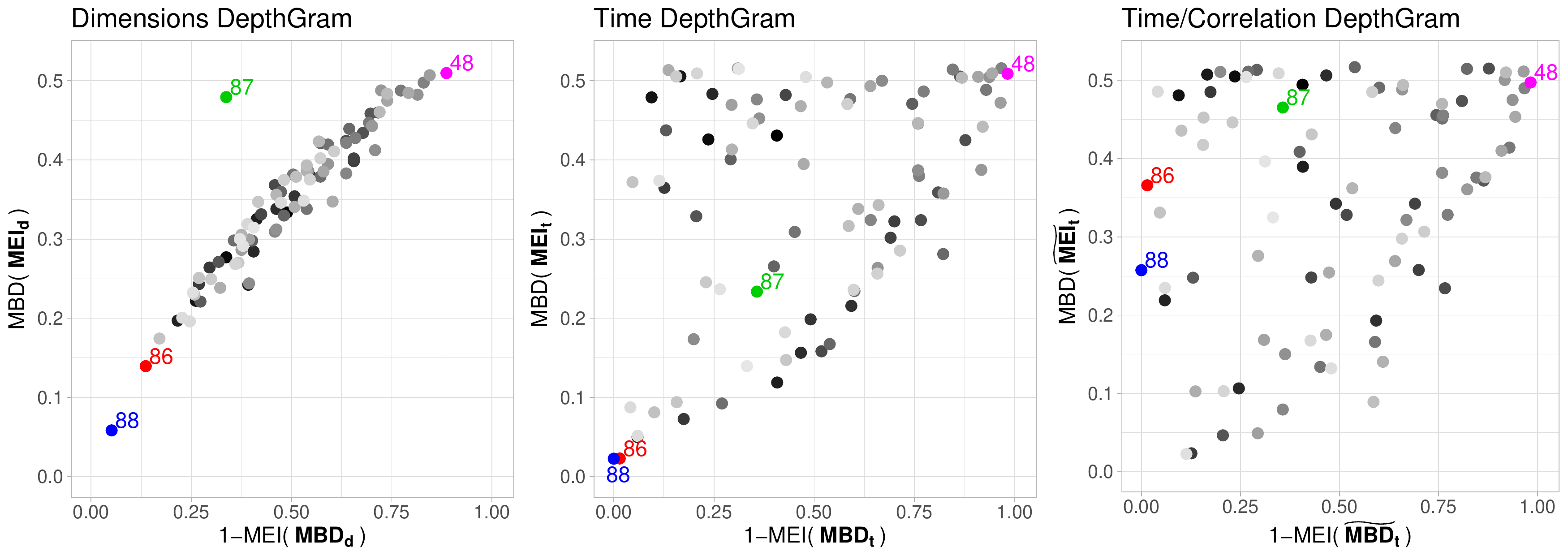}
	\end{center}
	\caption{In the first row, a three-variate functional data set with independent components is presented. The same color is used to draw the three components of the same observation. The observation coded in red could be considered as a joint outlier, since low values in the first dimension are associated to low values in the second and third dimension, whereas the rest of the curves behave independently across dimensions. The same happens for the blue observation, which, in addition, is a magnitude outlier in the first dimension. 
In the second row the three DepthGrams, on dimensions, time and time/correlation are presented. There is a spurious negative correlation between the second and third components of the data set, so the time/correlation DepthGram is different from the time DepthGram. However, since no clear correlation pattern is present in the sample, both exhibit the same dispersed appearance.}\label{Ex_Independence}
\end{figure}

Another situation in which our procedure will fail to identify joint outliers is the case where an observation has a clear association pattern across its components and the rest of the observations are independent dimension-wise. However, in this case, traditional multivariate depths applied marginally on the observations for each time point will fail too. Such a situation with independent dimensions will tend to yield a very spread representation of the $1-MEI(\mathbf{MBD}_t(\mathbf{x}))$ versus $MBD (\mathbf{MEI}_t(\mathbf{x}))$, with a similar behavior for the adjusted procedure (since in this case the dispersion is not related to oscillating correlations and can not be \emph{corrected}). See Figure \ref{Ex_Independence} for an example.

\subsection{DepthGrams}
We propose three graphical representations of the data based on the relationships between $MEI$ of $MBD$ and $MBD$ of $MEI$ as detailed before. On the one hand, the DepthGram on dimensions, where functional depths are first computed on the sample of curves $x^j_{1:n}$ observed over $\{t_1,\ldots,t_N\}$, for each $j=1,\ldots,p$, and the DepthGram on time, where functional depths are first computed on the sample of $p$-dimensional observations $x_{1:n}(t_k)$, for each time point $k=1,\ldots,N$. On the other hand, the \emph{time/correlation} DepthGram which is just the time DepthGram on a modified data set in order to \emph{correct} for negative association across dimensions.
\begin{itemize}
\item Dimensions DepthGram: scatter plot of points $DG^{d}_i=(DG^{d}_{i1},DG^{d}_{i2})$ where 
\begin{eqnarray*}
	DG^{d}_{i1}&=&1-MEI_{\{\mathbf{MBD}_d(\mathbf{x})_{1\cdot},\ldots,\mathbf{MBD}_d(\mathbf{x})_{n\cdot}\}} (\mathbf{MBD}_d(\mathbf{x})_{i\cdot})\\
	DG^{d}_{i2}&=&MBD_{\{\mathbf{MEI}_d(\mathbf{x})_{1\cdot},\ldots,\mathbf{MEI}_d(\mathbf{x})_{n\cdot}\}} (\mathbf{MEI}_d(\mathbf{x})_{i\cdot})
\end{eqnarray*}
\item Time DepthGram: scatter plot of points $DG^{t}_i=(DG^{t}_{i1},DG^{t}_{i2})$ where \begin{eqnarray*}
DG^{t}_{i1}&=&1-MEI_{\{\mathbf{MBD}_t(\mathbf{x})_{1\cdot},\ldots,\mathbf{MBD}_t(\mathbf{x})_{n\cdot}\}} (\mathbf{MBD}_t(\mathbf{x})_{i\cdot})\\
DG^{t}_{i2}&=&MBD_{\{\mathbf{MEI}_t(\mathbf{x})_{1\cdot},\ldots,\mathbf{MEI}_t(\mathbf{x})_{n\cdot}\}} (\mathbf{MEI}_t(\mathbf{x})_{i\cdot})
\end{eqnarray*}
\item Time/Correlation DepthGram: scatter plot of points $DG^{tc}_i=(DG^{tc}_{i1},DG^{tc}_{i2})$ where \begin{eqnarray*}
DG^{tc}_{i1}&=&1-MEI_{\{\mathbf{MBD}_t(\mathbf{\tilde{x}})_{1\cdot},\ldots,\mathbf{MBD}_t(\mathbf{\tilde{x}})_{n\cdot}\}} (\mathbf{MBD}_t(\mathbf{\tilde{x}})_{i\cdot})\\
DG^{tc}_{i2}&=&MBD_{\{\mathbf{MEI}_t(\mathbf{\tilde{x}})_{1\cdot},\ldots,\mathbf{MEI}_t(\mathbf{\tilde{x}})_{n\cdot}\}} (\mathbf{MEI}_t(\mathbf{\tilde{x}})_{i\cdot})
\end{eqnarray*}
with $\tilde{x}_i^j(t)=x_i^j(t) \prod_{k=2}^j sign(\rho(\mathbf{MEI}_d(\mathbf{x})_{\cdot\,k-1},\mathbf{MEI}_d(\mathbf{x})_{\cdot\,k}))$, where $\rho(\cdot,\cdot)$ is the Pearson's correlation coefficient function.
\end{itemize}
Indeed, the motivation for this third representation is the possible presence of negative association between some of the dimensions. The underlying idea is that of building up a new sample $\mathbf{x}$ with the same structure on the marginals but with positive association between dimensions, where here the association is understood as the linear correlation between curve ranks (in terms of $MEI$). Indeed, since the relative shape and position of the curves are preserved inside each component of the sample, depths computed marginally on each dimensions (those used to build the Dimension DepthGram) would be the same. However, since the transformed sample $\mathbf{\tilde{x}}$ has a positive association pattern along dimensions, the time DepthGram computed on it will be different from that built on $\mathbf{x}$ in that now multivariate observations at any time point will tend to have a more regular behavior for most of the observations and joint outliers will outstand in the parallel coordinate representation of these multivariate samples.

Unlike other existing tools based on outlyingness measures, the DepthGram representations are bounded in both horizontal ($MEI$ of $MBD$s) and vertical ($MBD$ of $MEI$s) axes, which eases interpretation. Indeed, it is not only a tool for visual identification of outliers, but for two-dimensional representation of the whole sample, which also allows to visualize central individuals and sample variability on time and dimensions.
Although the result of Proposition \ref{prop} establishes the conceptual basis for the definition of an outlier detection rule, as atypical observations will tend to lie above the parabola $g_n$, the determination of a threshold for this rule requires the approximation of the distribution of the distances to the parabola which is unfeasible in the high-dimensional setting that we consider (see section \ref{secLow} for an approximate non-optimized detection rule). 
However, the fact that scales are fixed on both axes of the DepthGram plots and that outliers are associated to particular values in these two-dimensional representations allows for visual identification of outliers. This is illustrated in Section \ref{simus}.

\section{Simulation study}\label{simus}

In this section we evaluate the performance of the proposed procedure via a simulation study in which we consider four different generating models in a high-dimensional functional setting. We do not compare our methodology to any competing method, since, up to our knowledge, alternative methods are restricted to low dimensional configurations. See section \ref{secLow} for a comparative study in such low dimensional settings.

Since the visualization tool proposed in this article does not provide an outlier detection rule that could be used to summarize performance through percentage of false and correctly identified outliers, we proceed otherwise to present our results. We want to show that over different simulation settings and different random replicates, the DepthGram on its three variations behaves as expected, that is, isolating different types of outliers on different areas of the DepthGram plot. For this, we have run the DepthGram on each synthetic data set and then represented together all the $(DG^k_{i1},DG^k_{i2})$, points from all data sets under the same simulation settings (for each DepthGram type, $k\in\{d,t,tc\}$). This works as a summary DepthGram plot in which 2-dimensional density contours are plotted according to the frequency of points in the plot area. The density contours are colored according to the type of observation they correspond to (non-outlying, or any of the three kinds of outliers) so that we can visually assess whether the procedure works at separating outlying from non-outlying observations and different type of outliers between them.
 
The simulation settings are as follows: we fix $n=100$ and $N=100$, we consider four different generating models, described below, and, for each model, we consider two values of the dimension of the data $p=10000, 50000$, and five values for the contamination rate, $c=0, 0.25, 0.5, 0.75, 1$. We generate 200 data sets under each of these simulation configurations. The level of contamination is defined in the following way: for every data set, we fix the number of outliers to $15$, with $5$ magnitude outliers, $5$ shape outliers and $5$ joint outliers. The parameter $c$ represents the proportion of dimensions on which the outlying curves are indeed outliers. That is, for $c=0$ there are no outlying curves in the sample, for $c=1$ there is a $15\%$ of outliers ($5\%$ of each kind) which are outliers (of each type) in every dimension, and for $c\in (0,1)$ there is a $15\%$ of outliers which are outliers only in a $100\cdot c \%$ of the dimensions. The choice of the dimensions in which these curves behave actually as outliers is done randomly and independently among the different curves.\\ 
 
 The general structure for the four models is the following: the $j$-th component, $j=1,\ldots,p$, of the $i$-th observation, $i=1,\ldots,n$ is given by
$$
X_i^j(t)=\left\{ \begin{array}{ll} 
X_i^0(t)h_j(t) + \varepsilon_{ij} (t), & \mbox{for } i,j \mbox{ non-outlying observ./component}\\
10+X_i^0(t)h_j(t) + \varepsilon_{ij} (t), &\mbox{for } i,j \mbox{ magnitude outlying observ./component}\\
X_i^{0s}(t)h_j(t) + \varepsilon_{ij} (t), &\mbox{for } i,j \mbox{ shape outlying observ./component}\\
X_{\ell_{ij}}^0(t)h_j(t) + \varepsilon_{ij} (t), &\mbox{for } i,j \mbox{ shape outlying observ./component}\\
\end{array}\right.
$$
with $t\in [0,1]$ and $\varepsilon_{ij}(t)$ are independent realizations of a Gaussian process with zero mean and covariance function $\gamma(s,t)=0.3 \exp \{-|s-t|/0.3\}$.\\
That is, the general model is a functional concurrent model or varying-coefficient model on each dimension from a reference data set $X^0_1,\ldots,X^0_n$ and with coefficient function $h_j(t)$ for the $j$-th dimension. Then, magnitude outliers are shifted upwards, shape outliers are generated with the same model but from a different reference set $X_i^{0s}$, and joint outliers are generated with the same model but applied, for each dimension on a different reference curve $X^0_{\ell_{ij}}$, where $\ell_{ij} \in \{1,\ldots,n\}$ is chosen in different ways depending on the particular model.

 \begin{itemize}
 \item[-] {\bf Model 1:} Let $t\in [0,1]$ and
 \begin{eqnarray*}
 X_i^0 (t) &=&sin(4\pi t) + \alpha_i, \quad i=1,\ldots,n\\
 X_i^{0s} (t) &=&cos(4\pi t+\pi/2) + \alpha_i , \quad \mbox{ for } i \mbox{ shape outlier}
\end{eqnarray*}
where $\alpha_i \sim N(0,1)$ are independent and identically distributed and let
$$h_j(t)=1+2t^{1+j/p}(1-t)^{2-j/p}, \quad j=1,\ldots,p.$$
 
For each pair of joint outlying observation/component $ij$, $\ell_{ij}$ is randomly chosen on the index subset of the non-outlying observations. That is, the $j$-th component of the $i$-th observation is linearly related to the realization of the reference process on a randomly chosen individual $\ell_{ij}$ instead of being related to $X_{i}^0(t)$.
This provides a way of introducing joint outliers that are not shape or magnitude marginal outliers in any dimension. Also, shape outliers are neither joint or magnitude outliers in the way they are generated, nor magnitude outliers are shape or joint outliers. This may help providing insight on how each different type of outlier is identified with our procedure. 
 \item[-] {\bf Model 2:} Let $t\in [0,1]$ and
  \begin{eqnarray*}
 X_i^0 (t) &=&4t + \alpha_i, \quad i=1,\ldots,n\\
 X_i^{0s} (t) &=&4t+2sin(4(t+0.5)\pi)+ \alpha_i , \quad \mbox{ for } i \mbox{ shape outlier}
\end{eqnarray*}
where $\alpha_i \sim N(0,1)$ are independent and identically distributed and let the varying coefficient functions be given by
$$h_j(t)=\left\{\begin{array}{ll} 1+2t^{1+j/p}(1-t)^{2-j/p} & \mbox{if } j \mbox{ is odd}\\
-1-2t^{1+j/p}(1-t)^{2-j/p} & \mbox{if } j \mbox{ is even}
 \end{array}\right. \quad j=1,\ldots,p.$$
Thus, between odd and even dimensions there is a negative correlation in the ordering of the curves. Notice that magnitude outliers are also joint outliers in this setting. The indexes $\ell_{ij}$ are chosen as in Model 1.
 \item[-] {\bf Model 3:} This model is the same as Model 1 except for the definition of $\ell_{ij}$. In this model the indexes $i$ for joint outliers are randomly chosen among the observations with lowest (for approximately half of them) and highest (for the other half) reference curves (in terms of the $\alpha_i$ values). Then, if $\alpha_i$ approximately corresponds to the $1-u$ sample quantile of $\mathbf{\alpha}$, $u\in(0,1)$, $\ell_{ij}$ is defined as
$$
\ell_{ij}=\left\{\begin{array}{ll}i \mbox{ if } j \mbox{ is odd} \\r_i \mbox{ if } j \mbox{ is even}  \end{array}\right.
$$
where $r_i$ is the index of a non-outlying observation whose $\alpha_{r_i}$ value is approximately the $1-u$ sample quantile of $\mathbf{\alpha}$.
That is, instead of having components that are independent, now the components of joint outliers exhibit an association pattern that is the opposed to the general one.
 \item[-] {\bf Model 4:} This model is the same as Model 2 but with $\ell_{ij}$ as defined in Model 3.
 \end{itemize}
 
\begin{figure}
\hspace*{-1cm}\includegraphics[width=18cm]{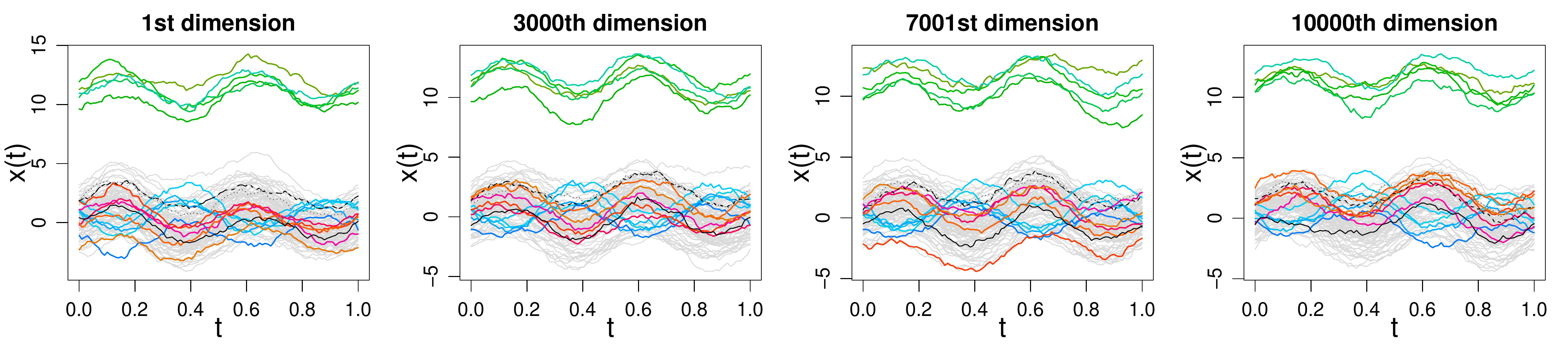}
\hspace*{-1cm}\includegraphics[width=18cm]{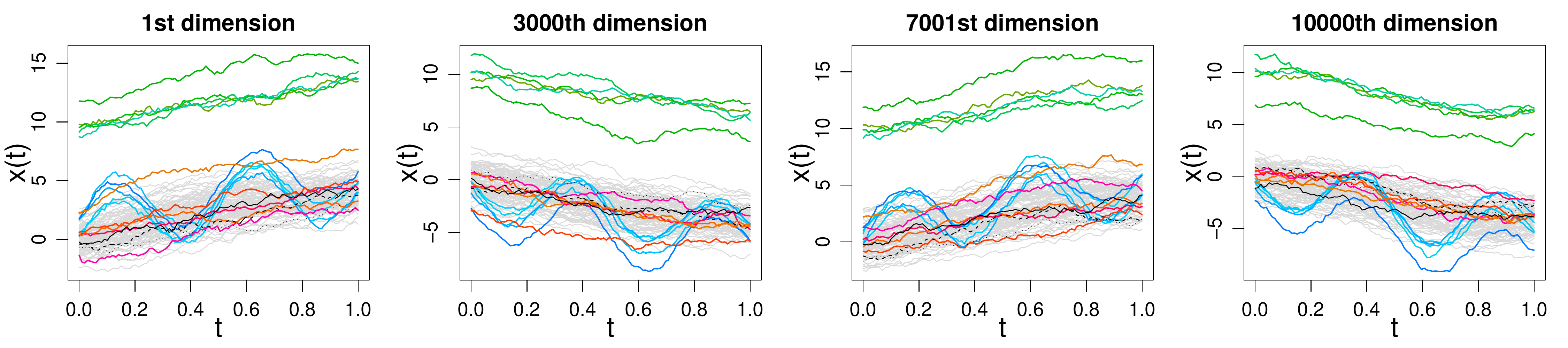}
\hspace*{-1cm}\includegraphics[width=18cm]{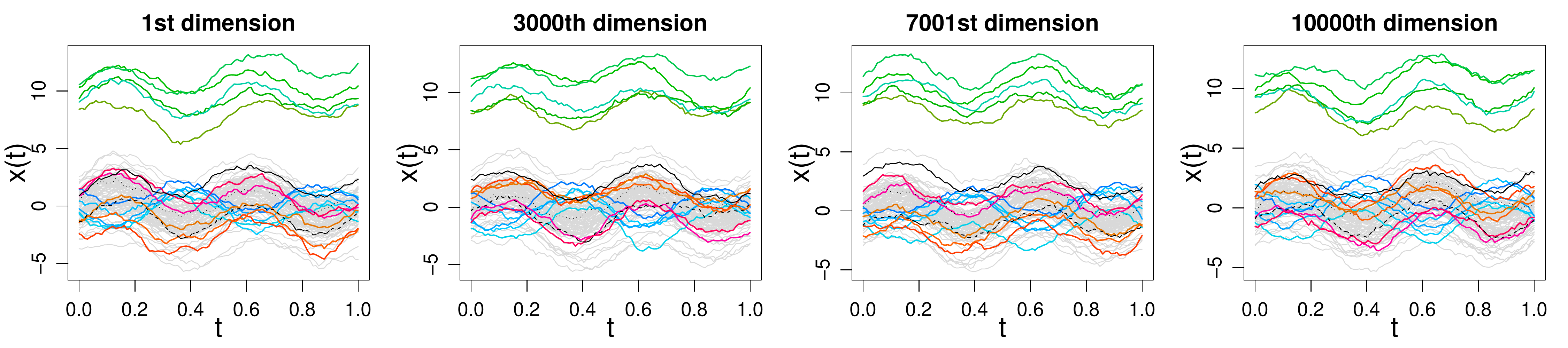}
\hspace*{-1cm}\includegraphics[width=18cm]{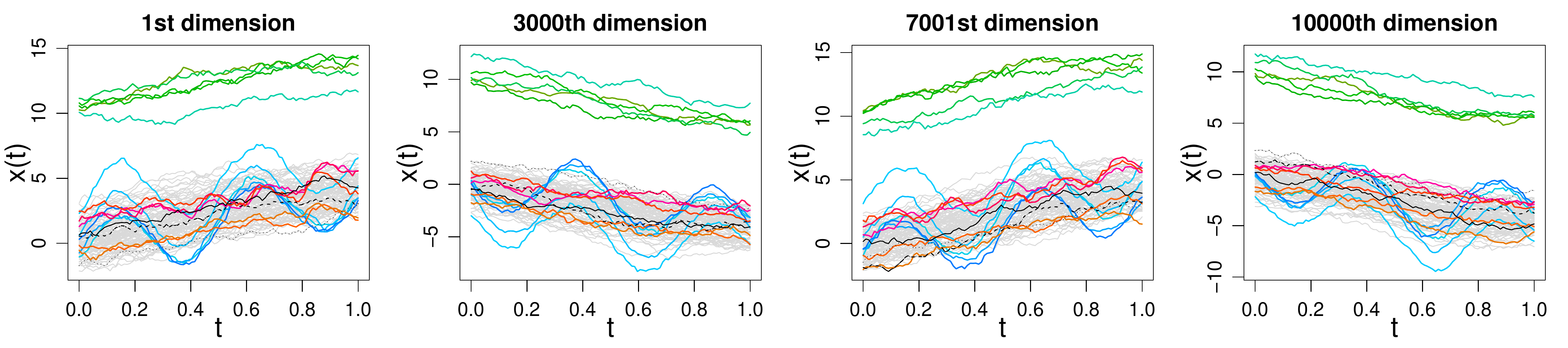}
\caption{Sample data sets of the four models and selected dimensions ($p=10000$, $c=1$). Each model is represented in a row, with row $i$ corresponding to the $i$-th model. Outliers are represented according to the following color code: magnitude outliers are displayed in green tones, shape outliers are displayed in blue tones and joint outliers are displayed in orange/red tones. The rest of the sample is drawn in gray. Three non-outlying curves are presented in black with solid, dashed and dotted lines to help visualize how the relative ordering between curves change among dimensions, which can be also noticed by looking at the variation of color positions for magnitude and shape outliers. In Models 1 and 3, the ordering is preserved across dimensions whereas in Models 2 and 4 the ordering is inverted in odd and even dimensions. In Models 1 and 2, the relative ordering of joint outliers across dimensions is random, whereas in Models 3 and 4, it follows a pattern opposed to that of the rest of the sample.}\label{Sim1234}
\end{figure}

\begin{figure}
\includegraphics[width=16cm]{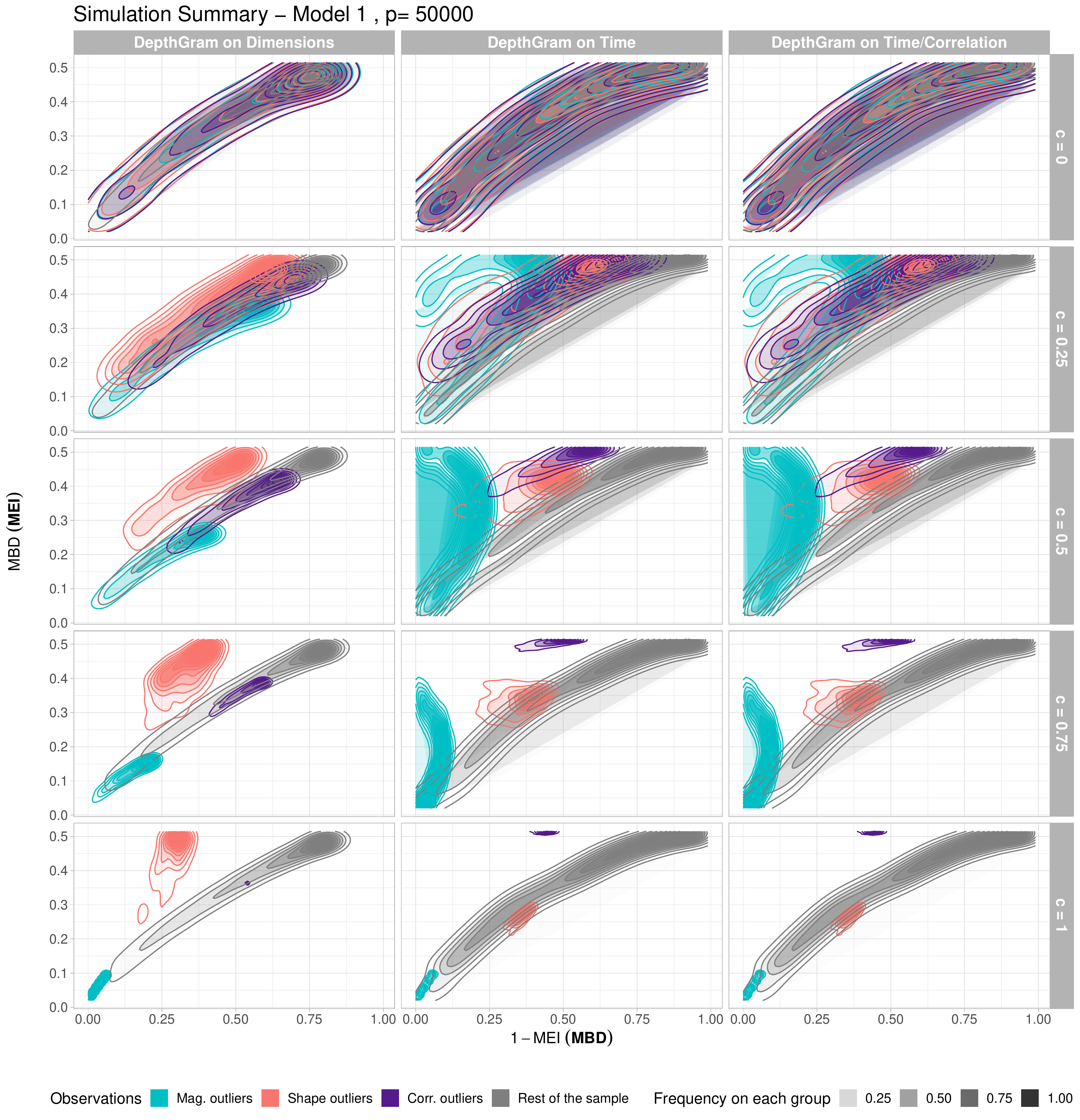}
\caption{Summary of 200 simulation runs under Model 1, with $p=50000$, and different contamination rates $c$. Summary DepthGrams are obtained as the density contours of mbd(epi) and 1-epi(mbd) points over the 200 simulated data sets. Colors stand for outlier classification (including non-outlying observations).}\label{Sim_mod1_p50}
\end{figure}

\begin{figure}
\includegraphics[width=16cm]{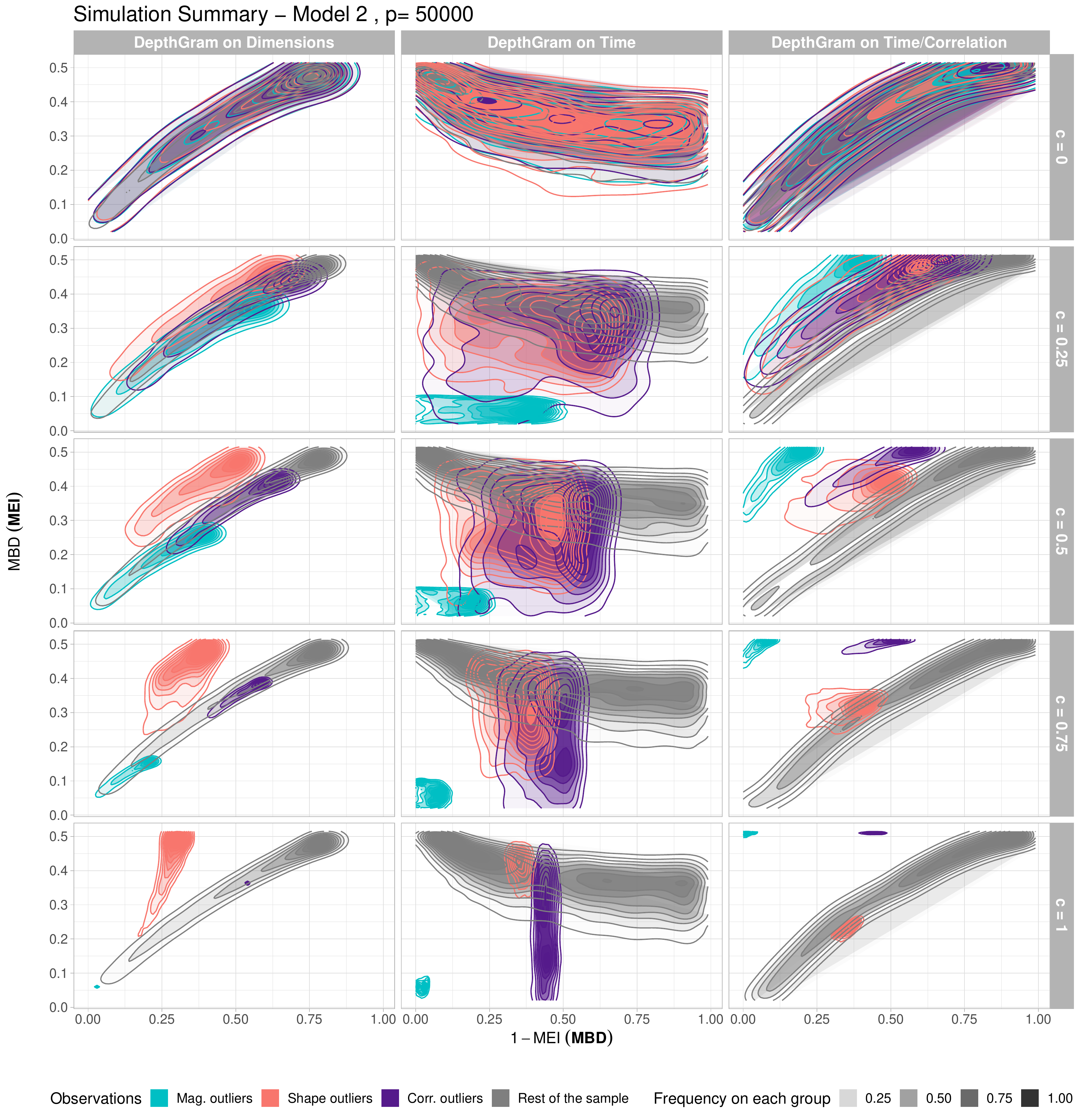}
\caption{Summary of 200 simulation runs under Model 2, with $p=50000$, and different contamination rates $c$. Summary DepthGrams are obtained as the density contours of mbd(epi) and 1-epi(mbd) points over the 200 simulated data sets. Colors stand for outlier classification (including non-outlying observations).}\label{Sim_mod2_p50}
\end{figure}

\begin{figure}
\includegraphics[width=16cm]{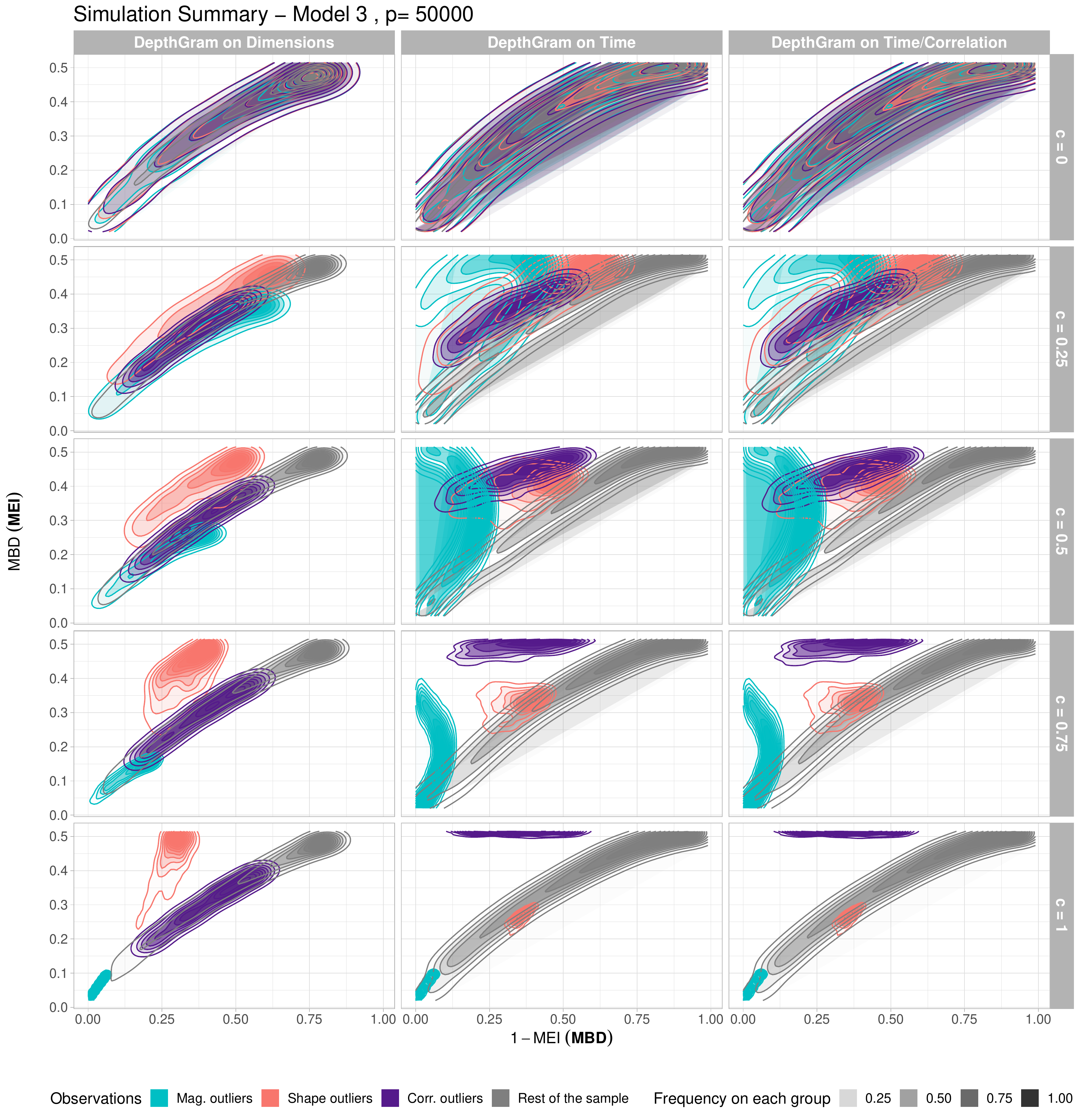}
\caption{Summary of 200 simulation runs under Model 3, with $p=50000$, and different contamination rates $c$. Summary DepthGrams are obtained as the density contours of mbd(epi) and 1-epi(mbd) points over the 200 simulated data sets. Colors stand for outlier classification (including non-outlying observations).}\label{Sim_mod3_p50}
\end{figure}

\begin{figure}
\includegraphics[width=16cm]{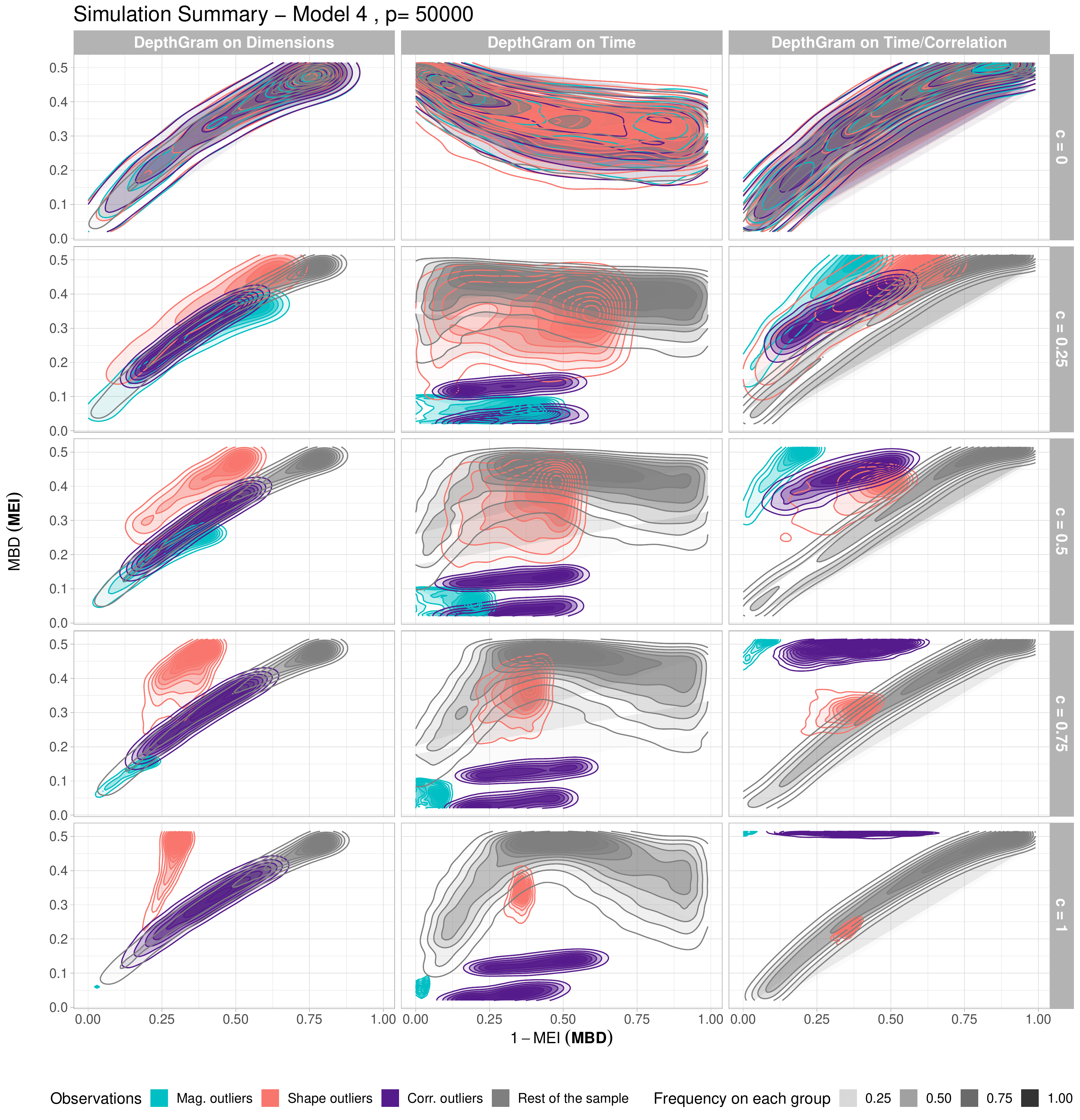}
\caption{Summary of 200 simulation runs under Model 4, with $p=50000$, and different contamination rates $c$. Summary DepthGrams are obtained as the density contours of mbd(epi) and 1-epi(mbd) points over the 200 simulated data sets. Colors stand for outlier classification (including non-outlying observations).}\label{Sim_mod4_p50}
\end{figure}

In Figure \ref{Sim1234} we present sample data sets generated under the four models.
The results of the simulations are presented in Figures \ref{Sim_mod1_p50} to \ref{Sim_mod4_p50}, were for conciseness only the results with $p=50000$ are shown. The DepthGram representations obtained with $p=10000$ look very similar to those presented here and can be found in the supplementary materials.\\ 
We can notice several things. First of all, in all four models, as the contamination rate $c$ decreases, all the observations (outliers of the different kinds and the rest of the sample) are mixed, which is an expected behavior. However, even for $c=0.25$ we can already find some separation between different classes of observations. Notice that in Models 2 and 4, magnitude outliers are also joint outliers, so it is not strange to find them together in the Time DepthGrams. In general, magnitude outliers are the most difficult to identify (except when they also behave as joint outliers), since even if they are always found at the left bottom corner of the Dimensions DepthGram, there might not be any separation at all between them and the rest of the typical observations. Indeed, this is true in this setting in which the order of the curves is very well preserved among dimensions. In other situations, non-outlying observations would have average ranks over dimension much smaller than magnitude outliers, and the separation would be more evident. However, the detection of magnitude and shape outliers is more of a marginal problem, as we explain below, and we are less concerned about it in the analysis of the DepthGram summaries.
We can also point out that in Model 4, where two different groups of joint outliers are generated (those taking high and low values in the reference set $X^0$), we found them as two separate cloud points in the time DepthGram. In Models 1 and 3, for which a positive correlation for curves ordering among dimensions exist, the time DepthGram and the time/dimension DepthGram are identical, as expected.

Additionally to the DepthGram analysis, in each data set we have conducted a marginal outlier detection through standard methods for magnitude and shape univariate functional outlier detection. We have used the functional boxplot \cite{SunGenton11} for magnitude outliers and the outliergram \cite{Outliergram} for shape outlier detection on each dimension of every high-dimensional functional data set. Notice that this can be easily incorporated to the DepthGram algorithm since both procedures rely on the same modified band depth and modified epigraph index quantities that are already computed dimension-wise for the DepthGram. But more interestingly, the incorporation of this step is also desirable since, by definition, magnitude and shape outliers in multivariate functional data sets are eminently marginal outliers, and informing the dimensions on which they are actually having an atypical behavior, and not only reporting an average outlyingness measure over all dimensions, is an advantage.
However, because of the computational burden of the high-dimensional setting, through the simulation study we used unoptimized versions the outliergram and the functional boxplot. The outlier detection rule in both cases mimics that of the univariate boxplot in which a factor value $F$ (typically $F=1.5$) needs to be specified so that the outlying observations are those lying below (resp. above) the first (resp. third) quartile minus (resp. plus) the interquartile range times $F$. Both methods are recommended to be used with data-driven choice of $F$ (see \citep{SunGenton12} for the adjusted version of the functional boxplot), which significantly increases their performances. However, because of time constraints when performing an extensive high-dimensional simulation study, the sub-efficient rule given by $F=1.5$ is used. Nonetheless, results are satisfactory as shown in table \ref{TabresultsSimus}, although the use of the optimized detection rule is feasible (and encouraged) when analyzing a single high-dimensional data set.

\begin{table}
\caption{Mean and standard deviation (in parentheses) of the proportion of correctly and falsely identified magnitude and shape outliers in the four simulation models over 200 simulation runs. Proportions are calculated considering, for each data set, the number of outlying and non-outlying curves as the sum over dimensions of the corresponding numbers on each dimension.
\label{TabresultsSimus} }
\hspace*{-1cm}{\small
\begin{tabular}{lcccccccc}
\hline
$p=10000$ &&&&&&&&\\
Magnitude outliers &\multicolumn{2}{c}{Model 1} &\multicolumn{2}{c}{Model 2}&\multicolumn{2}{c}{Model 3}&\multicolumn{2}{c}{Model 4}\\
\hline
Cont. level & $p_c$ & $p_f$ &  $p_c$ &$p_f$ & $p_c$ & $p_f$ & $p_c$ & $p_f$ \\
  \hline
$c=0$ &- & 0.005(0.004) & - &0.005(0.005)& - & 0.005(0.005) & - &0.005(0.005)\\                               
$c=0.25$ &1(0) & 0.006(0.004) &1(0) & 0.005(0.004)& 1(0) & 0.005(0.004) &1(0) & 0.006(0.004)\\                
$c=0.5$ &1(0) & 0.006(0.005) &1(0) & 0.007(0.005)& 1(0) & 0.006(0.005) &1(0) & 0.006(0.005)\\                 
$c=0.75$ &1(0.01) & 0.007(0.005) &1(0.01) & 0.007(0.005)& 1(0) & 0.007(0.005) &1(0.01) & 0.006(0.005)\\       
$c=1$ &1(0.01) & 0.007(0.007) &1(0.01) & 0.007(0.006)& 1(0.01) & 0.007(0.006) &1(0.01) & 0.007(0.006)\\       
\hline
$p=10000$&&&&&&&&\\
Shape outliers &\multicolumn{2}{c}{Model 1} &\multicolumn{2}{c}{Model 2}&\multicolumn{2}{c}{Model 3}&\multicolumn{2}{c}{Model 4}\\
\hline
Cont. level & $p_c$ & $p_f$ &  $p_c$ &$p_f$ & $p_c$ & $p_f$ & $p_c$ & $p_f$ \\
  \hline
$c=0$ &- & 0.039(0.003) & - &0.039(0.003)& - & 0.039(0.003) & - &0.04(0.003)\\                                
$c=0.25$ &0.955(0) & 0.036(0.002) &0.956(0) & 0.036(0.003)& 0.957(0) & 0.036(0.002) &0.961(0) & 0.036(0.002)\\
$c=0.5$ &0.956(0) & 0.033(0.002) &0.95(0) & 0.032(0.002)& 0.954(0) & 0.032(0.002) &0.947(0) & 0.033(0.002)\\  
$c=0.75$ &0.949(0) & 0.029(0.002) &0.952(0) & 0.029(0.002)& 0.95(0) & 0.029(0.002) &0.954(0) & 0.029(0.002)\\ 
$c=1$ &0.94(0) & 0.026(0.002) &0.943(0) & 0.026(0.002)& 0.951(0) & 0.026(0.002) &0.941(0) & 0.026(0.002)\\    
\hline
$p=50000$&&&&&&&&\\
Magnitude outliers &\multicolumn{2}{c}{Model 1} &\multicolumn{2}{c}{Model 2}&\multicolumn{2}{c}{Model 3}&\multicolumn{2}{c}{Model 4}\\
\hline
Cont. level & $p_c$ & $p_f$ &  $p_c$ &$p_f$ & $p_c$ & $p_f$ & $p_c$ & $p_f$ \\
  \hline
$c=0$ &- & 0.005(0.004) & - &0.005(0.004)& - & 0.004(0.004) & - &0.004(0.004)\\                               
$c=0.25$ &1(0) & 0.005(0.004) &1(0) & 0.006(0.005)& 1(0) & 0.006(0.004) &1(0) & 0.005(0.004)\\                
$c=0.5$ &1(0) & 0.006(0.004) &1(0) & 0.006(0.004)& 1(0) & 0.007(0.005) &1(0) & 0.006(0.005)\\                 
$c=0.75$ &1(0) & 0.007(0.005) &1(0) & 0.006(0.005)& 1(0.01) & 0.007(0.005) &1(0.01) & 0.007(0.005)\\          
$c=1$ &1(0.01) & 0.007(0.005) &1(0.01) & 0.007(0.005)& 1(0.01) & 0.007(0.006) &1(0) & 0.006(0.005)\\          
  \hline
$p=50000$&&&&&&&&\\
Shape outliers &\multicolumn{2}{c}{Model 1} &\multicolumn{2}{c}{Model 2}&\multicolumn{2}{c}{Model 3}&\multicolumn{2}{c}{Model 4}\\
\hline
Cont. level & $p_c$ & $p_f$ &  $p_c$ &$p_f$ & $p_c$ & $p_f$ & $p_c$ & $p_f$ \\
  \hline
$c=0$ &- & 0.04(0.002) & - &0.04(0.002)& - & 0.04(0.002) & - &0.04(0.002)\\                                   
$c=0.25$ &0.949(0) & 0.036(0.002) &0.949(0) & 0.036(0.002)& 0.953(0) & 0.036(0.002) &0.954(0) & 0.036(0.003)\\
$c=0.5$ &0.951(0) & 0.032(0.002) &0.948(0) & 0.032(0.002)& 0.95(0) & 0.032(0.002) &0.952(0) & 0.033(0.002)\\  
$c=0.75$ &0.952(0) & 0.029(0.002) &0.951(0) & 0.029(0.002)& 0.945(0) & 0.029(0.002) &0.955(0) & 0.029(0.002)\\
$c=1$ &0.948(0) & 0.026(0.002) &0.947(0) & 0.026(0.002)& 0.948(0) & 0.025(0.002) &0.954(0) & 0.026(0.002)\\ \hline
\end{tabular}}
\end{table}

\subsection{Low dimensional setting}\label{secLow}
In order to establish a reference with respect to other existing methods, in this section we compare the DepthGram with the \emph{Functional Outlier Map}, FOM \citep{Huberetal2015,Roussetal18}, and the \emph{Magnitude-Shape plot}, MS-plot \citep{DaiGenton18} as a tool for outlier detection. \\
The FOM is a two dimensional graphical representation of the data that can be used with any functional (integrated) depth or outlyingness measure. It displays the functional depth of each observation (which is obtained as an aggregation of multivariate depths over the observation domain) versus a measure of variability of the multivariate depth values for each time point. We consider here the FOM used with the \emph{functional directional outlyingness, fDO} as introduced in \citep{Roussetal18} for which an outlier detection rule is defined based on the distribution of the Euclidean distances of the FOM points to the origin, after scaling. \\
The MS-plot is based on an alternative definition of a \emph{directional outlyingness}, which assigns to a functional $p$-variate observation a $p$-dimensional vectorial (directional) outlyingness value. The MS-plot maps the multivariate functional data to multivariate points by representing each observation with its mean directional outlyingness vector and measure of its variability. When $p>2$, the graphical representation can be done by just representing the norm of the mean outlyingness vector versus its variability. The outlier detection rule relies on the approximation of the distribution of the robust Mahalanobis distance of the $(p+1)$-dimensional points of the MS-plot.\\
Both outlier detection procedures are designed for low dimensional settings and will fail if $p > n$. Indeed, FOM is used with fDO, and the multivariate outlyingness from which fDO is obtained by integrating over the time domain, is calculated through an approximate algorithm that relies on the assumption $p<n$. On the other hand, the MS-plot can be obtained for any value of $p$ but the associated outlier detection rule, in particular the approximation of the distribution of the robust Mahalanobis distance of the points, requires $p<n$. That is why for this second simulation study we will use the same four models and settings as before except for the values of $p$ which are now set to $p=10$ and $p=50$.\\
The two alternative methods considered have an important computational burden as $p$ increases, since they rely on the computation, over each point of the observation domain, of different $p$-variate outlyingness measures. To lighten this burden, we propose an alternative way to apply these outlier detection techniques, by considering the synthetic functional multivariate data sets as functional univariate data sets defined on a multivariate domain. That is, for each individual $i$ we can consider its observed realization as a $p$-variate function $x_i:{\cal I}
\longrightarrow \R^p$, $x_i(t)=(x_i^1(t),\ldots,x_i^p(t)$, or as surface or volume $x_i:{\cal I}\times D
\longrightarrow \R$, where $D$ is a continuous domain for which in practice the process is only observed at $p$ discretized points. Treating the data in this way the multivariate structure that may help detecting joint outliers is missed. However, we expect to identify this kind of outliers as shape outliers in the new functional univariate data set. Notice that for the DepthGram both approaches are equivalent and yield the same results. Indeed, the depth-related quantities involved in the construction of the DepthGram are computed over the $p\times N$ grid of all the dimensions and observation points.

In order to be able to establish a direct comparison in terms of detection rates, we define an empirical outlier detection rule for the DepthGram as follows:\begin{itemize}
\item For each one of the three DepthGram representations obtain $d^k_i=DG^k_{i2}-g_n(DG^k_{i1})$, $i=1,\ldots,n$, $k\in\{d,t,tc\}$.
\item Define three sets of outlying observations as $O^k=\{i=1,\ldots,n | d^k_i >Q_3(d^k)+ F\cdot IQR(d^k)\}$, $k\in\{d,t,tc\}$, where $Q_3$ and $IQR$ denote the sample third quartile and interquartile range, respectively.
\item Define the global set of outlying observations as $O= O^d\cup O^t \cup O^{tc}$.
\end{itemize}
The factor $F$ is set to 1.5 as in the classical boxplot rule. For this procedure, which is inspired by the results of Proposition \ref{prop}, to yield accurate results, we will need to use a data-driven estimation approach to approximate $F$, as it is done in the adjusted Outliergram \citep{Outliergram}. This, however, will require to approximate the distribution of $d^k$, $k\in\{d,t,tc\}$, which is unfeasible in a high-dimensional setting and is out of the scope of this paper. The objective here is only to provide a simple an approximate rule to be able to conduct the comparative analysis. Notice that the kind of graphical summaries that have been used in the high-dimensional simulation study to visually assess the performance of the DepthGram can not be used with the other two methods because the quantities that are represented in both the FOM and the MS-plot are mean and dispersion of outlyingness values, which are not bounded and might even exhibit very different ranges across simulation runs.\\

\begin{figure}
\includegraphics[width=16cm]{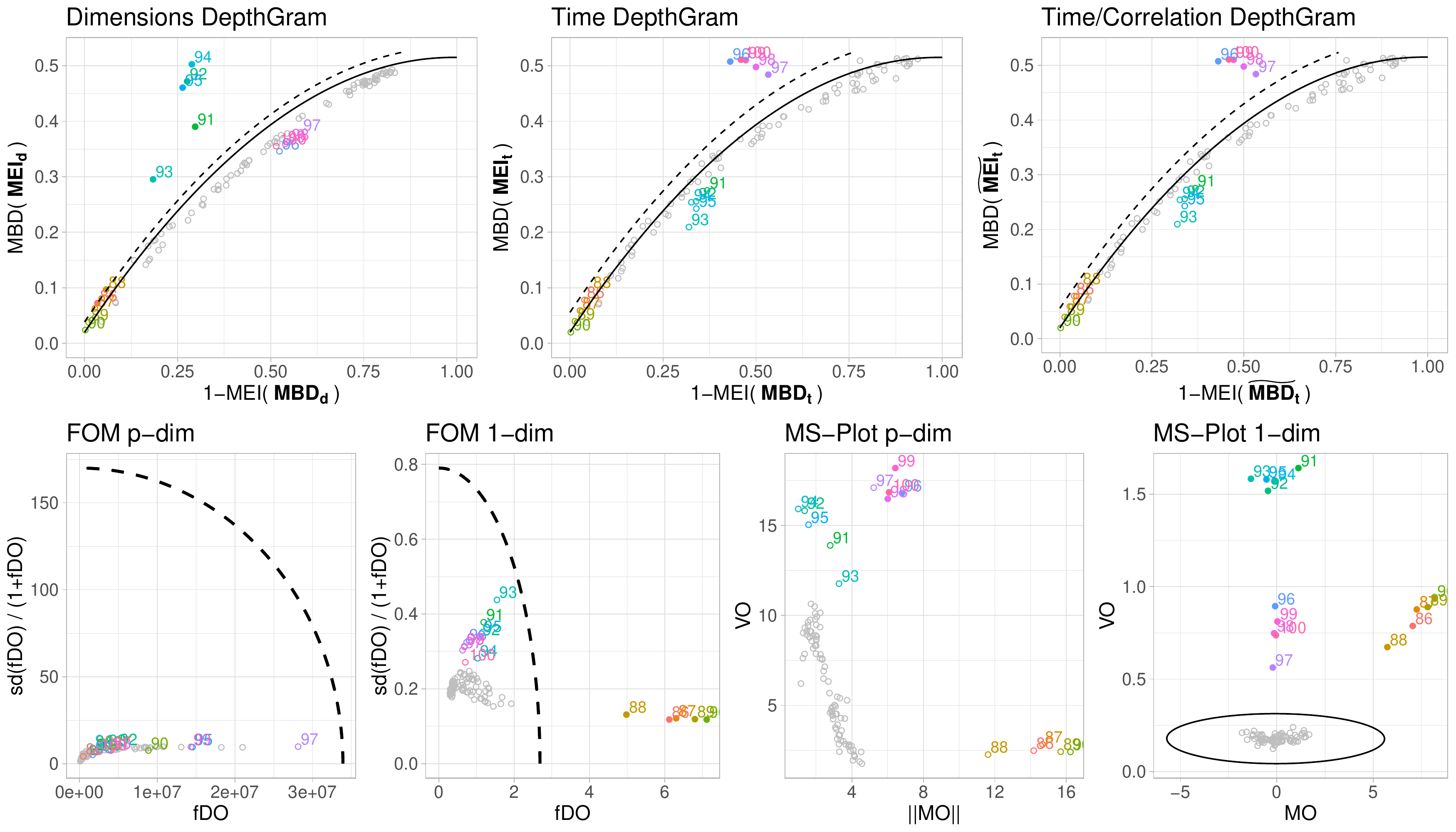}
\caption{Results for a single simulation run under Model 1, with $p=50$, and $c=1$. In the top row we present the three DepthGram representations. In the bottom row, we present the FOM and the MS-plot in their $p$-dimensional and $1$-dimensional versions. Except for the $p$-dimensional MS-plot, the boundary dividing the outlying and non-outlying observations is drawn (a dashed line for the DepthGram and FOM and a solid ellipse for the MS-plot). In all the plots, detected outliers are marked with a bullet while the rest of the observations are represented with a circle. True outliers are represented in color while non-outlying observations are drawn in gray. }\label{FigresultsSimusLowDimp50_Mod1}
\end{figure}

\begin{figure}
\includegraphics[width=16cm]{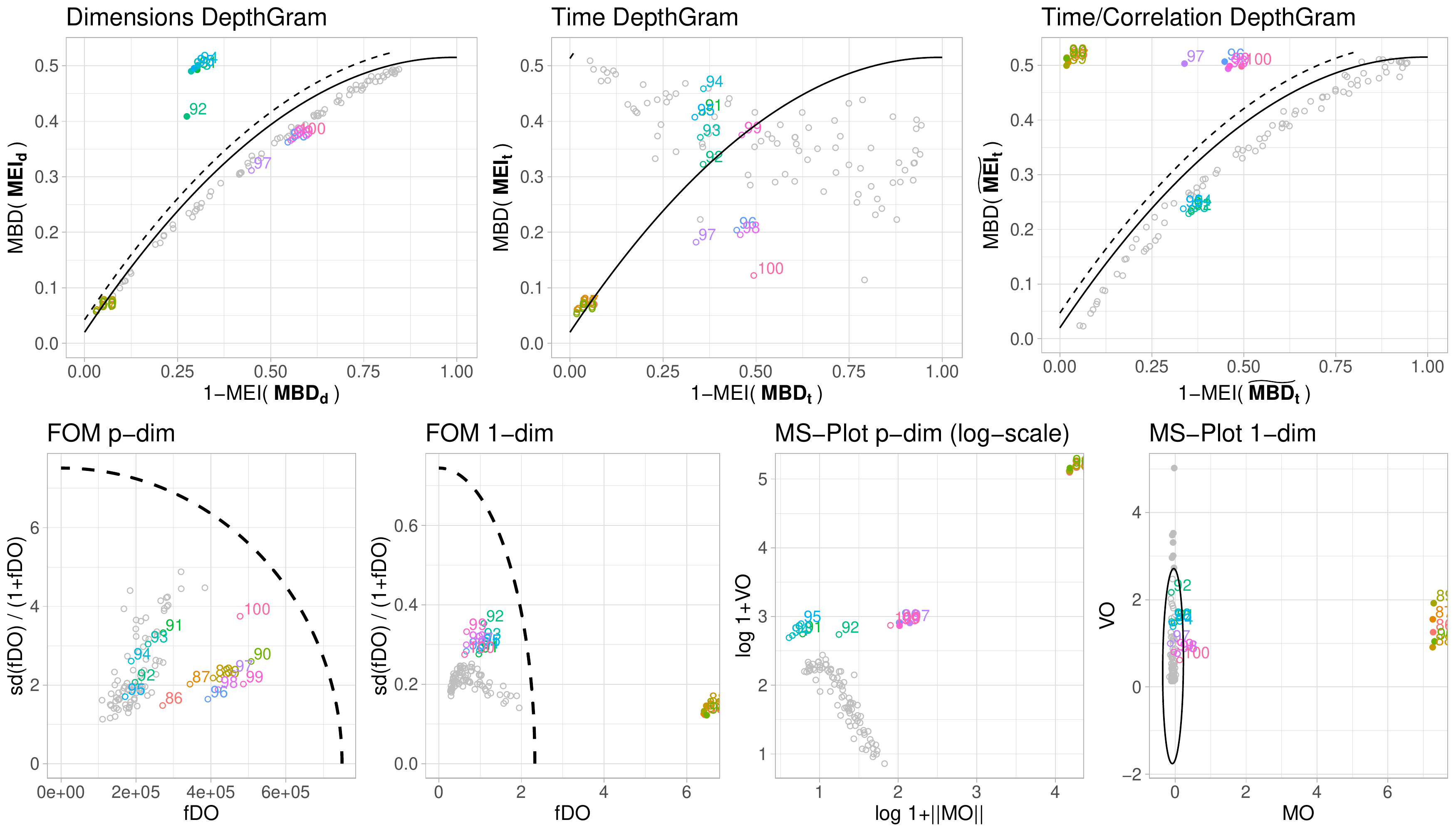}
\caption{Results for a single simulation run under Model 2, with $p=50$, and $c=1$. In the top row we present the three DepthGram representations. In the bottom row, we present the FOM and the MS-plot in their $p$-dimensional and $1$-dimensional versions. Except for the $p$-dimensional MS-plot, the boundary dividing the outlying and non-outlying observations is drawn (a dashed line for the DepthGram and FOM and a solid ellipse for the MS-plot). In all the plots, detected outliers are marked with a bullet while the rest of the observations are represented with a circle. True outliers are represented in color while non-outlying observations are drawn in gray. The $p$-dimensional MS-plot is presented in logarithmic scale to ease the visualization. For this particular simulation run, this method correctly detects as outliers all the magnitude outliers (86 to 90) and three joint outliers (96, 97 and 99).}\label{FigresultsSimusLowDimp50_Mod2}
\end{figure}

In Figures \ref{FigresultsSimusLowDimp50_Mod1} and \ref{FigresultsSimusLowDimp50_Mod2} we present the five graphical tools for two different simulation runs with $p=50$ from models 1 and 2 respectively. We can find similar graphical representations for models 3 and 4 in the supplementary material. In all simulation runs, outliers are coded as observations $86$ to $100$. In particular, observations $86$ to $90$ (colored in an orange/brown scale in the figures) are magnitude outliers, observations $91$ to $95$ (colored in a green/blue scale in the figures) are shape outliers and observations $96$ to $100$ are joint outliers (colored in a blue/purple scale in the figures). 
Tables 1 and 2 in the supplementary materials contain the results, in terms of the proportion of correctly identified and falsely identified outliers by each one of the methods, over the 200 simulation runs for the four models and different values of $p$. We can draw the following conclusions. The DepthGram behavior for both $p=10$ and $p=50$ is very similar to the one observed in a high-dimensional setting, with slightly better detection rates for $p=50$: shape outliers are detected through the Time DepthGram, joint outliers are detected through the Time/Correlation DepthGram and magnitude outliers are only detected when they are also joint outliers (models 2 and 4 with negative association among components). The FOM exhibits in general a low detection rate, with better performance in the lower dimension case ($p=10$) and in its $p$-dimensional version, except for magnitude outliers, which are always captured with this technique. For the MS-plot, the detection rates are very high in general, with better performance for $p=50$ and in its $1$-dimensional version. For this configuration ($p=50$, MS-plot $1$-dim) the method as a full detection capacity in models with positive association among components (models 1 and 3). However, for models 2 and 4, there is a high false positive detection rate (around $10\%$) and a low sensitivity for shape (both models) and joint outliers (only Model 2).

As we have mentioned in the previous section, marginal (shape and magnitude) outliers can be detected very efficiently with univariate functional detection methods. Moreover, these have the advantage, over multivariate methods, of identifying the components in which the outlying behavior happens. Thus, multivariate methods should focus on the detection of joint outliers, that, as we have seen, might be difficult in the presence of different types of outlying observations and, especially, negative association among dimensions. 

Another point for comparison is the computational complexity of the different methods. The DepthGram is a very efficient procedure since it is based on the computation of MBD and MEI which only requires the ranking of the observations at any time point and dimension. The outlyingness measures used for the FOM and MS-plot representations are heavier from a computational point of view, and even in their $1$-dimensional configuration in which they are computed over samples of real numbers at every time point and dimension, the computation times are significantly higher than those of the DepthGram (see the supplementary materials for details). 


\section{Task fMRI data exploration}\label{fMRIdata}

In this section we analyze two task fMRI experiments conducted on the same $n=100$ healthy individuals. Data (T1-weighted, T1w, and two tfMRI)  were obtained from the HCP database (\url{https://db.humanconnectome.org/}) and are described in detail in \cite{tfMRIData3}. Only the tfMRI acquired during two different functional tasks were selected for this study. The first stack of tfMRIs were acquired during a motor task where some visual cues asked the participants to either tap their left or right fingers. The second acquisition was performed during a language task where different stories or arithmetic operations were presented to the participants by means of an audio record and, after having listened to them, they were asked a question about what was heard and two possible answers were offered to be selected by pushing a button. Task and resting periods were alternated during a total duration of $T = 284$ and $316$ seconds for both motor and language experiments respectively.

Alongside the native T1w and tfMRI images, HCP provided the minimal pre-processed images \citep{tfMRIData2} which includes the tfMRI images spatially registered to a stereotactic space (MNI, Montreal Neurological Institute). These normalized images ($FOV = 91 \times 109 \times 91$ voxels of $2 mm$ isotropic resolution) were the ones used in this study. A binary image (1=brain, 0=background) defined in MNI space was used to select 192631 voxels.

The final data sets are composed by the brain activity measurements of $n=100$ subjects over $T=284$ and $T=316$ seconds. For each experiment, brain activity is recorded in $p=192631$ points of the brain, corresponding to those voxels of the $91\times 109 \times 91$ cube defining the common brain mask of the 100 individuals. Specifically, we have two sets of $100$ $192631-$dimensional curves observed over $284$ and $316$ time points respectively. The aim of this analysis is to help visualizing this high dimensional functional data set and to detect individuals with central and outlying brain activity patterns.

\begin{figure}
\includegraphics[trim = 0mm 5mm 0mm 15mm, clip, width=16cm]{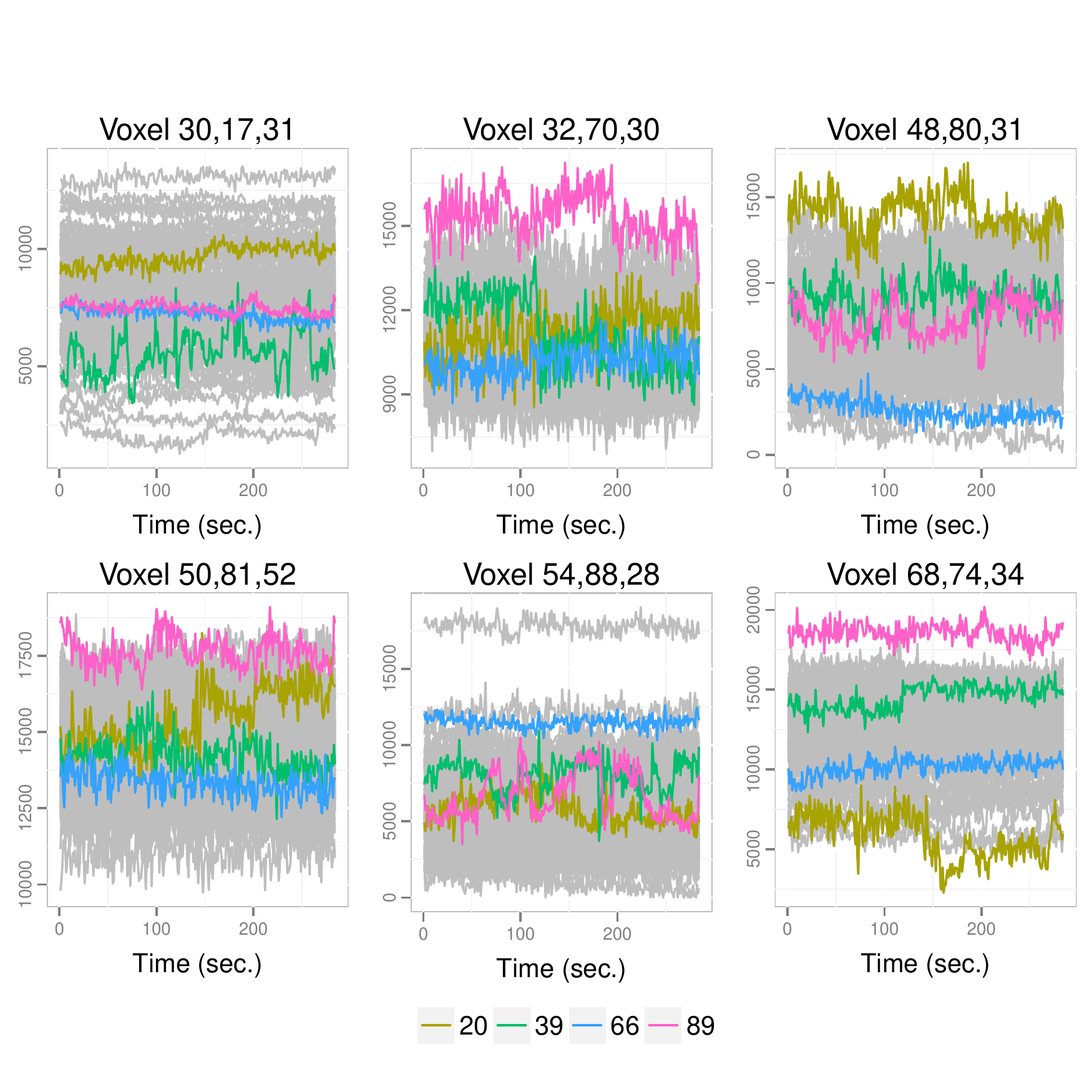}
\caption{Brain activity over time of $n=100$ individuals in 6 of the $p=192631$ voxels for the motor experiment. The activity brain of individuals $20$, $39$, $66$ and $89$ is highlighted.}\label{brain1}
\end{figure}

As an illustration of the kind of signals analyzed, in Figure \ref{brain1} we present the brain activity of all the individuals in 6 selected voxels for the motor experiment.

\begin{figure}
\begin{center}
\hspace*{-0.5cm}\includegraphics[width=18cm]{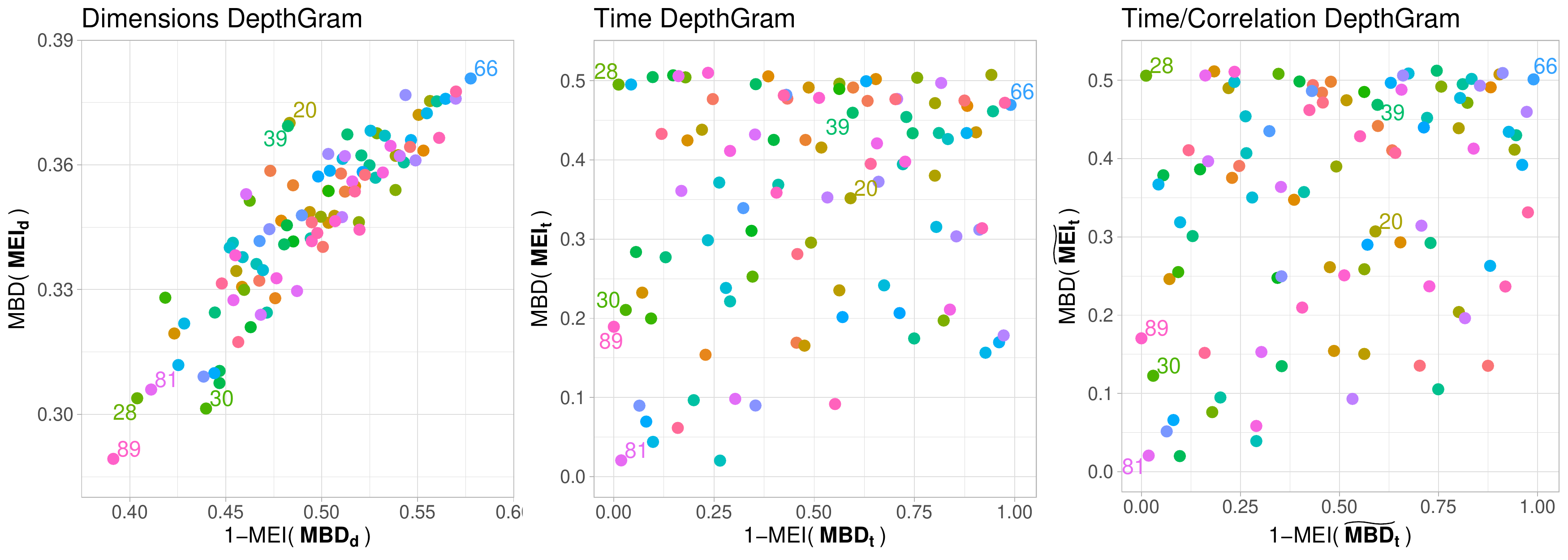}
\end{center}
\caption{The three DepthGrams for the motor experiment.}\label{motor1}
\end{figure}

The first step to analyse this data set is to obtain the $MEI$ and $MBD$ for each individual on each voxel and on each time point. Then, we proceed to compute the $MBD$ of $MEI$'s and $MEI$ of $MBD$'s to finally obtain the Depthgrams representation. DepthGrams for the motor experiment are shown in Figure \ref{motor1}. As we can observe, the time DepthGram is very spread on both dimensions, meaning that there is a lot of mixing/crossing of individuals across dimensions (voxels). When considering the time/correlation DepthGram, point coordinates are different but the global structure is the same. This heterogeneity across voxels does not follow an structured pattern and is rather the result of independent components. This might be due to the fact that only very specific regions of the brain are involved, and the expected to be activated, in the motor task, so the signals in the rest of voxels outside of these regions act as noise in this experiment, inducing this independence pattern across voxels.

\begin{figure}
	\begin{center}
		\hspace*{-0.5cm}\includegraphics[width=18cm]{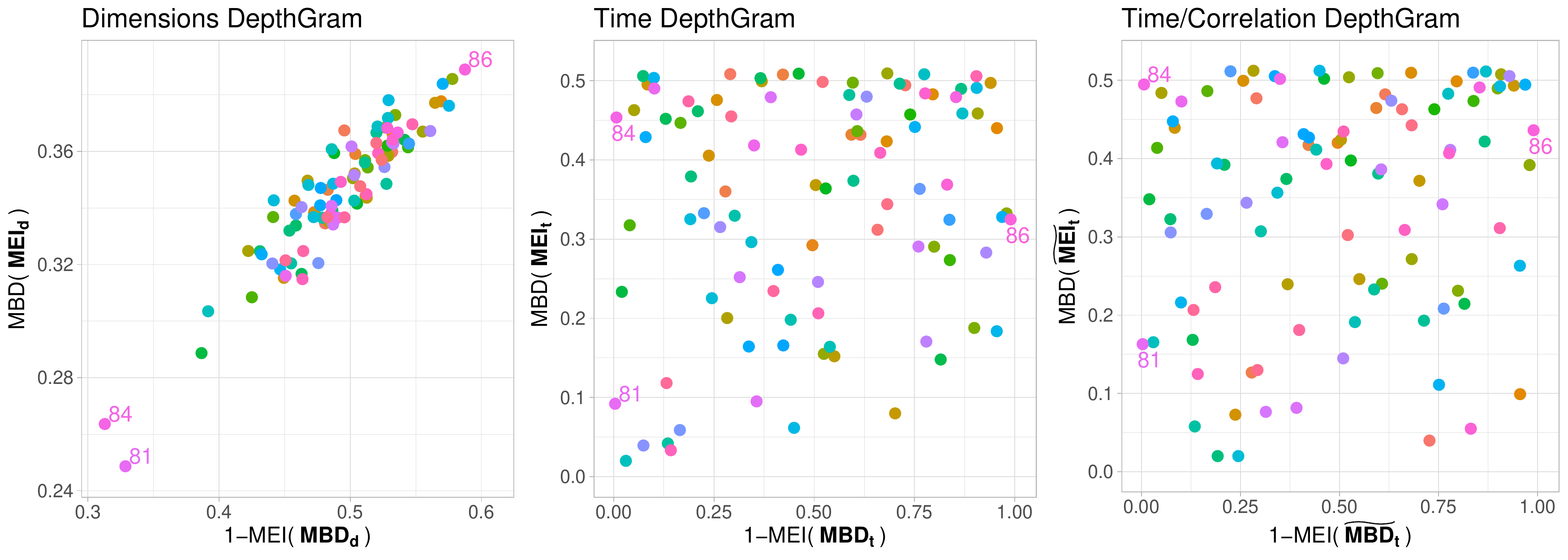}
	\end{center}
	\caption{The three DepthGrams for the language experiment.}\label{lang1}
\end{figure}

In Figure \ref{lang1} we show the same views but now for the language experiment. We can appreciate how the time and time/correlation DepthGrams follow the same unstructured pattern as in the motor experiment, since again, the regions involved in the language task represent a small part of the whole brain.

Regarding outlier detection, let us first point out that for the data set analysed meets quality standards in the fied since clinical diagnosis for several mental conditions were considered to exclude subjects from the experiment and standard fMRI techniques for artifact removal had been applied to the signals as a preprocessing step. Nevertheless, the Depthgrams have allowed to identify outlying patients. In the motor experiment, individual 39 is located in the north-west area of the voxel Depthgram as a potential shape outlier. After posterior examination, this subject happens to have moved more than what is reasonable during the experiment, and thus should have been removed from the sample. The same happens with subject 84 in the language experiment, that would be classified as a magnitude outlier. This different consideration of the outlier type in these two cases, which seem to share the source of noise, might be due to the different nature of the task performed in each experiment and how motion interferes with it. Moreover, individual 81, which appears to be a magnitude outlier in the language experiment, has been shown, by a posterior examination, to suffer a mild form of schizophrenia that had not been diagnosed.


\section{Discussion}\label{disc}

This article proposes the DepthGram as a tool for representing high-dimensional functional observations in the plane. Unlike current approaches that deal with depth notions for multivariate functional data through an aggregation over dimensions or an integration over the time domain of suitable functional or multivariate depth measures, our methodology relies on the \emph{depth of depths}. Indeed, the variables that define the 2-dimensional representation of the data are depth measures/indexes on the \emph{pseudo-functional} data sets obtained by computing depth on each dimension of the resulting multivariate data set for each time point. This approach allows to identify different types of outliers in different parts of the plot, including joint outliers. It is computationally efficient in the high-dimensional setting and, unlike procedures relying on outlyingness measures, it also allows to provide a global overview of the sample composition.

There are three versions of the DepthGram: The dimensions DepthGram, the time DepthGram and the time/correlation DepthGram. They are designed to be used together since they provide complementary information. The first one is most useful at identifying shape and magnitude outliers. The time DepthGram aims at identifying joint outliers, that is, those observations that are not marginal outliers in any of the dimensions but have a dependency pattern among dimensions different from the rest of the sample. The time/correlation DepthGram is designed to do the same in situations in which the general association pattern among dimensions is highly variable, and the time DepthGram fails to provide an structured representation of the sample. Indeed, the comparison of the time DepthGram and the time/correlation DepthGram sheds light on association patterns across dimensions, where here association is understood as linear correlation among the curve ranks given by the modified epigraph index. In general, for the three DepthGrams, the more similar and smooth the curves, and the more regular the association pattern in the sample, the more structured the DepthGram representations. That is, we can also get insight on the regularity or homogeneity of the sample by the spread of the DepthGrams representations.
Finally, we suggest to combine the DepthGram with specific tools for the detection of marginal outliers across the different univariate functional samples. Indeed, the challenge in a high-dimensional functional setting and what can not be achieved with existing tools for univariate functional data is to be able to detect atypical joint behaviour. If marginal outliers are present in the sample, not only it is more efficient to use specific methods marginally, but also it is more useful to identify the dimensions in which the observation exhibits an outlying behaviour than just classify the whole observation as an outlier. For this purpose, we recommend the use of the functional boxplot \citep{SunGenton11} and the outliergram \citep{Outliergram}, since they rely on the same depth tools than the DepthGram and can be computed simultaneously and efficiently.

\section*{Acknowledgments}

The authors are grateful to Luis Marcos Vidal and Daniel Martín de Blas for their insight on the tfMRI dataset.\\
Ana Arribas-Gil, Antonio Elías and Juan Romo acknowledge financial support from grant ECO2015-66593-P of the Ministerio de Economía y Competitividad, Spain. Ana Arribas-Gil also acknowledges financial support from grant MTM2014-56535-R of the same funding agency.

\bibliographystyle{plainnat}
\bibliography{bib_DepthGram}

\begin{thebibliography}{20}
\providecommand{\natexlab}[1]{#1}
\providecommand{\url}[1]{\texttt{#1}}
\expandafter\ifx\csname urlstyle\endcsname\relax
  \providecommand{\doi}[1]{doi: #1}\else
  \providecommand{\doi}{doi: \begingroup \urlstyle{rm}\Url}\fi

\bibitem[Arribas-Gil and Romo(2014)]{Outliergram}
A.~Arribas-Gil and J.~Romo.
\newblock Shape outlier detection and visualization for functional data: the
  outliergram.
\newblock \emph{Biostatistics}, 15(4):\penalty0 603--619, 2014.

\bibitem[Arribas-Gil and Romo(2015)]{Ana_DiscussionHuber}
A.~Arribas-Gil and J.~Romo.
\newblock Discussion of ``{M}ultivariate functional outlier detection''.
\newblock \emph{Statistical Methods and Applications}, 24:\penalty0 263--267,
  2015.

\bibitem[Barch et~al.(2013)Barch, Burgess, Harms, Petersen, Schlaggar,
  Corbetta, Glasser, Curtiss, Dixit, Feldt, Nolan, Bryant, Hartley, Footer,
  Bjork, Poldrack, Smith, Johansen-Berg, Snyder, Van~Essen, and
  Consortium]{tfMRI}
D.~M. Barch, G.~C. Burgess, M.~P. Harms, S.~E. Petersen, B.~L. Schlaggar,
  M.~Corbetta, M.~F. Glasser, S.~Curtiss, S.~Dixit, C.~Feldt, D.~Nolan,
  E.~Bryant, T.~Hartley, O.~Footer, J.~M. Bjork, R.~Poldrack, S.~Smith,
  H.~Johansen-Berg, A.~Z. Snyder, D.~C. Van~Essen, and WU-Minn~HCP Consortium.
\newblock Function in the human connectome: task-fmri and individual
  differences in behavior.
\newblock \emph{NeuroImage}, 80:\penalty0 169--189, 2013.

\bibitem[Chiou and M\"{u}ller(2014)]{flies}
J.-M. Chiou and H.-G. M\"{u}ller.
\newblock Linear manifold modelling of multivariate functional data.
\newblock \emph{Journal of the Royal Statistical Society, series B},
  76:\penalty0 605--626, 2014.

\bibitem[Claeskens et~al.(2014)Claeskens, Hubert, L., and
  Vakili]{Claeskens_JASA14}
G.~Claeskens, M.~Hubert, Slaets L., and K.~Vakili.
\newblock Multivariate functional halfspace depth.
\newblock \emph{Journal of the American Statistical Association}, 109
  (505):\penalty0 411--423, 2014.

\bibitem[Cook and Swayne(2007)]{ParaCoord}
D.~Cook and D.~F. Swayne.
\newblock \emph{{{I}nteractive and {D}ynamic {G}raphics for {D}ata {A}nalysis
  With {R} and {GG}obi}}.
\newblock {Springer}, 2007.

\bibitem[Dai and Genton(2018)]{DaiGenton18}
W.~Dai and M.~G. Genton.
\newblock Multivariate functional data visualization and outlier detection.
\newblock \emph{Journal of Computational and Graphical Statistics},
  27:\penalty0 923--934, 2018.

\bibitem[Glasser et~al.(2016)Glasser, Smith, Marcus, Andersson, Auerbach,
  Behrens, Coalson, Harms, Jenkinson, Moeller, Robinson, Sotiropoulos, Xu,
  Yacoub, Ugurbil, and Van~Essen]{tfMRIData2}
M.~F. Glasser, S.~M. Smith, D.~S. Marcus, J.~L. Andersson, E.~J. Auerbach,
  T.~E. Behrens, T.~S. Coalson, M.~P. Harms, M.~Jenkinson, S.~Moeller, E.~C.
  Robinson, S.~N. Sotiropoulos, J.~Xu, E.~Yacoub, K.~Ugurbil, and D.~C.
  Van~Essen.
\newblock The human connectome project's neuroimaging approach.
\newblock \emph{Nature Neuroscience}, 19:\penalty0 1175--1187, 2016.

\bibitem[Hodge et~al.(2016)Hodge, Horton, Brown, Herrick, Olsen, Hileman,
  McKay, Archie, Cler, Harms, Burgess, Glasser, Elam, Curtiss, Barch,
  Oostenveld, Larson-Prior, Ugurbil, Van~Essen, and Marcus]{tfMRIData3}
M.~R. Hodge, W.~Horton, T.~Brown, R.~Herrick, T.~Olsen, M.~E. Hileman,
  M.~McKay, K.~A. Archie, E.~Cler, M.~P. Harms, G.~C. Burgess, M.~F. Glasser,
  J.~S. Elam, S.~W. Curtiss, D.~M. Barch, R.~Oostenveld, L.~J. Larson-Prior,
  K.~Ugurbil, D.~C. Van~Essen, and D.~S. Marcus.
\newblock Connectomedb - sharing human brain connectivity data.
\newblock \emph{NeuroImage}, 124 B:\penalty0 1102–1107, 2016.

\bibitem[Hubert et~al.(2015)Hubert, Rousseeuw, and Segaert]{Huberetal2015}
M.~Hubert, P.~Rousseeuw, and P.~Segaert.
\newblock Multivariate functional outlier detection.
\newblock \emph{Statistical Methods and Applications}, 24:\penalty0 177--202,
  2015.

\bibitem[Ieva and Paganoni(2013)]{Ieva_Paganoni}
F.~Ieva and A.~M. Paganoni.
\newblock Depth measures for multivariate functional data.
\newblock \emph{Communications in Statistics - Theory and Methods},
  42(7):\penalty0 1265--1276, 2013.

\bibitem[Ieva and Paganoni(2020)]{Ieva_Paganoni2}
F.~Ieva and A.~M. Paganoni.
\newblock Component-wise outlier detection methods for robustifying
  multivariate functional samples.
\newblock \emph{Statistical Papers}, 61:\penalty0 595–614, 2020.
\newblock \doi{10.1007/s00362-017-0953-1}.

\bibitem[L\'{o}pez-Pintado and Romo(2009)]{BandDepth}
S.~L\'{o}pez-Pintado and J.~Romo.
\newblock On the concept of depth for functional data.
\newblock \emph{Journal of the American Statistical Association},
  104(486):\penalty0 718--734, 2009.

\bibitem[L\'{o}pez-Pintado and Romo(2011)]{HalfRegion11}
S.~L\'{o}pez-Pintado and J.~Romo.
\newblock A half-region depth for functional data.
\newblock \emph{Computational Statistics \& Data Analysis}, 55:\penalty0
  1679--1695, 2011.

\bibitem[Nieto-Reyes and Cuesta-Albertos(2015)]{NietoCuesta_DiscussionHuber}
A.~Nieto-Reyes and J.~A. Cuesta-Albertos.
\newblock {M}. {H}ubert, {P}. {R}ousseeuw and {P}. {S}egaert: {M}ultivariate
  functional outlier detection.
\newblock \emph{Statistical Methods and Applications}, 24:\penalty0 237--243,
  2015.

\bibitem[Rousseeuw et~al.(2018)Rousseeuw, Raymaekers, and Hubert]{Roussetal18}
P.~J. Rousseeuw, J.~Raymaekers, and M.~Hubert.
\newblock A measure of directional outlyingness with applications to image data
  and video.
\newblock \emph{Journal of Computational and Graphical Statistics}, 27\penalty0
  (2):\penalty0 345--359, 2018.

\bibitem[Sun and Genton(2011)]{SunGenton11}
Y.~Sun and M.~G. Genton.
\newblock Functional boxplots.
\newblock \emph{Journal of Computational and Graphical Statistics},
  20:\penalty0 316--334, 2011.

\bibitem[Sun and Genton(2012)]{SunGenton12}
Y.~Sun and M.~G. Genton.
\newblock Adjusted functional boxplots for spatio-temporal data visualization
  and outlier detection.
\newblock \emph{Environmetrics}, 23:\penalty0 54--64, 2012.

\bibitem[Sun et~al.(2012)Sun, Genton, and Nychka]{FastBD}
Y.~Sun, M.~G. Genton, and D.~C. Nychka.
\newblock Exact fast computation of band depth for large functional datasets:
  How quickly can one million curves be ranked?
\newblock \emph{Stat}, 1:\penalty0 68--74, 2012.

\bibitem[Zhang et~al.(2016)Zhang, Li, Lv, Jiang, Guo, and Liu]{tfMRI2}
S.~Zhang, X.~Li, J.~Lv, X.~Jiang, L.~Guo, and T.~Liu.
\newblock Characterizing and differentiating task-based and resting state fmri
  signals via two-stage sparse representations.
\newblock \emph{Brain imaging and behavior}, 10:\penalty0 21--32, 2016.

\end{thebibliography}

\appendix

This document is a supplement to the main text. It contains the proof of Proposition 1, additional figures and tables about the high-dimensional and low-dimensional simulation studies as well as a comparative analysis of computation times of the DepthGram and alternative methods.

\section{Appendix: Proof of Proposition 1}
\begin{dem}
	\setcounter{equation}{3}
	To simplify the notation, let us denote by $U=(u_{ij})^{j=1,\ldots,p}_{i=1,\ldots,n}$ the matrix $\mathbf{MEI}_d(\mathbf{x})$, whose rows $u_i$ will be considered as functional observations recorded as discrete points $j=1,\ldots,p$. In the same way, let us denote by by $V=(v_{ij})^{j=1,\ldots,p}_{i=1,\ldots,n}$ the matrix $\mathbf{MBD}_d(\mathbf{x})$.  Under assumption \emph{a)}, that is, if the original curves \textbf{x} do not cross in any of the dimensions $1,\ldots,p$, then, by (1) (in the main text), it holds that
	$$v_{ij}=MBD_{\{x^j_1,\dots,x^j_n\}} (x^j_i) = f_n\left(  MEI_{\{x^j_1,\dots,x^j_n\}} (x^j_i)\right) =f_n(u_{ij}), \quad i=1,\ldots,n,\,j=1,\ldots,p. $$
	Moreover, if  \emph{b)} also holds, that is, if for any time point, the order of individual curves across dimensions is preserved, then both $U$ and $V$ \emph{functional data sets} consist on constant functions, since the values of MBD and MEI will be constant across dimensions. In that case, again by (1) (in the main text) applied to the data set of non-crossing curves $U$, we have that
	\begin{equation}\label{ep}MBD_{\{u_1,\dots,u_n\}} (u_i) = f_n\left(  MEI_{\{u_1,\dots,u_n\}} (u_i)\right), \quad i=1,\ldots,n.\end{equation}
	But since $v_{i}=f_n(u_{i})$, where $f_n$ is now applied to all the components of $u_i$, then, by the symmetry around $x_0=\frac{n+1}{2n}$ of the parabola $f_n$ and its monotonicity in $(-\infty,x_0)$ and $[x_0,\infty)$ we get $MEI_{\{v_1,\dots,v_n\}} (v_i) = \frac{1}{n}+2|MEI_{\{u_1,\dots,u_n\}} (u_i) -x_0|$, $i=1,\ldots,n$. Equivalently, we get
	$$MEI_{\{u_1,\dots,u_n\}} (u_i) =\left\{ \begin{array}{ll} x_0+\frac{1}{2}\left(MEI_{\{v_1,\dots,v_n\}} (v_i)-\frac{1}{n}\right), & \mbox{if }  MEI_{\{u_1,\dots,u_n\}} (u_i) \geq x_0 \\
	x_0-\frac{1}{2}\left(MEI_{\{v_1,\dots,v_n\}} (v_i)-\frac{1}{n}\right), & \mbox{if }  MEI_{\{u_1,\dots,u_n\}} (u_i) < x_0
	\end{array}\right. .$$
	Because of the symmetry of $f_n$ around $x_0$, replacing this last expression in (\ref{ep}) yields
	$$MBD_{\{u_1,\dots,u_n\}} (u_i) = f_n\left(  x_0-\frac{1}{2}\left(MEI_{\{v_1,\dots,v_n\}} (v_i)-\frac{1}{n}\right)\right)
	= g_n\left(1-MEI_{\{v_1,\dots,v_n\}} (v_i) \right), \quad i=1,\ldots,n$$
	which, by switching back to the original notation, is the stated result.\\
	Notice that if \emph{b)} does not hold, then (\ref{ep}) becomes an inequality and so does the final result.
\end{dem}

\section{Appendix: High dimensional simulation study}
In this section we present the graphical summaries for the case $p=10000$ in the high-dimensional simulation study (section 3).
\begin{figure}[h!]
	\hspace*{-1cm}\includegraphics[width=18cm]{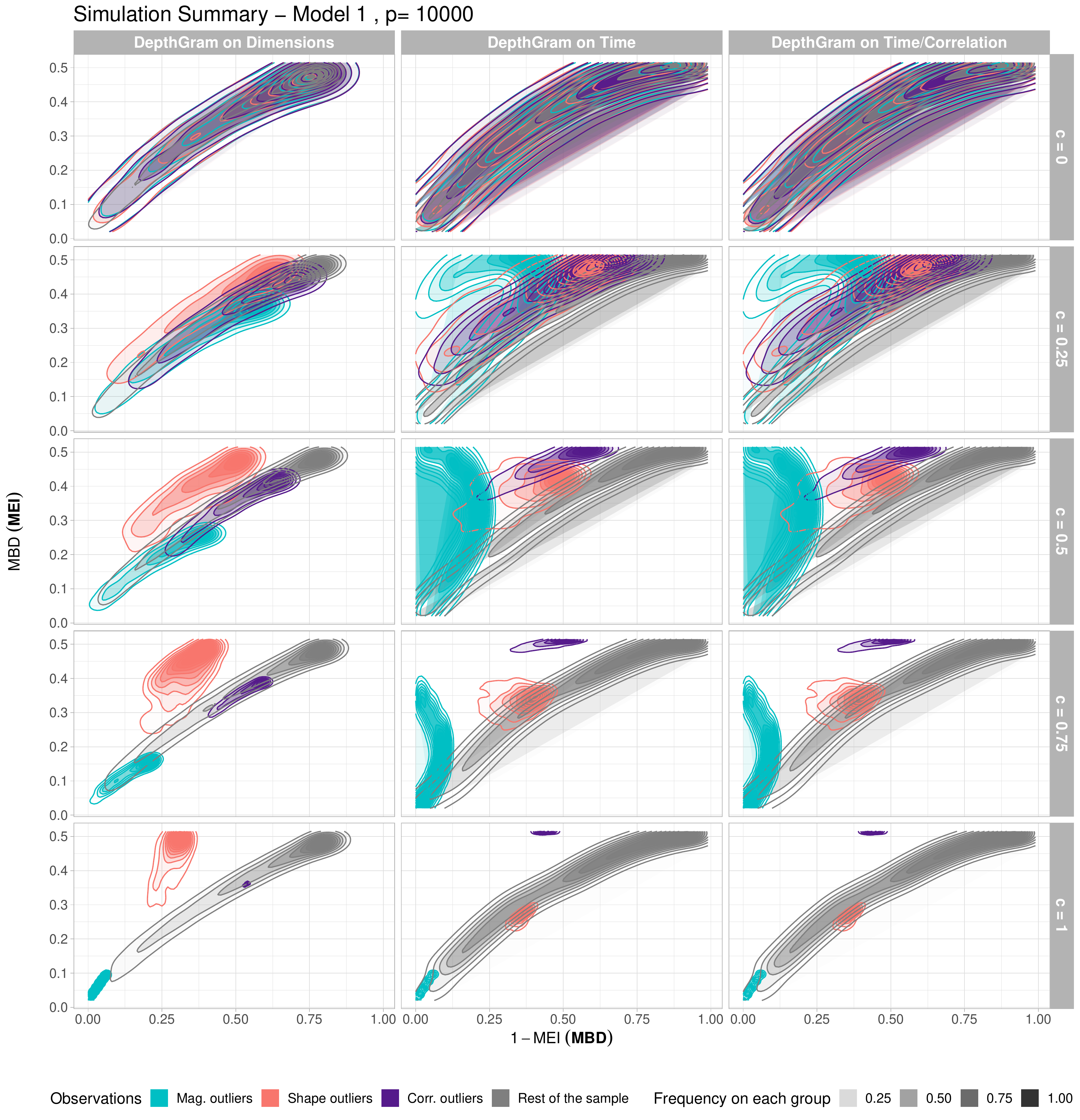}
	\caption{Summary of 200 simulation runs under Model 1, with $p=10000$, and different contamination rates $c$. Summary DepthGrams are obtained as the density contours of mbd(epi) and 1-epi(mbd) points over the 200 simulated data sets. Colors stand for outlier classification (including non-outlying observations).}\label{Sim_mod1_p10}
\end{figure}
\begin{figure}[h!]
	\hspace*{-1cm}\includegraphics[width=18cm]{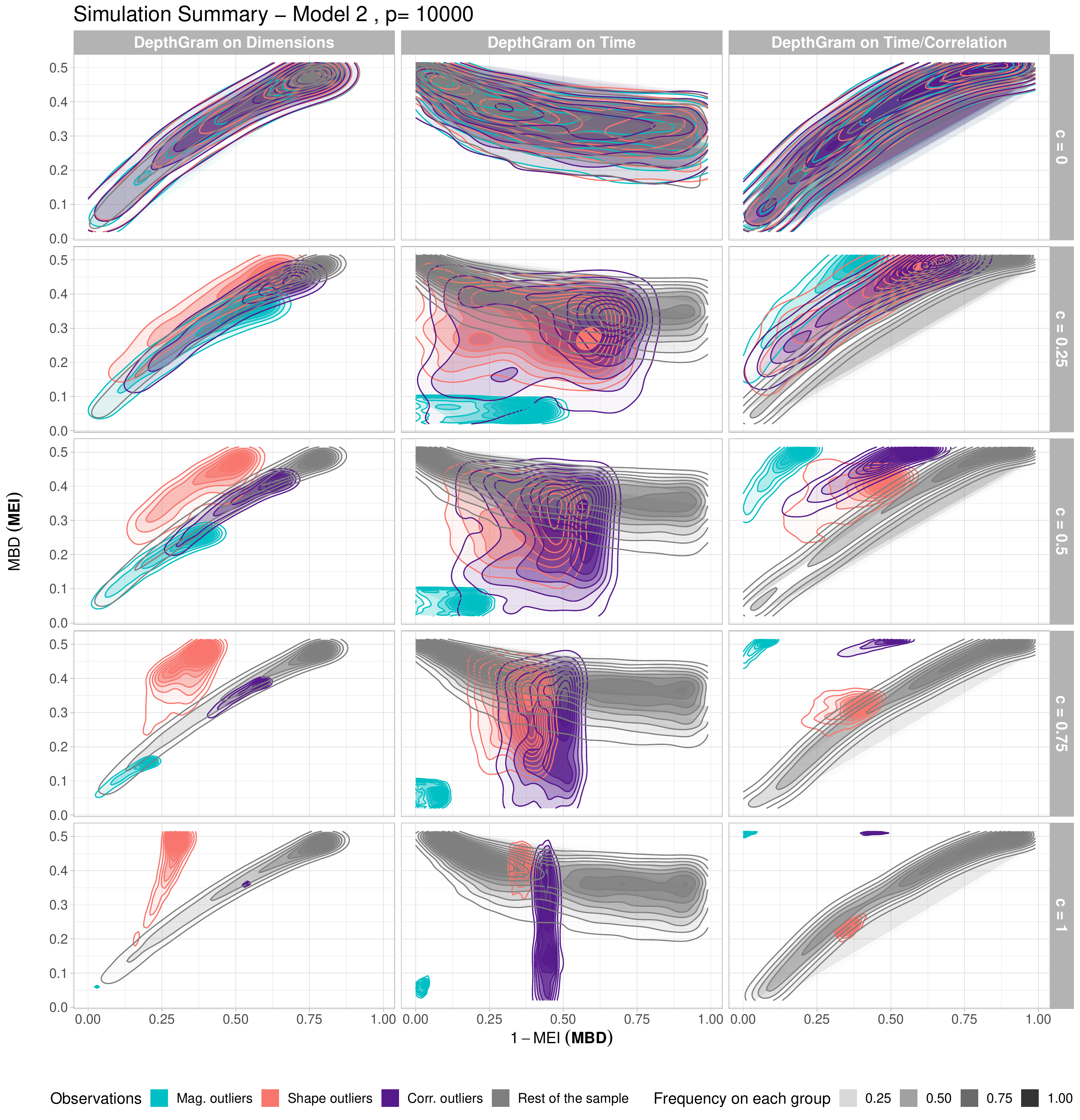}
	\caption{Summary of 200 simulation runs under Model 2, with $p=10000$, and different contamination rates $c$. Summary DepthGrams are obtained as the density contours of mbd(epi) and 1-epi(mbd) points over the 200 simulated data sets. Colors stand for outlier classification (including non-outlying observations).}\label{Sim_mod2_p10}
\end{figure}
\begin{figure}[h!]
	\hspace*{-1cm}\includegraphics[width=18cm]{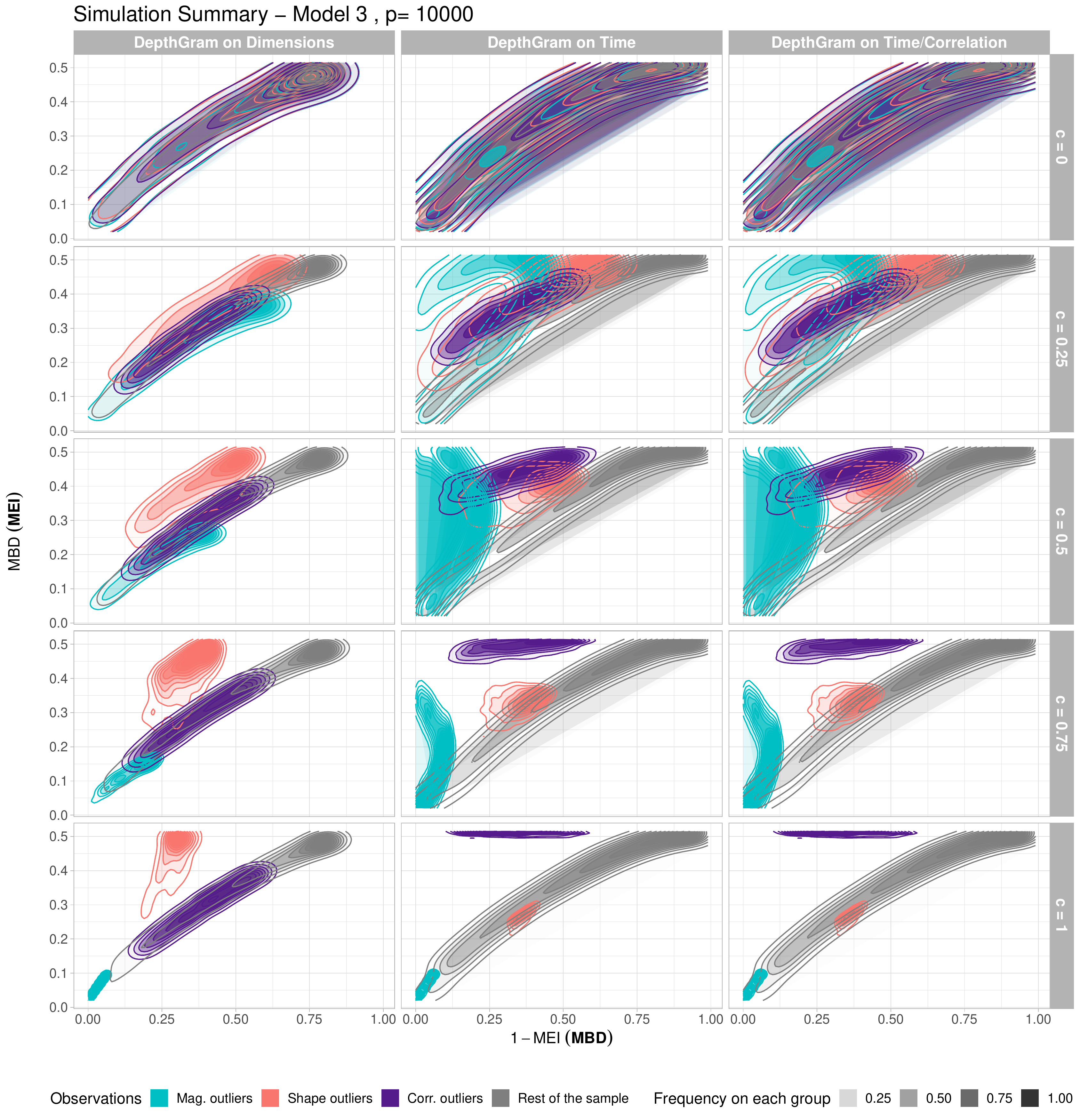}
	\caption{Summary of 200 simulation runs under Model 3, with $p=10000$, and different contamination rates $c$. Summary DepthGrams are obtained as the density contours of mbd(epi) and 1-epi(mbd) points over the 200 simulated data sets. Colors stand for outlier classification (including non-outlying observations).}\label{Sim_mod3_p10}
\end{figure}
\begin{figure}[h!]
	\hspace*{-1cm}\includegraphics[width=18cm]{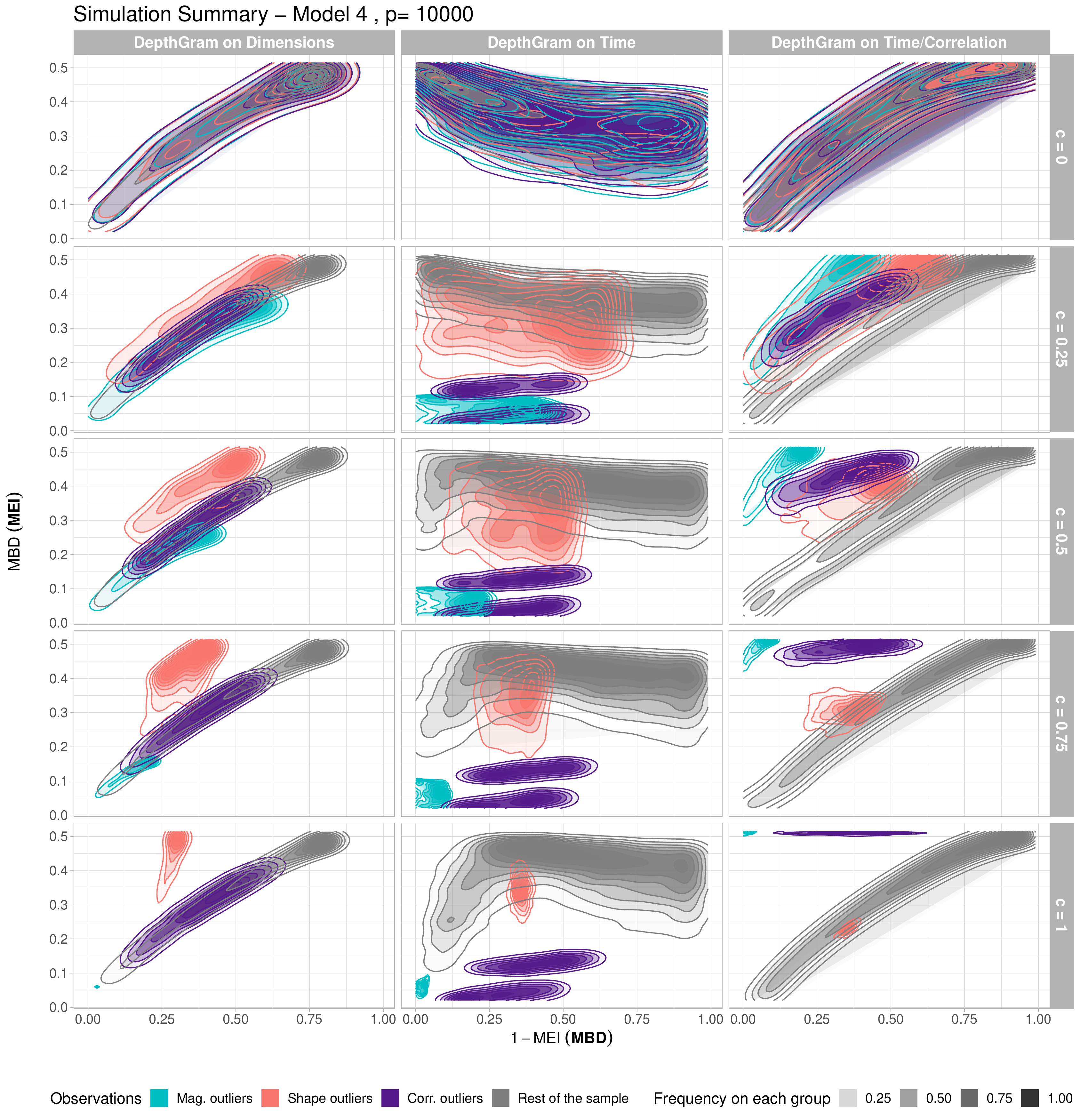}
	\caption{Summary of 200 simulation runs under Model 4, with $p=10000$, and different contamination rates $c$. Summary DepthGrams are obtained as the density contours of mbd(epi) and 1-epi(mbd) points over the 200 simulated data sets. Colors stand for outlier classification (including non-outlying observations).}\label{Sim_mod4_p10}
\end{figure}
\phantom{a}\vspace{2cm}\\
\\
\\
\newpage
\section{Appendix: Low dimensional simulation study}
We present the full results of the low dimensional simulation study (section 3.1).

\begin{figure}[h!]
	\includegraphics[width=16cm]{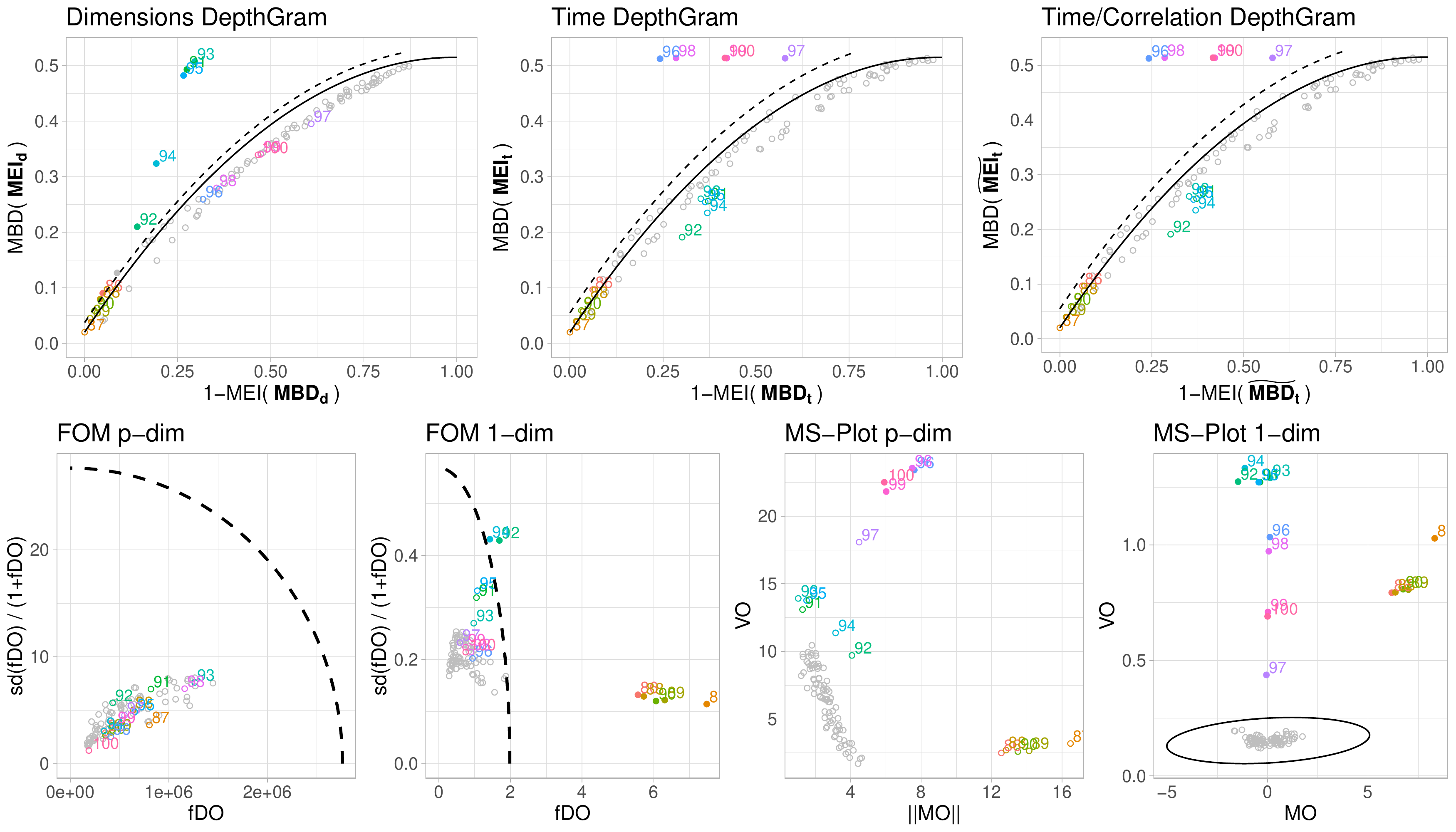}
	\caption{Results for a single simulation run under Model 3, with $p=50$, and $c=1$. In the top row we present the three DepthGram representations. In the bottom row, we present the FOM and the MS-plot in their $p$-dimensional and $1$-dimensional versions. Except for the $p$-dimensional MS-plot, the boundary dividing the outlying and non-outlying observations is drawn (a dashed line for the DepthGram and FOM and a solid ellipse for the MS-plot). In all the plots, detected outliers are marked with a bullet while the rest of the observations are represented with a circle. True outliers are represented in color while non-outlying observations are drawn in gray. }\label{FigresultsSimusLowDimp50_Mod3}
\end{figure}
\begin{figure}[h!]
	\includegraphics[width=16cm]{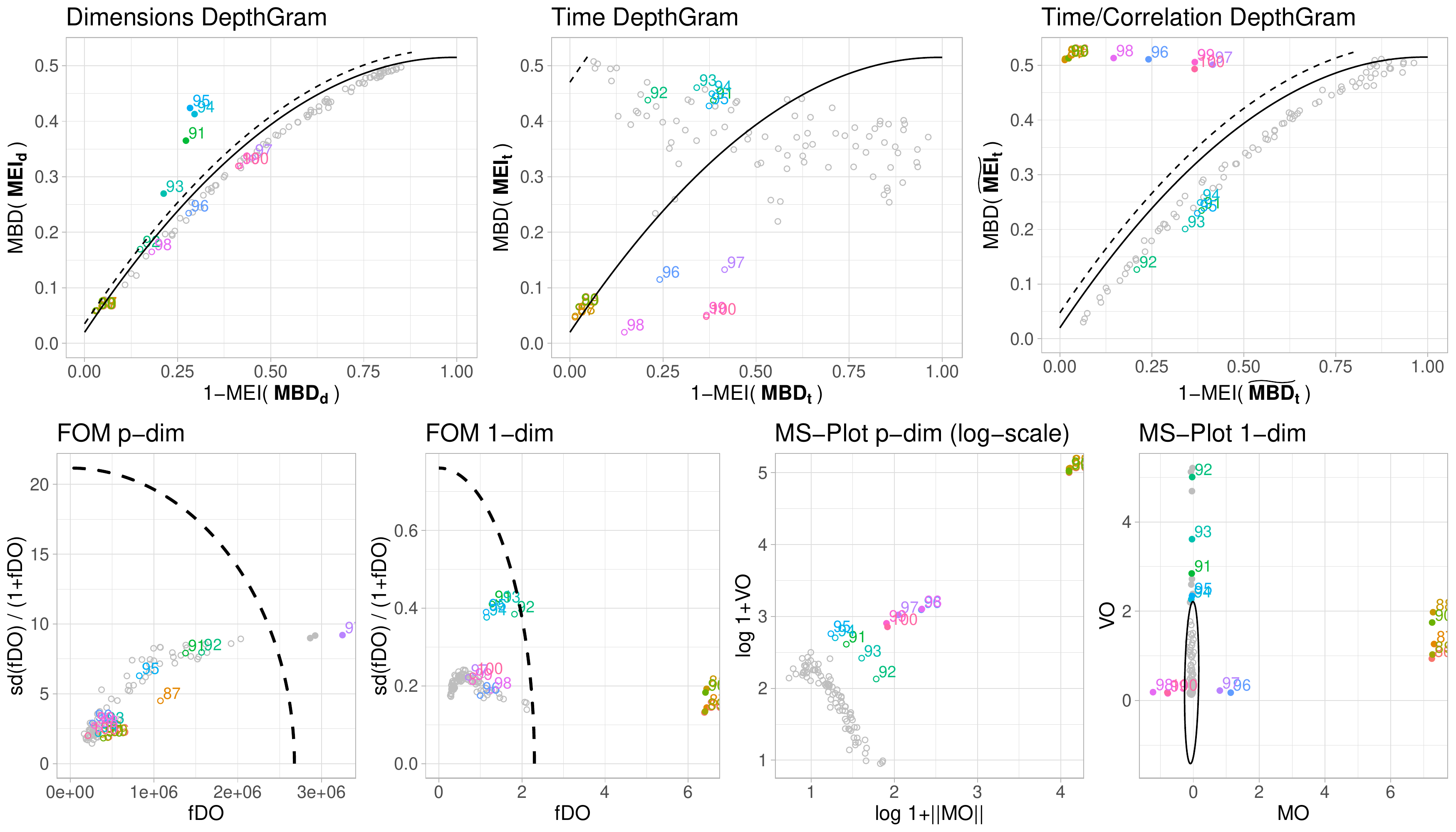}
	\caption{Results for a single simulation run under Model 4, with $p=50$, and $c=1$. In the top row we present the three DepthGram representations. In the bottom row, we present the FOM and the MS-plot in their $p$-dimensional and $1$-dimensional versions. Except for the $p$-dimensional MS-plot, the boundary dividing the outlying and non-outlying observations is drawn (a dashed line for the DepthGram and FOM and a solid ellipse for the MS-plot). In all the plots, detected outliers are marked with a bullet while the rest of the observations are represented with a circle. True outliers are represented in color while non-outlying observations are drawn in gray. The $p$-dimensional MS-plot is presented in logarithmic scale to ease the visualization.}\label{FigresultsSimusLowDimp50_Mod4}
\end{figure}

\newpage
\thispagestyle{empty}
\begin{sidewaystable}[h!]
	\caption{Mean and standard deviation (in parentheses) of the proportion of correctly (by type: magnitude, shape, joint) and falsely identified outliers in the four simulation models in low dimension ($p=10$) over 200 simulation runs. The DepthGram is compared with the Functional outlier map (FOM) both in its $p$-dimensional and its one dimensional versions and with the Magnitude-Shape plot (MS-plot)  both in its $p$-dimensional and its one dimensional versions.
		\label{TabresultsSimusLowDimp10} }
	\hspace*{-3cm}{\footnotesize
		\begin{tabular}{lcccccccccccccccc}
			\hline                                                                                                                                                                                                     
			$p=10$ &&&&&&&&&&&&&&&&\\                                                                                                                                                                                   
			DepthGram&\multicolumn{4}{c}{Model 1} &\multicolumn{4}{c}{Model 2}&\multicolumn{4}{c}{Model 3}&\multicolumn{4}{c}{Model 4}\\                                                                            
			\hline                                                                                                                                                                                                     
			Cont. level & $p^m_c$ & $p^s_c$ &$p^j_c$ & $p_f$ &  $p^m_c$ & $p^s_c$ &$p^j_c$ & $p_f$ & $p^m_c$ & $p^s_c$ &$p^j_c$ & $p_f$ & $p^m_c$ & $p^s_c$ &$p^j_c$ & $p_f$ \\                                         
			\hline                                                                                                                                                                                                     
			$c=0\!\!\!\!\!\!\!\!\!\!$ &- &- &- & 0.01(0.01)\!\!\!\!\!\!\!\! &- &- &- &0.01(0.01)\!\!\!\!\!\!\!\! &- &- &- &0.01(0.01)\!\!\!\!\!\!\!\! &- &- &- &0.01(0.01)\!\!\!\!\!\!\!\!\\                                                                                                                
			\hline                                                                                                                                                                                                     
			$c=0.25\!\!\!\!\!\!\!\!\!\!$ &0.66(0.2)\!\!\!\!\!\!\!\! & 0.71(0.22)\!\!\!\!\!\!\!\! & 0.38(0.2)\!\!\!\!\!\!\!\! & 0(0)\!\!\!\!\!\!\!\! & 0.84(0.16)\!\!\!\!\!\!\!\! & 0.71(0.21)\!\!\!\!\!\!\!\! & 0.33(0.21)\!\!\!\!\!\!\!\! & 0(0)\!\!\!\!\!\!\!\! & 0.62(0.21)\!\!\!\!\!\!\!\! & 0.73(0.21)\!\!\!\!\!\!\!\! & 0.5(0.23)\!\!\!\!\!\!\!\! & 0(0)\!\!\!\!\!\!\!\! & 0.83(0.16)\!\!\!\!\!\!\!\! & 0.71(0.21)\!\!\!\!\!\!\!\! & 0.55(0.22)\!\!\!\!\!\!\!\! & 0(0)\!\!\!\!\!\!\!\!\\              
			\hline                                                                                                                                                                                                     
			$c=0.5\!\!\!\!\!\!\!\!\!\!$ &0.64(0.2)\!\!\!\!\!\!\!\! & 0.93(0.1)\!\!\!\!\!\!\!\! & 0.65(0.2)\!\!\!\!\!\!\!\! & 0(0)\!\!\!\!\!\!\!\! & 1(0.02)\!\!\!\!\!\!\!\! & 0.93(0.11)\!\!\!\!\!\!\!\! & 0.54(0.2)\!\!\!\!\!\!\!\! & 0(0)\!\!\!\!\!\!\!\! & 0.63(0.22)\!\!\!\!\!\!\!\! & 0.91(0.13)\!\!\!\!\!\!\!\! & 0.77(0.18)\!\!\!\!\!\!\!\! & 0(0)\!\!\!\!\!\!\!\! & 1(0.02)\!\!\!\!\!\!\!\! & 0.93(0.11)\!\!\!\!\!\!\!\! & 0.71(0.19)\!\!\!\!\!\!\!\! & 0(0)\!\!\!\!\!\!\!\!\\                      
			\hline                                                                                                                                                                                                     
			$c=0.75\!\!\!\!\!\!\!\!\!\!$ &0.58(0.2)\!\!\!\!\!\!\!\! & 0.96(0.09)\!\!\!\!\!\!\!\! & 0.84(0.17)\!\!\!\!\!\!\!\! & 0(0.01)\!\!\!\!\!\!\!\! & 1(0)\!\!\!\!\!\!\!\! & 0.96(0.09)\!\!\!\!\!\!\!\! & 0.71(0.2)\!\!\!\!\!\!\!\! & 0(0)\!\!\!\!\!\!\!\! & 0.58(0.18)\!\!\!\!\!\!\!\! & 0.95(0.09)\!\!\!\!\!\!\!\! & 0.95(0.1)\!\!\!\!\!\!\!\! & 0(0.01)\!\!\!\!\!\!\!\! & 1(0)\!\!\!\!\!\!\!\! & 0.96(0.09)\!\!\!\!\!\!\!\! & 0.87(0.15)\!\!\!\!\!\!\!\! & 0(0)\!\!\!\!\!\!\!\!\\                    
			\hline                                                                                                                                                                                                     
			$c=1\!\!\!\!\!\!\!\!\!\!$ &0.06(0.13)\!\!\!\!\!\!\!\! & 0.98(0.06)\!\!\!\!\!\!\!\! & 0.95(0.1)\!\!\!\!\!\!\!\! & 0(0)\!\!\!\!\!\!\!\! & 1(0)\!\!\!\!\!\!\!\! & 0.98(0.06)\!\!\!\!\!\!\!\! & 0.91(0.13)\!\!\!\!\!\!\!\! & 0(0)\!\!\!\!\!\!\!\! & 0.09(0.16)\!\!\!\!\!\!\!\! & 0.98(0.06)\!\!\!\!\!\!\!\! & 1(0.03)\!\!\!\!\!\!\!\! & 0(0)\!\!\!\!\!\!\!\! & 1(0)\!\!\!\!\!\!\!\! & 0.98(0.06)\!\!\!\!\!\!\!\! & 0.98(0.07)\!\!\!\!\!\!\!\! & 0(0)\!\!\!\!\!\!\!\!\\                              
			\hline                                                                                                                                                                                                     
			$p=10$ &&&&&&&&&&&&&&&&\\                                                                                                                                                                                   
			FOM $p$-dim&\multicolumn{4}{c}{Model 1} &\multicolumn{4}{c}{Model 2}&\multicolumn{4}{c}{Model 3}&\multicolumn{4}{c}{Model 4}\\                                                                          
			\hline                                                                                                                                                                                                     
			Cont. level & $p^m_c$ & $p^s_c$ &$p^j_c$ & $p_f$ &  $p^m_c$ & $p^s_c$ &$p^j_c$ & $p_f$ & $p^m_c$ & $p^s_c$ &$p^j_c$ & $p_f$ & $p^m_c$ & $p^s_c$ &$p^j_c$ & $p_f$ \\                                         
			\hline                                                                                                                                                                                                     
			$c=0\!\!\!\!\!\!\!\!\!\!$ &- &- &- & 0.01(0.02)\!\!\!\!\!\!\!\! &- &- &- &0.01(0.01)\!\!\!\!\!\!\!\! &- &- &- &0.01(0.01)\!\!\!\!\!\!\!\! &- &- &- &0.01(0.01)\!\!\!\!\!\!\!\!\\                                                                                                                
			\hline                                                                                                                                                                                                     
			$c=0.25\!\!\!\!\!\!\!\!\!\!$ &1(0)\!\!\!\!\!\!\!\! & 0.93(0.13)\!\!\!\!\!\!\!\! & 0.28(0.19)\!\!\!\!\!\!\!\! & 0(0.01)\!\!\!\!\!\!\!\! & 1(0)\!\!\!\!\!\!\!\! & 0.9(0.14)\!\!\!\!\!\!\!\! & 0.25(0.2)\!\!\!\!\!\!\!\! & 0(0.01)\!\!\!\!\!\!\!\! & 1(0)\!\!\!\!\!\!\!\! & 0.93(0.13)\!\!\!\!\!\!\!\! & 0.36(0.23)\!\!\!\!\!\!\!\! & 0(0.01)\!\!\!\!\!\!\!\! & 1(0)\!\!\!\!\!\!\!\! & 0.91(0.14)\!\!\!\!\!\!\!\! & 0.37(0.23)\!\!\!\!\!\!\!\! & 0(0.01)\!\!\!\!\!\!\!\!\\                         
			\hline                                                                                                                                                                                                     
			$c=0.5\!\!\!\!\!\!\!\!\!\!$ &1(0)\!\!\!\!\!\!\!\! & 1(0.02)\!\!\!\!\!\!\!\! & 0.55(0.23)\!\!\!\!\!\!\!\! & 0(0.01)\!\!\!\!\!\!\!\! & 1(0)\!\!\!\!\!\!\!\! & 1(0.03)\!\!\!\!\!\!\!\! & 0.55(0.26)\!\!\!\!\!\!\!\! & 0(0.01)\!\!\!\!\!\!\!\! & 1(0)\!\!\!\!\!\!\!\! & 1(0.04)\!\!\!\!\!\!\!\! & 0.67(0.24)\!\!\!\!\!\!\!\! & 0(0)\!\!\!\!\!\!\!\! & 1(0)\!\!\!\!\!\!\!\! & 0.99(0.05)\!\!\!\!\!\!\!\! & 0.62(0.23)\!\!\!\!\!\!\!\! & 0(0.01)\!\!\!\!\!\!\!\!\\                                    
			\hline                                                                                                                                                                                                     
			$c=0.75\!\!\!\!\!\!\!\!\!\!$ &1(0)\!\!\!\!\!\!\!\! & 0.99(0.06)\!\!\!\!\!\!\!\! & 0.68(0.22)\!\!\!\!\!\!\!\! & 0(0.01)\!\!\!\!\!\!\!\! & 1(0)\!\!\!\!\!\!\!\! & 0.97(0.09)\!\!\!\!\!\!\!\! & 0.65(0.22)\!\!\!\!\!\!\!\! & 0(0)\!\!\!\!\!\!\!\! & 1(0)\!\!\!\!\!\!\!\! & 0.98(0.07)\!\!\!\!\!\!\!\! & 0.75(0.21)\!\!\!\!\!\!\!\! & 0(0)\!\!\!\!\!\!\!\! & 1(0)\!\!\!\!\!\!\!\! & 0.97(0.08)\!\!\!\!\!\!\!\! & 0.66(0.24)\!\!\!\!\!\!\!\! & 0(0)\!\!\!\!\!\!\!\!\\                                
			\hline                                                                                                                                                                                                     
			$c=1\!\!\!\!\!\!\!\!\!\!$ &0.99(0.04)\!\!\!\!\!\!\!\! & 0(0.03)\!\!\!\!\!\!\!\! & 0.83(0.18)\!\!\!\!\!\!\!\! & 0(0.01)\!\!\!\!\!\!\!\! & 1(0)\!\!\!\!\!\!\!\! & 0.02(0.05)\!\!\!\!\!\!\!\! & 0.78(0.21)\!\!\!\!\!\!\!\! & 0(0.01)\!\!\!\!\!\!\!\! & 0.99(0.04)\!\!\!\!\!\!\!\! & 0.01(0.04)\!\!\!\!\!\!\!\! & 0.83(0.2)\!\!\!\!\!\!\!\! & 0(0.01)\!\!\!\!\!\!\!\! & 1(0)\!\!\!\!\!\!\!\! & 0.02(0.06)\!\!\!\!\!\!\!\! & 0.64(0.25)\!\!\!\!\!\!\!\! & 0(0.01)\!\!\!\!\!\!\!\!\\                  
			\hline                                                                                                                                                                                                     
			$p=10$ &&&&&&&&&&&&&&&&\\                                                                                                                                                                                   
			FOM $1$-dim&\multicolumn{4}{c}{Model 1} &\multicolumn{4}{c}{Model 2}&\multicolumn{4}{c}{Model 3}&\multicolumn{4}{c}{Model 4}\\                                                                          
			\hline                                                                                                                                                                                                     
			Cont. level & $p^m_c$ & $p^s_c$ &$p^j_c$ & $p_f$ &  $p^m_c$ & $p^s_c$ &$p^j_c$ & $p_f$ & $p^m_c$ & $p^s_c$ &$p^j_c$ & $p_f$ & $p^m_c$ & $p^s_c$ &$p^j_c$ & $p_f$ \\                                         
			\hline                                                                                                                                                                                                     
			$c=0\!\!\!\!\!\!\!\!\!\!$ &- &- &- & 0.02(0.02)\!\!\!\!\!\!\!\! &- &- &- &0.02(0.02)\!\!\!\!\!\!\!\! &- &- &- &0.02(0.02)\!\!\!\!\!\!\!\! &- &- &- &0.02(0.02)\!\!\!\!\!\!\!\!\\                                                                                                                
			\hline                                                                                                                                                                                                     
			$c=0.25\!\!\!\!\!\!\!\!\!\!$ &0.99(0.04)\!\!\!\!\!\!\!\! & 0.01(0.03)\!\!\!\!\!\!\!\! & 0(0.02)\!\!\!\!\!\!\!\! & 0.01(0.01)\!\!\!\!\!\!\!\! & 1(0)\!\!\!\!\!\!\!\! & 0.01(0.05)\!\!\!\!\!\!\!\! & 0.01(0.04)\!\!\!\!\!\!\!\! & 0.01(0.01)\!\!\!\!\!\!\!\! & 1(0.03)\!\!\!\!\!\!\!\! & 0(0.01)\!\!\!\!\!\!\!\! & 0(0)\!\!\!\!\!\!\!\! & 0.01(0.01)\!\!\!\!\!\!\!\! & 1(0)\!\!\!\!\!\!\!\! & 0.01(0.05)\!\!\!\!\!\!\!\! & 0(0)\!\!\!\!\!\!\!\! & 0.01(0.01)\!\!\!\!\!\!\!\!\\                    
			\hline                                                                                                                                                                                                     
			$c=0.5\!\!\!\!\!\!\!\!\!\!$ &1(0)\!\!\!\!\!\!\!\! & 0(0.03)\!\!\!\!\!\!\!\! & 0(0.02)\!\!\!\!\!\!\!\! & 0.01(0.01)\!\!\!\!\!\!\!\! & 1(0)\!\!\!\!\!\!\!\! & 0(0.02)\!\!\!\!\!\!\!\! & 0(0)\!\!\!\!\!\!\!\! & 0.01(0.01)\!\!\!\!\!\!\!\! & 1(0)\!\!\!\!\!\!\!\! & 0.01(0.04)\!\!\!\!\!\!\!\! & 0(0)\!\!\!\!\!\!\!\! & 0.01(0.01)\!\!\!\!\!\!\!\! & 1(0)\!\!\!\!\!\!\!\! & 0(0.04)\!\!\!\!\!\!\!\! & 0(0)\!\!\!\!\!\!\!\! & 0.01(0.01)\!\!\!\!\!\!\!\!\\                                          
			\hline                                                                                                                                                                                                     
			$c=0.75\!\!\!\!\!\!\!\!\!\!$ &1(0)\!\!\!\!\!\!\!\! & 0(0.03)\!\!\!\!\!\!\!\! & 0(0)\!\!\!\!\!\!\!\! & 0.01(0.01)\!\!\!\!\!\!\!\! & 1(0)\!\!\!\!\!\!\!\! & 0.01(0.04)\!\!\!\!\!\!\!\! & 0(0)\!\!\!\!\!\!\!\! & 0(0.01)\!\!\!\!\!\!\!\! & 1(0)\!\!\!\!\!\!\!\! & 0.01(0.05)\!\!\!\!\!\!\!\! & 0(0)\!\!\!\!\!\!\!\! & 0.01(0.01)\!\!\!\!\!\!\!\! & 1(0)\!\!\!\!\!\!\!\! & 0.01(0.05)\!\!\!\!\!\!\!\! & 0(0)\!\!\!\!\!\!\!\! & 0.01(0.01)\!\!\!\!\!\!\!\!\\                                         
			\hline                                                                                                                                                                                                     
			$c=1\!\!\!\!\!\!\!\!\!\!$ &1(0)\!\!\!\!\!\!\!\! & 0.03(0.08)\!\!\!\!\!\!\!\! & 0(0)\!\!\!\!\!\!\!\! & 0.01(0.01)\!\!\!\!\!\!\!\! & 1(0)\!\!\!\!\!\!\!\! & 0.01(0.05)\!\!\!\!\!\!\!\! & 0(0)\!\!\!\!\!\!\!\! & 0(0.01)\!\!\!\!\!\!\!\! & 1(0)\!\!\!\!\!\!\!\! & 0.03(0.08)\!\!\!\!\!\!\!\! & 0(0)\!\!\!\!\!\!\!\! & 0.01(0.01)\!\!\!\!\!\!\!\! & 1(0)\!\!\!\!\!\!\!\! & 0.02(0.07)\!\!\!\!\!\!\!\! & 0(0)\!\!\!\!\!\!\!\! & 0.01(0.01)\!\!\!\!\!\!\!\!\\                                         
			\hline                                                                                                                                                                                                     
			$p=10$ &&&&&&&&&&&&&&&&\\                                                                                                                                                                                   
			MS-plot $p$-dim&\multicolumn{4}{c}{Model 1} &\multicolumn{4}{c}{Model 2}&\multicolumn{4}{c}{Model 3}&\multicolumn{4}{c}{Model 4}\\                                                                      
			\hline                                                                                                                                                                                                     
			Cont. level & $p^m_c$ & $p^s_c$ &$p^j_c$ & $p_f$ &  $p^m_c$ & $p^s_c$ &$p^j_c$ & $p_f$ & $p^m_c$ & $p^s_c$ &$p^j_c$ & $p_f$ & $p^m_c$ & $p^s_c$ &$p^j_c$ & $p_f$ \\                                         
			\hline                                                                                                                                                                                                     
			$c=0\!\!\!\!\!\!\!\!\!\!$ &- &- &- & 0(0)\!\!\!\!\!\!\!\! &- &- &- &0(0)\!\!\!\!\!\!\!\! &- &- &- &0(0)\!\!\!\!\!\!\!\! &- &- &- &0.02(0.01)\!\!\!\!\!\!\!\!\\                                                                                                                                  
			\hline                                                                                                                                                                                                     
			$c=0.25\!\!\!\!\!\!\!\!\!\!$ &1(0)\!\!\!\!\!\!\!\! & 0.46(0.24)\!\!\!\!\!\!\!\! & 0.24(0.18)\!\!\!\!\!\!\!\! & 0(0)\!\!\!\!\!\!\!\! & 1(0)\!\!\!\!\!\!\!\! & 0.45(0.22)\!\!\!\!\!\!\!\! & 0.22(0.19)\!\!\!\!\!\!\!\! & 0(0)\!\!\!\!\!\!\!\! & 1(0)\!\!\!\!\!\!\!\! & 0.47(0.22)\!\!\!\!\!\!\!\! & 0.28(0.2)\!\!\!\!\!\!\!\! & 0(0)\!\!\!\!\!\!\!\! & 1(0)\!\!\!\!\!\!\!\! & 0.5(0.27)\!\!\!\!\!\!\!\! & 0.08(0.12)\!\!\!\!\!\!\!\! & 0(0.01)\!\!\!\!\!\!\!\!\\                                  
			\hline                                                                                                                                                                                                     
			$c=0.5\!\!\!\!\!\!\!\!\!\!$ &1(0)\!\!\!\!\!\!\!\! & 0.89(0.15)\!\!\!\!\!\!\!\! & 0.63(0.22)\!\!\!\!\!\!\!\! & 0(0)\!\!\!\!\!\!\!\! & 1(0)\!\!\!\!\!\!\!\! & 0.87(0.16)\!\!\!\!\!\!\!\! & 0.62(0.2)\!\!\!\!\!\!\!\! & 0(0)\!\!\!\!\!\!\!\! & 1(0)\!\!\!\!\!\!\!\! & 0.87(0.16)\!\!\!\!\!\!\!\! & 0.7(0.2)\!\!\!\!\!\!\!\! & 0(0)\!\!\!\!\!\!\!\! & 1(0)\!\!\!\!\!\!\!\! & 1(0)\!\!\!\!\!\!\!\! & 0.19(0.17)\!\!\!\!\!\!\!\! & 0(0)\!\!\!\!\!\!\!\!\\                                             
			\hline                                                                                                                                                                                                     
			$c=0.75\!\!\!\!\!\!\!\!\!\!$ &1(0)\!\!\!\!\!\!\!\! & 0.82(0.2)\!\!\!\!\!\!\!\! & 0.78(0.2)\!\!\!\!\!\!\!\! & 0(0)\!\!\!\!\!\!\!\! & 1(0)\!\!\!\!\!\!\!\! & 0.82(0.18)\!\!\!\!\!\!\!\! & 0.77(0.19)\!\!\!\!\!\!\!\! & 0(0)\!\!\!\!\!\!\!\! & 1(0)\!\!\!\!\!\!\!\! & 0.8(0.2)\!\!\!\!\!\!\!\! & 0.84(0.17)\!\!\!\!\!\!\!\! & 0(0)\!\!\!\!\!\!\!\! & 1(0)\!\!\!\!\!\!\!\! & 1(0)\!\!\!\!\!\!\!\! & 0.18(0.2)\!\!\!\!\!\!\!\! & 0(0)\!\!\!\!\!\!\!\!\\                                              
			\hline                                                                                                                                                                                                     
			$c=1\!\!\!\!\!\!\!\!\!\!$ &0(0.02)\!\!\!\!\!\!\!\! & 0(0.02)\!\!\!\!\!\!\!\! & 0.93(0.12)\!\!\!\!\!\!\!\! & 0(0)\!\!\!\!\!\!\!\! & 1(0)\!\!\!\!\!\!\!\! & 0(0)\!\!\!\!\!\!\!\! & 0.89(0.15)\!\!\!\!\!\!\!\! & 0(0)\!\!\!\!\!\!\!\! & 0(0.03)\!\!\!\!\!\!\!\! & 0(0.02)\!\!\!\!\!\!\!\! & 0.96(0.11)\!\!\!\!\!\!\!\! & 0(0)\!\!\!\!\!\!\!\! & 1(0)\!\!\!\!\!\!\!\! & 1(0)\!\!\!\!\!\!\!\! & 0.18(0.19)\!\!\!\!\!\!\!\! & 0(0)\!\!\!\!\!\!\!\!\\                                                  
			\hline                                                                                                                                                                                                     
			$p=10$ &&&&&&&&&&&&&&&&\\                                                                                                                                                                                   
			MS-plot $1$-dim&\multicolumn{4}{c}{Model 1} &\multicolumn{4}{c}{Model 2}&\multicolumn{4}{c}{Model 3}&\multicolumn{4}{c}{Model 4}\\                                                                      
			\hline                                                                                                                                                                                                     
			Cont. level & $p^m_c$ & $p^s_c$ &$p^j_c$ & $p_f$ &  $p^m_c$ & $p^s_c$ &$p^j_c$ & $p_f$ & $p^m_c$ & $p^s_c$ &$p^j_c$ & $p_f$ & $p^m_c$ & $p^s_c$ &$p^j_c$ & $p_f$ \\                                         
			\hline                                                                                                                                                                                                     
			$c=0\!\!\!\!\!\!\!\!\!\!$ &- &- &- & 0(0)\!\!\!\!\!\!\!\! &- &- &- &0.12(0.05)\!\!\!\!\!\!\!\! &- &- &- &0(0)\!\!\!\!\!\!\!\! &- &- &- &0.12(0.05)\!\!\!\!\!\!\!\!\\                                                                                                                            
			\hline                                                                                                                                                                                                     
			$c=0.25\!\!\!\!\!\!\!\!\!\!$ &1(0)\!\!\!\!\!\!\!\! & 0.92(0.14)\!\!\!\!\!\!\!\! & 0.38(0.19)\!\!\!\!\!\!\!\! & 0(0)\!\!\!\!\!\!\!\! & 1(0)\!\!\!\!\!\!\!\! & 0.12(0.14)\!\!\!\!\!\!\!\! & 0.08(0.11)\!\!\!\!\!\!\!\! & 0.09(0.04)\!\!\!\!\!\!\!\! & 1(0)\!\!\!\!\!\!\!\! & 0.93(0.12)\!\!\!\!\!\!\!\! & 0.51(0.24)\!\!\!\!\!\!\!\! & 0(0)\!\!\!\!\!\!\!\! & 1(0)\!\!\!\!\!\!\!\! & 0.11(0.14)\!\!\!\!\!\!\!\! & 0.08(0.13)\!\!\!\!\!\!\!\! & 0.09(0.04)\!\!\!\!\!\!\!\!\\                       
			\hline                                                                                                                                                                                                     
			$c=0.5\!\!\!\!\!\!\!\!\!\!$ &1(0)\!\!\!\!\!\!\!\! & 1(0)\!\!\!\!\!\!\!\! & 0.72(0.21)\!\!\!\!\!\!\!\! & 0(0)\!\!\!\!\!\!\!\! & 1(0)\!\!\!\!\!\!\!\! & 0.14(0.18)\!\!\!\!\!\!\!\! & 0.05(0.09)\!\!\!\!\!\!\!\! & 0.07(0.04)\!\!\!\!\!\!\!\! & 1(0)\!\!\!\!\!\!\!\! & 1(0)\!\!\!\!\!\!\!\! & 0.84(0.18)\!\!\!\!\!\!\!\! & 0(0)\!\!\!\!\!\!\!\! & 1(0)\!\!\!\!\!\!\!\! & 0.15(0.18)\!\!\!\!\!\!\!\! & 0.23(0.2)\!\!\!\!\!\!\!\! & 0.08(0.05)\!\!\!\!\!\!\!\!\\                                     
			\hline                                                                                                                                                                                                     
			$c=0.75\!\!\!\!\!\!\!\!\!\!$ &1(0)\!\!\!\!\!\!\!\! & 1(0)\!\!\!\!\!\!\!\! & 0.84(0.17)\!\!\!\!\!\!\!\! & 0(0)\!\!\!\!\!\!\!\! & 1(0)\!\!\!\!\!\!\!\! & 0.2(0.22)\!\!\!\!\!\!\!\! & 0.05(0.1)\!\!\!\!\!\!\!\! & 0.07(0.04)\!\!\!\!\!\!\!\! & 1(0)\!\!\!\!\!\!\!\! & 1(0)\!\!\!\!\!\!\!\! & 0.92(0.13)\!\!\!\!\!\!\!\! & 0(0)\!\!\!\!\!\!\!\! & 1(0)\!\!\!\!\!\!\!\! & 0.22(0.22)\!\!\!\!\!\!\!\! & 0.47(0.29)\!\!\!\!\!\!\!\! & 0.08(0.04)\!\!\!\!\!\!\!\!\\                                     
			\hline                                                                                                                                                                                                     
			$c=1\!\!\!\!\!\!\!\!\!\!$ &1(0)\!\!\!\!\!\!\!\! & 1(0)\!\!\!\!\!\!\!\! & 0.9(0.15)\!\!\!\!\!\!\!\! & 0(0)\!\!\!\!\!\!\!\! & 1(0)\!\!\!\!\!\!\!\! & 0.26(0.24)\!\!\!\!\!\!\!\! & 0.04(0.1)\!\!\!\!\!\!\!\! & 0.08(0.04)\!\!\!\!\!\!\!\! & 1(0)\!\!\!\!\!\!\!\! & 1(0)\!\!\!\!\!\!\!\! & 0.96(0.1)\!\!\!\!\!\!\!\! & 0(0)\!\!\!\!\!\!\!\! & 1(0)\!\!\!\!\!\!\!\! & 0.31(0.29)\!\!\!\!\!\!\!\! & 0.85(0.22)\!\!\!\!\!\!\!\! & 0.09(0.05)\!\!\!\!\!\!\!\!\\                                         
			\hline   
	\end{tabular}}
\end{sidewaystable}

\newpage
\thispagestyle{empty}
\begin{sidewaystable}[h!]
	\caption{Mean and standard deviation (in parentheses) of the proportion of correctly (by type: magnitude, shape, joint) and falsely identified outliers in the four simulation models in low dimension ($p=50$) over 200 simulation runs. The DepthGram is compared with the Functional outlier map (FOM) both in its $p$-dimensional and its one dimensional versions and with the Magnitude-Shape plot (MS-plot)  both in its $p$-dimensional and its one dimensional versions.
		\label{TabresultsSimusLowDimp50} }
	\hspace*{-3cm}{\footnotesize
		\begin{tabular}{lcccccccccccccccc}
			\hline                                                                                                                                                                                                     
			$p=50$ &&&&&&&&&&&&&&&&\\                                                                                                                                                                                   
			DepthGram&\multicolumn{4}{c}{Model 1} &\multicolumn{4}{c}{Model 2}&\multicolumn{4}{c}{Model 3}&\multicolumn{4}{c}{Model 4}\\                                                                            
			\hline                                                                                                                                                                                                     
			Cont. level & $p^m_c$ & $p^s_c$ &$p^j_c$ & $p_f$ &  $p^m_c$ & $p^s_c$ &$p^j_c$ & $p_f$ & $p^m_c$ & $p^s_c$ &$p^j_c$ & $p_f$ & $p^m_c$ & $p^s_c$ &$p^j_c$ & $p_f$ \\                                         
			\hline                                                                                                                                                                                                     
			$c=0\!\!\!\!\!\!\!\!\!\!$ &- &- &- & 0.01(0.02)\!\!\!\!\!\!\!\! &- &- &- &0.01(0.02)\!\!\!\!\!\!\!\! &- &- &- &0.01(0.02)\!\!\!\!\!\!\!\! &- &- &- &0.01(0.01)\!\!\!\!\!\!\!\!\\                                                                                                                
			\hline                                                                                                                                                                                                     
			$c=0.25\!\!\!\!\!\!\!\!\!\!$ &0.67(0.21)\!\!\!\!\!\!\!\! & 0.89(0.14)\!\!\!\!\!\!\!\! & 0.5(0.23)\!\!\!\!\!\!\!\! & 0(0.01)\!\!\!\!\!\!\!\! & 0.99(0.05)\!\!\!\!\!\!\!\! & 0.88(0.14)\!\!\!\!\!\!\!\! & 0.46(0.23)\!\!\!\!\!\!\!\! & 0(0.01)\!\!\!\!\!\!\!\! & 0.65(0.22)\!\!\!\!\!\!\!\! & 0.89(0.13)\!\!\!\!\!\!\!\! & 0.67(0.22)\!\!\!\!\!\!\!\! & 0(0.01)\!\!\!\!\!\!\!\! & 0.99(0.04)\!\!\!\!\!\!\!\! & 0.9(0.13)\!\!\!\!\!\!\!\! & 0.62(0.24)\!\!\!\!\!\!\!\! & 0(0.01)\!\!\!\!\!\!\!\!\\ 
			\hline                                                                                                                                                                                                     
			$c=0.5\!\!\!\!\!\!\!\!\!\!$ &0.68(0.19)\!\!\!\!\!\!\!\! & 0.96(0.1)\!\!\!\!\!\!\!\! & 0.86(0.16)\!\!\!\!\!\!\!\! & 0(0.01)\!\!\!\!\!\!\!\! & 1(0)\!\!\!\!\!\!\!\! & 0.95(0.1)\!\!\!\!\!\!\!\! & 0.73(0.2)\!\!\!\!\!\!\!\! & 0(0)\!\!\!\!\!\!\!\! & 0.66(0.19)\!\!\!\!\!\!\!\! & 0.95(0.1)\!\!\!\!\!\!\!\! & 0.9(0.13)\!\!\!\!\!\!\!\! & 0(0.01)\!\!\!\!\!\!\!\! & 1(0)\!\!\!\!\!\!\!\! & 0.94(0.11)\!\!\!\!\!\!\!\! & 0.8(0.17)\!\!\!\!\!\!\!\! & 0(0)\!\!\!\!\!\!\!\!\\                        
			\hline                                                                                                                                                                                                     
			$c=0.75\!\!\!\!\!\!\!\!\!\!$ &0.57(0.2)\!\!\!\!\!\!\!\! & 0.97(0.07)\!\!\!\!\!\!\!\! & 1(0.02)\!\!\!\!\!\!\!\! & 0(0.01)\!\!\!\!\!\!\!\! & 1(0)\!\!\!\!\!\!\!\! & 0.97(0.08)\!\!\!\!\!\!\!\! & 0.98(0.07)\!\!\!\!\!\!\!\! & 0(0)\!\!\!\!\!\!\!\! & 0.57(0.2)\!\!\!\!\!\!\!\! & 0.97(0.07)\!\!\!\!\!\!\!\! & 1(0.01)\!\!\!\!\!\!\!\! & 0(0.01)\!\!\!\!\!\!\!\! & 1(0)\!\!\!\!\!\!\!\! & 0.97(0.07)\!\!\!\!\!\!\!\! & 0.97(0.09)\!\!\!\!\!\!\!\! & 0(0)\!\!\!\!\!\!\!\!\\                         
			\hline                                                                                                                                                                                                     
			$c=1\!\!\!\!\!\!\!\!\!\!$ &0.2(0.26)\!\!\!\!\!\!\!\! & 0.99(0.05)\!\!\!\!\!\!\!\! & 1(0)\!\!\!\!\!\!\!\! & 0(0.01)\!\!\!\!\!\!\!\! & 1(0)\!\!\!\!\!\!\!\! & 0.99(0.05)\!\!\!\!\!\!\!\! & 1(0.01)\!\!\!\!\!\!\!\! & 0(0)\!\!\!\!\!\!\!\! & 0.24(0.24)\!\!\!\!\!\!\!\! & 0.98(0.06)\!\!\!\!\!\!\!\! & 1(0)\!\!\!\!\!\!\!\! & 0(0.01)\!\!\!\!\!\!\!\! & 1(0)\!\!\!\!\!\!\!\! & 0.98(0.06)\!\!\!\!\!\!\!\! & 1(0.03)\!\!\!\!\!\!\!\! & 0(0)\!\!\!\!\!\!\!\!\\                                       
			\hline                                                                                                                                                                                                     
			$p=50$ &&&&&&&&&&&&&&&&\\                                                                                                                                                                                   
			FOM $p$-dim&\multicolumn{4}{c}{Model 1} &\multicolumn{4}{c}{Model 2}&\multicolumn{4}{c}{Model 3}&\multicolumn{4}{c}{Model 4}\\                                                                          
			\hline                                                                                                                                                                                                     
			Cont. level & $p^m_c$ & $p^s_c$ &$p^j_c$ & $p_f$ &  $p^m_c$ & $p^s_c$ &$p^j_c$ & $p_f$ & $p^m_c$ & $p^s_c$ &$p^j_c$ & $p_f$ & $p^m_c$ & $p^s_c$ &$p^j_c$ & $p_f$ \\                                         
			\hline                                                                                                                                                                                                     
			$c=0\!\!\!\!\!\!\!\!\!\!$ &- &- &- & 0.01(0.01)\!\!\!\!\!\!\!\! &- &- &- &0.01(0.01)\!\!\!\!\!\!\!\! &- &- &- &0.01(0.02)\!\!\!\!\!\!\!\! &- &- &- &0(0.01)\!\!\!\!\!\!\!\!\\                                                                                                                   
			\hline                                                                                                                                                                                                     
			$c=0.25\!\!\!\!\!\!\!\!\!\!$ &0.8(0.29)\!\!\!\!\!\!\!\! & 0.06(0.11)\!\!\!\!\!\!\!\! & 0.04(0.09)\!\!\!\!\!\!\!\! & 0(0.01)\!\!\!\!\!\!\!\! & 0.81(0.28)\!\!\!\!\!\!\!\! & 0.06(0.11)\!\!\!\!\!\!\!\! & 0.03(0.08)\!\!\!\!\!\!\!\! & 0(0.02)\!\!\!\!\!\!\!\! & 0.78(0.32)\!\!\!\!\!\!\!\! & 0.06(0.11)\!\!\!\!\!\!\!\! & 0.04(0.09)\!\!\!\!\!\!\!\! & 0(0.01)\!\!\!\!\!\!\!\! & 0.8(0.29)\!\!\!\!\!\!\!\! & 0.06(0.13)\!\!\!\!\!\!\!\! & 0.05(0.11)\!\!\!\!\!\!\!\! & 0(0)\!\!\!\!\!\!\!\!\\    
			\hline                                                                                                                                                                                                     
			$c=0.5\!\!\!\!\!\!\!\!\!\!$ &0.85(0.26)\!\!\!\!\!\!\!\! & 0.08(0.14)\!\!\!\!\!\!\!\! & 0.07(0.13)\!\!\!\!\!\!\!\! & 0(0.01)\!\!\!\!\!\!\!\! & 0.87(0.25)\!\!\!\!\!\!\!\! & 0.08(0.13)\!\!\!\!\!\!\!\! & 0.05(0.11)\!\!\!\!\!\!\!\! & 0(0)\!\!\!\!\!\!\!\! & 0.83(0.27)\!\!\!\!\!\!\!\! & 0.08(0.13)\!\!\!\!\!\!\!\! & 0.08(0.14)\!\!\!\!\!\!\!\! & 0(0)\!\!\!\!\!\!\!\! & 0.86(0.27)\!\!\!\!\!\!\!\! & 0.08(0.13)\!\!\!\!\!\!\!\! & 0.06(0.13)\!\!\!\!\!\!\!\! & 0(0)\!\!\!\!\!\!\!\!\\         
			\hline                                                                                                                                                                                                     
			$c=0.75\!\!\!\!\!\!\!\!\!\!$ &0.8(0.3)\!\!\!\!\!\!\!\! & 0.07(0.13)\!\!\!\!\!\!\!\! & 0.11(0.17)\!\!\!\!\!\!\!\! & 0(0)\!\!\!\!\!\!\!\! & 0.86(0.26)\!\!\!\!\!\!\!\! & 0.08(0.14)\!\!\!\!\!\!\!\! & 0.09(0.16)\!\!\!\!\!\!\!\! & 0(0)\!\!\!\!\!\!\!\! & 0.8(0.28)\!\!\!\!\!\!\!\! & 0.07(0.14)\!\!\!\!\!\!\!\! & 0.07(0.13)\!\!\!\!\!\!\!\! & 0(0.01)\!\!\!\!\!\!\!\! & 0.8(0.3)\!\!\!\!\!\!\!\! & 0.06(0.12)\!\!\!\!\!\!\!\! & 0.03(0.09)\!\!\!\!\!\!\!\! & 0(0)\!\!\!\!\!\!\!\!\\             
			\hline                                                                                                                                                                                                     
			$c=1\!\!\!\!\!\!\!\!\!\!$ &0.01(0.04)\!\!\!\!\!\!\!\! & 0(0.02)\!\!\!\!\!\!\!\! & 0.19(0.23)\!\!\!\!\!\!\!\! & 0(0)\!\!\!\!\!\!\!\! & 0.16(0.35)\!\!\!\!\!\!\!\! & 0(0.02)\!\!\!\!\!\!\!\! & 0.16(0.2)\!\!\!\!\!\!\!\! & 0(0.01)\!\!\!\!\!\!\!\! & 0.01(0.07)\!\!\!\!\!\!\!\! & 0.01(0.03)\!\!\!\!\!\!\!\! & 0.02(0.07)\!\!\!\!\!\!\!\! & 0(0.01)\!\!\!\!\!\!\!\! & 0.05(0.17)\!\!\!\!\!\!\!\! & 0(0.03)\!\!\!\!\!\!\!\! & 0(0.03)\!\!\!\!\!\!\!\! & 0(0.01)\!\!\!\!\!\!\!\!\\                  
			\hline                                                                                                                                                                                                     
			$p=50$ &&&&&&&&&&&&&&&&\\                                                                                                                                                                                   
			FOM $1$-dim&\multicolumn{4}{c}{Model 1} &\multicolumn{4}{c}{Model 2}&\multicolumn{4}{c}{Model 3}&\multicolumn{4}{c}{Model 4}\\                                                                          
			\hline                                                                                                                                                                                                     
			Cont. level & $p^m_c$ & $p^s_c$ &$p^j_c$ & $p_f$ &  $p^m_c$ & $p^s_c$ &$p^j_c$ & $p_f$ & $p^m_c$ & $p^s_c$ &$p^j_c$ & $p_f$ & $p^m_c$ & $p^s_c$ &$p^j_c$ & $p_f$ \\                                         
			\hline                                                                                                                                                                                                     
			$c=0\!\!\!\!\!\!\!\!\!\!$ &- &- &- & 0.02(0.02)\!\!\!\!\!\!\!\! &- &- &- &0.02(0.02)\!\!\!\!\!\!\!\! &- &- &- &0.02(0.02)\!\!\!\!\!\!\!\! &- &- &- &0.02(0.02)\!\!\!\!\!\!\!\!\\                                                                                                                
			\hline                                                                                                                                                                                                     
			$c=0.25\!\!\!\!\!\!\!\!\!\!$ &1(0.04)\!\!\!\!\!\!\!\! & 0.01(0.04)\!\!\!\!\!\!\!\! & 0(0.01)\!\!\!\!\!\!\!\! & 0.01(0.01)\!\!\!\!\!\!\!\! & 1(0)\!\!\!\!\!\!\!\! & 0.01(0.04)\!\!\!\!\!\!\!\! & 0.01(0.03)\!\!\!\!\!\!\!\! & 0.01(0.01)\!\!\!\!\!\!\!\! & 1(0)\!\!\!\!\!\!\!\! & 0.01(0.04)\!\!\!\!\!\!\!\! & 0(0)\!\!\!\!\!\!\!\! & 0.01(0.01)\!\!\!\!\!\!\!\! & 1(0)\!\!\!\!\!\!\!\! & 0.01(0.03)\!\!\!\!\!\!\!\! & 0(0)\!\!\!\!\!\!\!\! & 0.01(0.01)\!\!\!\!\!\!\!\!\\                       
			\hline                                                                                                                                                                                                     
			$c=0.5\!\!\!\!\!\!\!\!\!\!$ &1(0)\!\!\!\!\!\!\!\! & 0.01(0.03)\!\!\!\!\!\!\!\! & 0(0)\!\!\!\!\!\!\!\! & 0.01(0.01)\!\!\!\!\!\!\!\! & 1(0)\!\!\!\!\!\!\!\! & 0(0.03)\!\!\!\!\!\!\!\! & 0(0.01)\!\!\!\!\!\!\!\! & 0.01(0.01)\!\!\!\!\!\!\!\! & 1(0)\!\!\!\!\!\!\!\! & 0.01(0.04)\!\!\!\!\!\!\!\! & 0(0)\!\!\!\!\!\!\!\! & 0.01(0.01)\!\!\!\!\!\!\!\! & 1(0)\!\!\!\!\!\!\!\! & 0(0.03)\!\!\!\!\!\!\!\! & 0(0)\!\!\!\!\!\!\!\! & 0.01(0.01)\!\!\!\!\!\!\!\!\\                                       
			\hline                                                                                                                                                                                                     
			$c=0.75\!\!\!\!\!\!\!\!\!\!$ &1(0)\!\!\!\!\!\!\!\! & 0.01(0.03)\!\!\!\!\!\!\!\! & 0(0)\!\!\!\!\!\!\!\! & 0(0.01)\!\!\!\!\!\!\!\! & 1(0)\!\!\!\!\!\!\!\! & 0.01(0.05)\!\!\!\!\!\!\!\! & 0(0)\!\!\!\!\!\!\!\! & 0.01(0.01)\!\!\!\!\!\!\!\! & 1(0)\!\!\!\!\!\!\!\! & 0.01(0.05)\!\!\!\!\!\!\!\! & 0(0)\!\!\!\!\!\!\!\! & 0.01(0.01)\!\!\!\!\!\!\!\! & 1(0)\!\!\!\!\!\!\!\! & 0.01(0.04)\!\!\!\!\!\!\!\! & 0(0)\!\!\!\!\!\!\!\! & 0.01(0.01)\!\!\!\!\!\!\!\!\\                                      
			\hline                                                                                                                                                                                                     
			$c=1\!\!\!\!\!\!\!\!\!\!$ &1(0)\!\!\!\!\!\!\!\! & 0.02(0.05)\!\!\!\!\!\!\!\! & 0(0)\!\!\!\!\!\!\!\! & 0(0.01)\!\!\!\!\!\!\!\! & 1(0)\!\!\!\!\!\!\!\! & 0.01(0.04)\!\!\!\!\!\!\!\! & 0(0)\!\!\!\!\!\!\!\! & 0(0.01)\!\!\!\!\!\!\!\! & 1(0)\!\!\!\!\!\!\!\! & 0.04(0.09)\!\!\!\!\!\!\!\! & 0(0)\!\!\!\!\!\!\!\! & 0.01(0.01)\!\!\!\!\!\!\!\! & 1(0)\!\!\!\!\!\!\!\! & 0.03(0.08)\!\!\!\!\!\!\!\! & 0(0)\!\!\!\!\!\!\!\! & 0.01(0.01)\!\!\!\!\!\!\!\!\\                                            
			\hline                                                                                                                                                                                                     
			$p=50$ &&&&&&&&&&&&&&&&\\                                                                                                                                                                                   
			MS-plot $p$-dim&\multicolumn{4}{c}{Model 1} &\multicolumn{4}{c}{Model 2}&\multicolumn{4}{c}{Model 3}&\multicolumn{4}{c}{Model 4}\\                                                                      
			\hline                                                                                                                                                                                                     
			Cont. level & $p^m_c$ & $p^s_c$ &$p^j_c$ & $p_f$ &  $p^m_c$ & $p^s_c$ &$p^j_c$ & $p_f$ & $p^m_c$ & $p^s_c$ &$p^j_c$ & $p_f$ & $p^m_c$ & $p^s_c$ &$p^j_c$ & $p_f$ \\                                         
			\hline                                                                                                                                                                                                     
			$c=0\!\!\!\!\!\!\!\!\!\!$ &- &- &- & 0(0)\!\!\!\!\!\!\!\! &- &- &- &0(0)\!\!\!\!\!\!\!\! &- &- &- &0(0)\!\!\!\!\!\!\!\! &- &- &- &0(0)\!\!\!\!\!\!\!\!\\                                                                                                                                        
			\hline                                                                                                                                                                                                     
			$c=0.25\!\!\!\!\!\!\!\!\!\!$ &1(0)\!\!\!\!\!\!\!\! & 0.06(0.1)\!\!\!\!\!\!\!\! & 0(0.02)\!\!\!\!\!\!\!\! & 0(0)\!\!\!\!\!\!\!\! & 1(0)\!\!\!\!\!\!\!\! & 0.08(0.12)\!\!\!\!\!\!\!\! & 0.01(0.04)\!\!\!\!\!\!\!\! & 0(0)\!\!\!\!\!\!\!\! & 1(0)\!\!\!\!\!\!\!\! & 0.06(0.11)\!\!\!\!\!\!\!\! & 0.01(0.03)\!\!\!\!\!\!\!\! & 0(0)\!\!\!\!\!\!\!\! & 1(0)\!\!\!\!\!\!\!\! & 0.02(0.06)\!\!\!\!\!\!\!\! & 0(0)\!\!\!\!\!\!\!\! & 0(0)\!\!\!\!\!\!\!\!\\                                             
			\hline                                                                                                                                                                                                     
			$c=0.5\!\!\!\!\!\!\!\!\!\!$ &1(0)\!\!\!\!\!\!\!\! & 0.12(0.15)\!\!\!\!\!\!\!\! & 0.08(0.12)\!\!\!\!\!\!\!\! & 0(0)\!\!\!\!\!\!\!\! & 1(0)\!\!\!\!\!\!\!\! & 0.16(0.16)\!\!\!\!\!\!\!\! & 0.08(0.12)\!\!\!\!\!\!\!\! & 0(0)\!\!\!\!\!\!\!\! & 1(0)\!\!\!\!\!\!\!\! & 0.12(0.14)\!\!\!\!\!\!\!\! & 0.15(0.18)\!\!\!\!\!\!\!\! & 0(0)\!\!\!\!\!\!\!\! & 1(0)\!\!\!\!\!\!\!\! & 0.4(0.36)\!\!\!\!\!\!\!\! & 0(0)\!\!\!\!\!\!\!\! & 0(0)\!\!\!\!\!\!\!\!\\                                           
			\hline                                                                                                                                                                                                     
			$c=0.75\!\!\!\!\!\!\!\!\!\!$ &1(0)\!\!\!\!\!\!\!\! & 0.07(0.11)\!\!\!\!\!\!\!\! & 0.2(0.21)\!\!\!\!\!\!\!\! & 0(0)\!\!\!\!\!\!\!\! & 1(0)\!\!\!\!\!\!\!\! & 0.05(0.1)\!\!\!\!\!\!\!\! & 0.2(0.21)\!\!\!\!\!\!\!\! & 0(0)\!\!\!\!\!\!\!\! & 1(0)\!\!\!\!\!\!\!\! & 0.05(0.1)\!\!\!\!\!\!\!\! & 0.37(0.22)\!\!\!\!\!\!\!\! & 0(0)\!\!\!\!\!\!\!\! & 1(0)\!\!\!\!\!\!\!\! & 0.93(0.2)\!\!\!\!\!\!\!\! & 0(0.01)\!\!\!\!\!\!\!\! & 0(0)\!\!\!\!\!\!\!\!\\                                           
			\hline                                                                                                                                                                                                     
			$c=1\!\!\!\!\!\!\!\!\!\!$ &0(0)\!\!\!\!\!\!\!\! & 0(0)\!\!\!\!\!\!\!\! & 0.72(0.28)\!\!\!\!\!\!\!\! & 0(0)\!\!\!\!\!\!\!\! & 1(0)\!\!\!\!\!\!\!\! & 0(0)\!\!\!\!\!\!\!\! & 0.45(0.3)\!\!\!\!\!\!\!\! & 0(0)\!\!\!\!\!\!\!\! & 0(0)\!\!\!\!\!\!\!\! & 0(0)\!\!\!\!\!\!\!\! & 0.68(0.25)\!\!\!\!\!\!\!\! & 0(0)\!\!\!\!\!\!\!\! & 1(0)\!\!\!\!\!\!\!\! & 1(0)\!\!\!\!\!\!\!\! & 0(0.02)\!\!\!\!\!\!\!\! & 0(0)\!\!\!\!\!\!\!\!\\                                                                  
			\hline                                                                                                                                                                                                     
			$p=50$ &&&&&&&&&&&&&&&&\\                                                                                                                                                                                   
			MS-plot $1$-dim&\multicolumn{4}{c}{Model 1} &\multicolumn{4}{c}{Model 2}&\multicolumn{4}{c}{Model 3}&\multicolumn{4}{c}{Model 4}\\                                                                      
			\hline                                                                                                                                                                                                     
			Cont. level & $p^m_c$ & $p^s_c$ &$p^j_c$ & $p_f$ &  $p^m_c$ & $p^s_c$ &$p^j_c$ & $p_f$ & $p^m_c$ & $p^s_c$ &$p^j_c$ & $p_f$ & $p^m_c$ & $p^s_c$ &$p^j_c$ & $p_f$ \\                                         
			\hline                                                                                                                                                                                                     
			$c=0\!\!\!\!\!\!\!\!\!\!$ &- &- &- & 0(0)\!\!\!\!\!\!\!\! &- &- &- &0.13(0.06)\!\!\!\!\!\!\!\! &- &- &- &0(0.01)\!\!\!\!\!\!\!\! &- &- &- &0.13(0.05)\!\!\!\!\!\!\!\!\\                                                                                                                         
			\hline                                                                                                                                                                                                     
			$c=0.25\!\!\!\!\!\!\!\!\!\!$ &1(0)\!\!\!\!\!\!\!\! & 1(0)\!\!\!\!\!\!\!\! & 0.82(0.17)\!\!\!\!\!\!\!\! & 0(0)\!\!\!\!\!\!\!\! & 1(0)\!\!\!\!\!\!\!\! & 0.14(0.17)\!\!\!\!\!\!\!\! & 0.07(0.12)\!\!\!\!\!\!\!\! & 0.09(0.04)\!\!\!\!\!\!\!\! & 1(0)\!\!\!\!\!\!\!\! & 1(0)\!\!\!\!\!\!\!\! & 0.91(0.13)\!\!\!\!\!\!\!\! & 0(0)\!\!\!\!\!\!\!\! & 1(0)\!\!\!\!\!\!\!\! & 0.12(0.16)\!\!\!\!\!\!\!\! & 0.33(0.27)\!\!\!\!\!\!\!\! & 0.09(0.04)\!\!\!\!\!\!\!\!\\                                   
			\hline                                                                                                                                                                                                     
			$c=0.5\!\!\!\!\!\!\!\!\!\!$ &1(0)\!\!\!\!\!\!\!\! & 1(0)\!\!\!\!\!\!\!\! & 0.99(0.03)\!\!\!\!\!\!\!\! & 0(0)\!\!\!\!\!\!\!\! & 1(0)\!\!\!\!\!\!\!\! & 0.18(0.2)\!\!\!\!\!\!\!\! & 0.06(0.12)\!\!\!\!\!\!\!\! & 0.08(0.04)\!\!\!\!\!\!\!\! & 1(0)\!\!\!\!\!\!\!\! & 1(0)\!\!\!\!\!\!\!\! & 0.99(0.04)\!\!\!\!\!\!\!\! & 0(0)\!\!\!\!\!\!\!\! & 1(0)\!\!\!\!\!\!\!\! & 0.17(0.19)\!\!\!\!\!\!\!\! & 0.91(0.17)\!\!\!\!\!\!\!\! & 0.08(0.04)\!\!\!\!\!\!\!\!\\                                     
			\hline                                                                                                                                                                                                     
			$c=0.75\!\!\!\!\!\!\!\!\!\!$ &1(0)\!\!\!\!\!\!\!\! & 1(0)\!\!\!\!\!\!\!\! & 1(0)\!\!\!\!\!\!\!\! & 0(0)\!\!\!\!\!\!\!\! & 1(0)\!\!\!\!\!\!\!\! & 0.21(0.22)\!\!\!\!\!\!\!\! & 0.05(0.1)\!\!\!\!\!\!\!\! & 0.07(0.04)\!\!\!\!\!\!\!\! & 1(0)\!\!\!\!\!\!\!\! & 1(0)\!\!\!\!\!\!\!\! & 1(0.01)\!\!\!\!\!\!\!\! & 0(0)\!\!\!\!\!\!\!\! & 1(0)\!\!\!\!\!\!\!\! & 0.25(0.25)\!\!\!\!\!\!\!\! & 1(0.03)\!\!\!\!\!\!\!\! & 0.09(0.05)\!\!\!\!\!\!\!\!\\                                                
			\hline                                                                                                                                                                                                     
			$c=1\!\!\!\!\!\!\!\!\!\!$ &1(0)\!\!\!\!\!\!\!\! & 1(0)\!\!\!\!\!\!\!\! & 1(0)\!\!\!\!\!\!\!\! & 0(0)\!\!\!\!\!\!\!\! & 1(0)\!\!\!\!\!\!\!\! & 0.25(0.25)\!\!\!\!\!\!\!\! & 0.04(0.09)\!\!\!\!\!\!\!\! & 0.07(0.04)\!\!\!\!\!\!\!\! & 1(0)\!\!\!\!\!\!\!\! & 1(0)\!\!\!\!\!\!\!\! & 1(0)\!\!\!\!\!\!\!\! & 0(0)\!\!\!\!\!\!\!\! & 1(0)\!\!\!\!\!\!\!\! & 0.33(0.29)\!\!\!\!\!\!\!\! & 1(0)\!\!\!\!\!\!\!\! & 0.09(0.04)\!\!\!\!\!\!\!\!\\                                                        
			\hline     
	\end{tabular}}
\end{sidewaystable}

\newpage
\section{Appendix: Computational complexity}
Regarding the alternative methods considered for the low dimensional simulation study (section 3.1), we present here a comparative analysis of the computation times of these procedures. Experiments have been carried out in R (version 3.4.4) in an Intel(R) Xeon(R) CPU E5-1650 v3 (x64) @ 3.50GHz with 128GiB of RAM under Windows 10.

\begin{figure}[h!]
	\includegraphics[width=16cm]{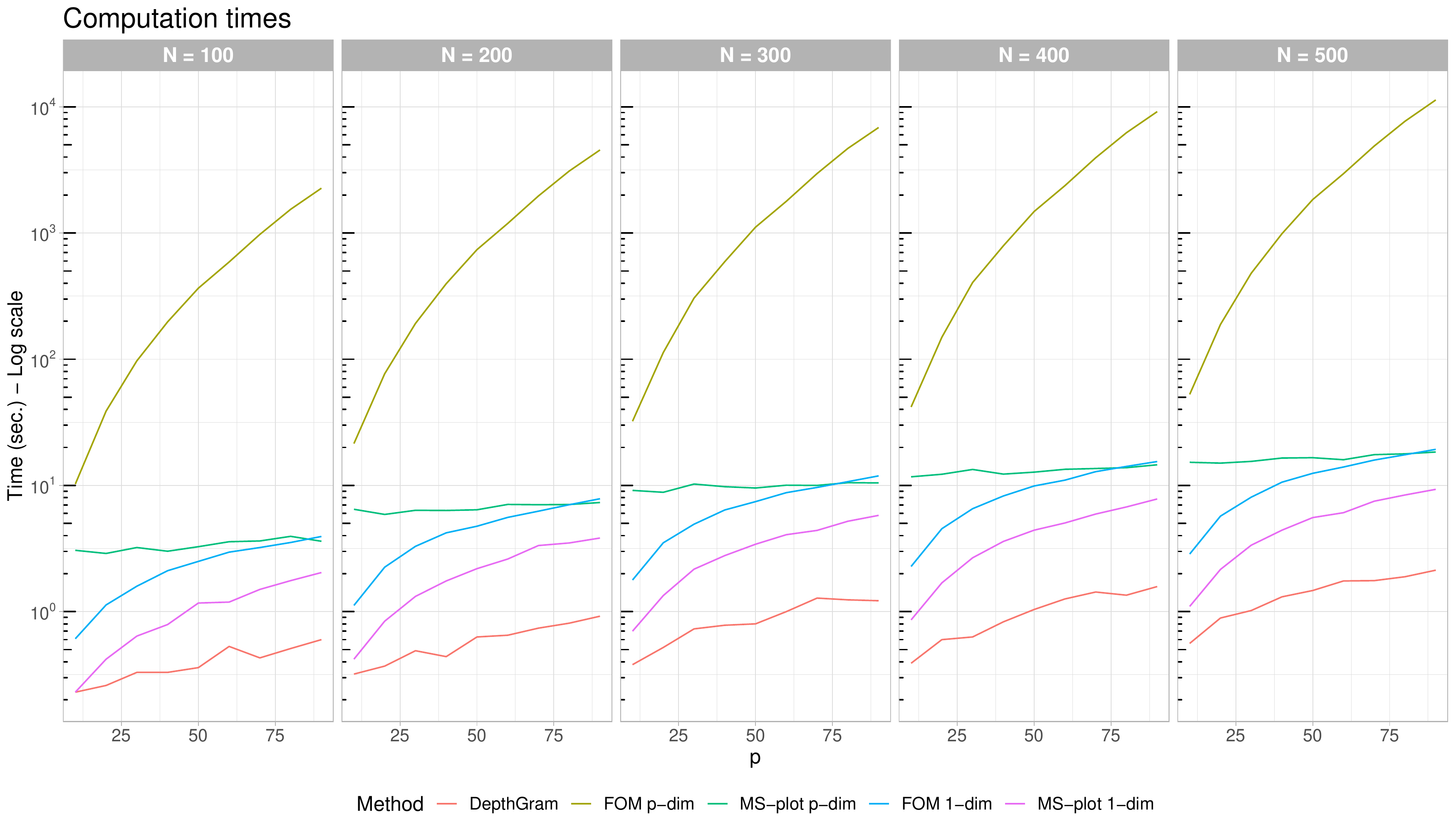}
	\caption{Time performance for the different algorithms on multivariate functional data sets of varying dimensions $p$, with $n=100$ observations and $N$ observation points.}\label{time1}
\end{figure}

\begin{figure}[h!]
	\includegraphics[width=16cm]{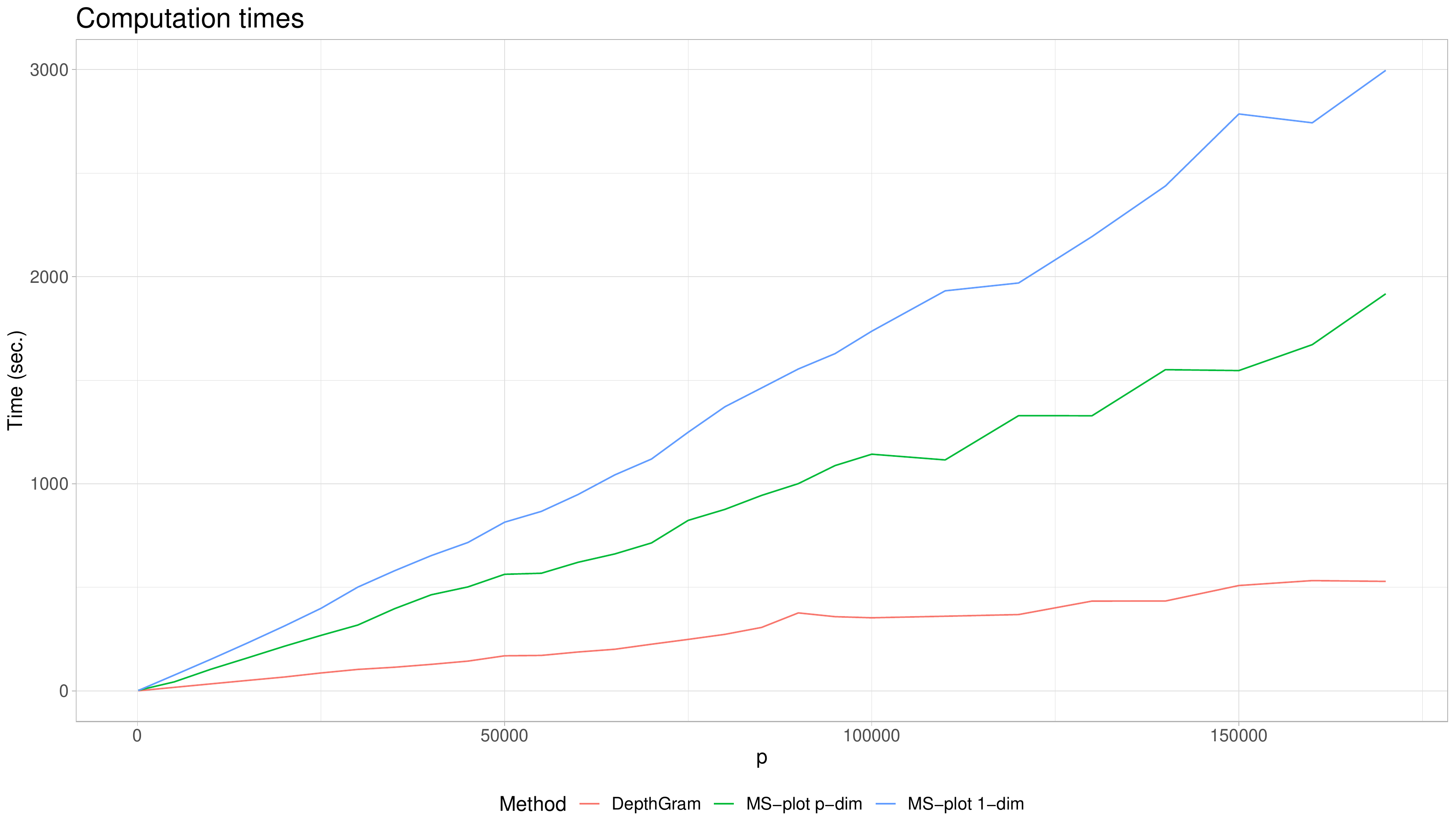}
	\caption{Time performance for the DepthGram and MS-plot on multivariate functional data sets of varying dimensions $p$, with $n=100$ observations and $N=100$ observation points.}\label{time2}
\end{figure}

For the DepthGram implementation, the code is provided (\texttt{DepthGram.R}), whereas for the FOM the R package \texttt{mrfDepth} has been used, and for the MS-plot the code is the one provided by the authors in the supplementary materials of their paper \url{https://www.tandfonline.com/doi/ref/10.1080/10618600.2018.1473781}. All the methods have been used with the options to obtain the limits of the non-outlying regions disabled. That is, only the times required to compute the two variables used in each of the two-dimensional representations are compared. FOM is used with the \emph{functional directional outlyingness} measure (\emph{fDO}) and MS-plot with the one based on the random projection depth for multivariate data (for the $p$-dimensional version). For the exact settings used, check the file \texttt{Computation\_times.R} that allows to reproduce the analysis whose results are summarized here.
In a first analysis, all five methods are compared in a low dimensional setting, restricted to $p<n$ so that the FOM representation can be obtained. In a second analysis, a comparison of the two versions of the MS-plot and the DepthGram is established in higher dimensions. Results are represented in Figures \ref{time1} and \ref{time2} where we can see how the $p$-dimensional version of FOM is computationally very heavy compared to the rest of the methods and how the DepthGram exhibits the best performance with a computation time around $3.5$ times faster than the $p$-dimensional version of the MS-plot and $6$ times faster than its $1$-dimensional version.

\end{document}